\documentclass[12pt]{article}
\usepackage{epsfig,graphicx}
\usepackage{amsmath,amsfonts,amssymb}
\usepackage{color,ulem}
\newcommand{\unit}{\leavevmode\hbox{\small1\kern-3.6pt\normalsize1}}

\def\mw{{M_W}}

\def\chsnsn{C_{\higgsi\snr\snr}}
\def\chisnsn{C_{\higgsi\snr\snr}}
\def\chjsnsn{C_{\higgsj\snr\snr}}

\def\chlsnsn{C_{\higgsl\snr\snr}}

\def\chaa{C_{\phiggsi\phiggsj\higgsk}}

\def\caasnsn{C_{\phiggsi\phiggsj\snr\snr}}

\def\cnnhi{C_{NN\higgsi}}

\def\cnnhl{C_{NN\higgsl}}

\def\csnnneui{C_{\snr N\neuti}}

\def\l{\lambda}
\def\k{\kappa}
\def\vevs{v_s}
\def\al{{A_\lambda}}
\def\ak{{A_\kappa}}

\def\ln{{\lambda_N}}
\def\aln{{A_{\lambda_N}}}
\def\mn{{m_{\tilde{N}}}}

\def\neuti{{{\tilde\chi}_{i}}}

\def\neutmass{{m_{{\tilde\chi}_{1}^0}}}

\def\neutcomps{{N^{\tilde\chi}_{15}}}

\def\neutcomps{{N^{5}_{\tilde\chi_1}}}

\def\snr{{\tilde N}}

\def\snmassr{{m_{\tilde N_1}}}
\def\snmassrsq{{m^2_{\tilde N_1}}}
\def\snmassrqb{{m^3_{\tilde N_1}}}

\def\rhn{{N}}
\def\rhnmass{{M_{N}}}

\def\higgsi{{H_i^0}}
\def\higgsj{{H_j^0}}
\def\higgsk{{H_k^0}}
\def\higgsl{{H_1^0}}

\def\hmassi{m_{H_i^0}}

\def\hmassl{m_{H_1^0}}
\def\hmassm{m_{H_2^0}}
\def\hmassh{m_{H_3^0}}
\def\hmasslsq{m^2_{H_1^0}}

\def\hmassisq{m^2_{H_i^0}}

\def\hcompld{S_{H_1^0}^1}
\def\hcomplu{S_{H_1^0}^2}
\def\hcompls{S_{H_1^0}^3}

\def\hcompid{S_{H_i^0}^1}
\def\hcompiu{S_{H_i^0}^2}
\def\hcompis{S_{H_i^0}^3}

\def\hcompjd{S_{H_j^0}^1}
\def\hcompju{S_{H_j^0}^2}

\def\phiggsi{{A_a^0}}
\def\phiggsj{{A_b^0}}

\def\phiggsl{{A_1^0}}

\def\phmassl{{m_{A_1^0}}}
\def\phmassh{{m_{A_2^0}}}

\newcommand{\captions}{\sf\caption}

\newcommand{\crosssec}{\sigma_{\snr p}^{\rm SI}}
\newcommand{\neutcrosssec}{\sigma_{\tilde\chi p}^{\rm SI}}
\newcommand{\nmh}{{\tt NMHDECAY}}
\newcommand{\snrelic}{{\Omega_\snr h^2}}

\newcommand{\bsg}{b\to s\gamma}
\newcommand{\bmumu}{B_S\to\mu^+\mu^-}
\newcommand{\asusy}{a^{\rm SUSY}_\mu}

\def\lsim{\raise0.3ex\hbox{$\;<$\kern-0.75em\raise-1.1ex\hbox{$\sim\;$}}}
\def\gsim{\raise0.3ex\hbox{$\;>$\kern-0.75em\raise-1.1ex\hbox{$\sim\;$}}}

\renewcommand{\thefootnote}{\fnsymbol{footnote}}
\allowdisplaybreaks

\begin{document}

\thispagestyle{empty}
\begin{flushright}
  FTUAM 11/54\\
  IFT-UAM/CSIC-11-58\\
  HGU-CAP 12\\

  \vspace*{2.mm}{4 August 2011}
\end{flushright}

\begin{center}
  {\Large \textbf{Very light right-handed sneutrino dark matter in the NMSSM
  } }  
  
  \vspace{0.5cm}
  David G.~Cerde\~no${}^{1}$,
  Ji-Haeng Huh\footnote{MultiDark Fellow}${}^{1}$,
  Miguel Peir\'o${}^{1}$,
  Osamu Seto${}^2$ \\[0.2cm] 
    
  {\textit{ ${}^1$ Departamento de F\'{\i}sica Te\'{o}rica,
      \& 
      Instituto de F\'{\i}sica Te\'{o}rica
      UAM/CSIC, \\[0pt] 
      Universidad Aut\'{o}noma de Madrid, 
      Cantoblanco, E-28049
      Madrid, Spain\\[0pt] 
      ${}^2$ 
      Department of Architecture and Building Engineering,
      Hokkai-Gakuen University,\\[0pt]
      Sapporo 062-8605, Japan
  }}
  
\vspace*{0.3cm}
\begin{abstract}
Very light right-handed (RH) sneutrinos in the Next-to-Minimal Supersymmetric Standard Model can be viable candidates for cold dark matter. 
We investigate the prospects for their direct detection, addressing their compatibility with the recent signal observed by the CoGeNT detector, and study the implications for Higgs phenomenology.
We find that in order to reproduce the correct relic abundance very light RH sneutrinos can annihilate into either a fermion-antifermion pair, very light pseudoscalar Higgses or RH neutrinos. 
If the main annihilation channel is into fermions, we point out that RH sneutrinos could naturally account for the CoGeNT signal. Furthermore, the lightest Higgs has a very large invisible decay width, and in some cases the second-lightest Higgs too.
On the other hand, if the RH sneutrino annihilates mostly into pseudoscalars or RH neutrinos the predictions for direct detection are below the current experimental sensitivities and satisfy the constraints set by CDMS and XENON. 
We also calculate the gamma ray flux from RH sneutrino annihilation in the
Galactic centre,
including as an interesting new possibility RH neutrinos in the final state. These are produced through a resonance with the Higgs and the resulting flux can exhibit a significant Breit-Wigner enhancement. 
\end{abstract}
\end{center}

	\newpage
	
\renewcommand{\thefootnote}{\arabic{footnote}}
\setcounter{footnote}{0}

\section{Introduction}
\label{sec:introduction}

Very light weakly-interacting massive particles (WIMPs) are currently receiving much attention as a potential solution to the dark matter problem. This has been motivated by some recent experimental results in direct detection experiments that might favour them over more conventional scenarios with heavier candidates.

In their search for the elastic scattering of dark matter particles, the CoGeNT collaboration observed an irreducible excess in their data \cite{Aalseth:2010vx} that, if interpreted in terms of WIMPs, would correspond to a very light particle, with mass in the range $7-12$~GeV, and a large elastic scattering cross section, of order $10^{-4}$~pb.
Furthermore, hints of an annual modulation in the CoGeNT experiment have also been observed \cite{Aalseth:2011wp}, which are also consistent with their previous results and narrow the range of WIMP masses down to $7-9$~GeV. Although, such a particle has similar properties to the candidate suggested to account for the annual modulation signal reported by the DAMA/LIBRA collaboration \cite{Bernabei:2003za,Bernabei:2008yi}, a joint explanation of both experimental results is not possible unless extreme assumptions are made for the different uncertainties \cite{Schwetz:2011xm,Hooper:2011hd,Farina:2011pw,McCabe:2011sr} such as the inclusion of large quenching factors or channeling effects.

These observations are, however, challenged by the negative results obtained in searches by other experimental collaborations. Most notably, CDMS \cite{Ahmed:2009zw}, XENON10 \cite{Angle:2011th}, XENON100 \cite{Aprile:2011hi} and recently, SIMPLE \cite{Felizardo:2011uw} have set upper bounds on the spin-independent part of the WIMP-proton cross section that are in tension with the regions of the parameter space compatible with DAMA/LIBRA and CoGeNT. 
The compatibility between these experimental results was tested
in Ref.\,\cite{Arina:2011si} using Bayesian statistical methods.
It was claimed that when uncertainties in the scintillation
efficiency of XENON100 are taken into account, the resulting
exclusion limit is not sufficient to rule out the CoGeNT region
(see also Ref.\,\cite{Collar:2011wq}). Also, from a theoretical
point of view, an unconventional dark matter candidate, coupling
differently to protons and neutrons, could account for the
CoGeNT signal while having escaped detection in XENON
\cite{Feng:2011vu,Frandsen:2011ts}.
Regarding the comparison of CoGeNT with CDMS data, there is no consensus between both collaborations on to which extent their spectra for low-energy events observed in both experiments disagree, see in this respect Refs.\,\cite{Ahmed:2010wy} and \cite{Collar:2011kf}. Although channeling effects in the CoGeNT crystals could help reconciling both results, it is not clear if this effect can be large enough \cite{Bozorgnia:2010ax}.

Various theoretical constructions with very light WIMP dark matter have been proposed in the literature. For example, in the case of supersymmetric models, it was realised that very light neutralinos were viable in the Minimal Supersymmetric Standard Model (MSSM) both in an effective low-energy description \cite{Hooper:2002nq,Bottino:2002ry,Bottino:2003cz,Bottino:2003iu,Belanger:2003wb} as well as with parameters described at the Grand Unification scale (see e.g., \cite{Cerdeno:2004zj}), provided the mass terms for both scalars and gauginos were non-universal. 
These neutralinos can lie within the DAMA/LIBRA region \cite{Bottino:2002ry,Bottino:2003cz,Bottino:2008mf,Fornengo:2010mk} when the correct relic density is required. However, this possibility relies on the use of light scalar and pseudoscalar Higgses in order to enhance the neutralino annihilation cross section, and a choice of parameters that leads to sizable contributions to low-energy observables.
In particular, the branching ratios of some rare decays (mainly $\bsg$ and $\bmumu$) impose very stringent constraints that make these solutions very fine-tuned \cite{Feldman:2010ke}.
Furthermore, large values of the ratio of the Higgs vacuum
expectation values ($\tan\beta$) are becoming increasingly
constrained by the recent results from the LHC
\cite{Chatrchyan:2011nx,Aad:2011rv}.

This problem can be in principle alleviated when extensions of the MSSM are considered. In particular, in the Next-to-MSSM, where an extra singlet field is included in order to provide a $\mu$ parameter of order of the Electroweak scale, the contribution to low-energy observables can be reduced.
Very light neutralinos are also possible within the NMSSM
\cite{Gunion:2005rw}. An increase in their annihilation cross
section can be obtained in the presence of either a very light
scalar or pseudoscalar Higgs. These can be viable if their singlet composition is large enough. 
The predictions for the neutralino-nucleus scattering cross section, which
are very sensitive to changes in the Higgs sector
\cite{Cerdeno:2004xw,Cerdeno:2007sn}, vary significantly and span several
orders of magnitude in the small mass region \cite{Aalseth:2008rx}. Recent
analyses of this scenario have been made
\cite{Draper:2010ew,Belikov:2010yi,Vasquez:2010ru,Cao:2011re,Vasquez:2011js,Cumberbatch:2011jp} in which the role of experimental constraints and the naturalness of the parameters have been thoroughly addressed in the light of the CoGeNT results.

The neutralino is not the only viable WIMP in supersymmetric theories. Another interesting possibility is the sneutrino \cite{Ibanez}. The sneutrino is excluded as dark matter in the MSSM, where it is a left-handed (LH) field, due to its large annihilation cross section and excessive scattering cross section off nuclei \cite{Falk:1994es}. Nevertheless, it can be viable in extended models where its coupling to the $Z$ boson is reduced. 
This can be achieved by introducing a right-handed (RH)
sneutrino superfield (which also entails the inclusion of a RH
neutrino and thus the possibility of having a see-saw mechanism
that accounts for the smallness of neutrino masses). The
sneutrino can thus be a mixed LH-RH field
\cite{ArkaniHamed:2000bq,Borzumati:2000mc,Hooper:2004dc,valle}
and reproduce the correct relic density, even if it is very light
\cite{Belanger:2010cd}. However this requires
a particular supersymmetry-breaking scheme with very large
trilinear terms which is not available with the standard
supergravity mediated paradigm. Another possibility is having a
pure RH sneutrino, which is generically non-thermal due to its
extremely small Yukawa coupling
\cite{Asaka:2005cn,Gopalakrishna:2006kr,McDonald:2006if,Page:2007sh}
unless it is somehow coupled to the observable sector, for
example via an extension of the gauge
\cite{Lee:2007mt,Allahverdi:2007wt,Bandyopadhyay:2011qm,Gao:2011ka}
or Higgs \cite{pilaftsis,Cerdeno:2008ep,Deppisch:2008bp,Cerdeno:2009dv} sectors. 

In this paper we will work with an extension of the MSSM that was presented in Ref.\,\cite{Cerdeno:2008ep} and in which two new singlet superfields were included, as in Refs.\,\cite{ko99,pilaftsis}. 
An extra singlet superfield $S$ addresses the $\mu$ problem in the same way as in the NMSSM and provides extra Higgs and neutralino states, while an extra singlet superfield $N$ accounts for RH neutrino and sneutrino states. The phenomenology of this construction was studied in \cite{Cerdeno:2009dv}, where the possibility for very light RH sneutrinos as candidates for cold dark matter was already pointed out. 

We will further investigate here the viability of very light RH sneutrinos in the light of the current experimental situation. In particular, in Section\,\ref{sec:verylight} we will review the conditions under which the correct sneutrino relic density is achieved. 
We find three possible scenarios, where RH sneutrinos annihilate mainly in a fermion-antifermion pair, very light pseudoscalar Higgses or RH neutrinos. 
We explore these annihilation channels and investigate the available parameter space in each case.
Then, in Section\,\ref{sec:direct} we compute the spin-independent part of the RH sneutrino-nucleon elastic scattering cross section. 
In order to contemplate all the possibilities, we will consider cases where the RH sneutrinos can have an elastic cross section off nuclei which is large enough to provide a WIMP interpretation of the CoGeNT signal but we will also study other scenarios in which the predictions are below the current exclusion limits set by CDMS and XENON.
In an attempt to find discriminating features of this scenario that distinguishes it from the case of very light neutralinos we will study some possible collider implications in Section\,\ref{sec:colliders}, computing the invisible decay width of the lightest and second-lightest CP-even Higgs bosons. 
After that, in Section\,\ref{sec:indirect} we compute the gamma ray flux
from RH sneutrino annihilation in the Galactic Centre (GC) for each of the
scenarios, and compare it to the current data by the Fermi satellite. 
As a novel feature, the channel with RH neutrinos in the final state is included in the analysis.
Finally, our conclusions will be presented in Section\,\ref{sec:conclusions}.


\section{Very light RH sneutrinos in the NMSSM}
\label{sec:verylight}

The details of the model with RH sneutrinos in the NMSSM were already introduced in Ref.~\cite{Cerdeno:2008ep} and  the phenomenology of this supersymmetric WIMP was studied in more detail by some of us in Ref.~\cite{Cerdeno:2009dv}.
In particular, we showed that the correct relic density could be obtained for a wide range of sneutrino masses and in extensive regions of the NMSSM parameter space. 

The superpotential of this construction is expressed as
\begin{eqnarray}
  W &=& W_{\rm NMSSM} + \lambda_N S N N + y_N L \cdot H_2 N,  \nonumber\\
  W_{\rm NMSSM} &=& Y_u H_2 \cdot Q u + Y_d H_1 \cdot Q d 
  + Y_e H_1 \cdot L e
  -\lambda S H_1 \cdot H_2 + \frac{1}{3}\kappa S^3 ,
  \label{superpotential}
\end{eqnarray}
where flavour indices are omitted and the dot denotes the $SU(2)_L$
antisymmetric product. As in the NMSSM, a global $Z_3$ symmetry is
imposed for each superfield, so that there are no supersymmetric mass
terms in the superpotential.  
After radiative Electroweak symmetry-breaking the Higgs fields get non-vanishing vacuum expectation values (VEV). The VEV of the singlet Higgs, $\vevs$, induces an effective $\mu$ parameter, $\mu=\lambda\vevs$, and a Majorana mass for RH neutrinos, $\rhnmass=2\ln\vevs$, both of order of the Electroweak scale.
The phenomenology of this model is largely dependent on the $SNN$ coupling and thus on the new parameter $\ln$.

The Lagrangian, with the corresponding soft-supersymmetry breaking terms reads
\begin{eqnarray}
  -{\cal L}_{\rm scalar \, mass}
  &=& m_{\tilde{Q}}^2 |\tilde{Q}|^2 + m_{\tilde{u}}^2 |\tilde{u}|^2 
  + m_{\tilde{d}}^2 |\tilde{d}|^2 + m_{\tilde{L}}^2 |\tilde{L}|^2 
  + m_{\tilde{e}}^2 |\tilde{e}|^2 \nonumber \\
  && + m_{H_1}^2 |H_1|^2 + m_{H_2}^2 |H_2|^2 + m_S^2 |S|^2 
  + m_{\tilde{N}}^2 |\tilde{N}|^2  ,
  \label{lagrangian_masses}\\
-{\cal L}_{\rm A-terms} &=&  
\left(A_u Y_u H_2 \tilde{Q}\tilde{u} + A_d Y_d H_1 \tilde{Q} \tilde{d} 
  + A_e Y_e H_1 \tilde{L}\tilde{e} + {\rm H.c.} \right) \nonumber \\
&& + \left(-\lambda A_{\lambda} S H_1 H_2 + \frac{1}{3}\kappa A_{\kappa} S^3
  + {\rm H.c.} \right) \nonumber \\
&& + \left( \lambda_N A_{\lambda_N} S \tilde{N}^2 
  + y_N A_{y_N} \tilde{L} H_2 \tilde{N}+ {\rm H.c.}  \right) ,
  \label{lagrangian_couplings}
\end{eqnarray}
and contains a soft mass term for the RH sneutrino, $m_{\tilde N}$, and new trilinear soft terms $\aln$ and $A_{y_N}$. 

The smallness of the neutrino Yukawa coupling implies that the sneutrino mass eigenstates have a negligible mixing and therefore can be identified with the LH and RH components. 
The lightest sneutrino is then a pure RH field and its mass can be expressed in terms of the NMSSM parameters as follows
\begin{equation}
	\snmassrsq= m_{\tilde{N}}^2 +|2\lambda_N \vevs|^2  + |y_N v_2|^2 
	+ 2 \lambda_N \left( A_{\lambda_N} \vevs+
	(\kappa \vevs^2-\lambda v_1 v_2 )^{\dagger} \right) .
	\label{eq:snmass}
\end{equation}

\begin{figure}[t!]
	\begin{center}  
	\epsfig{file=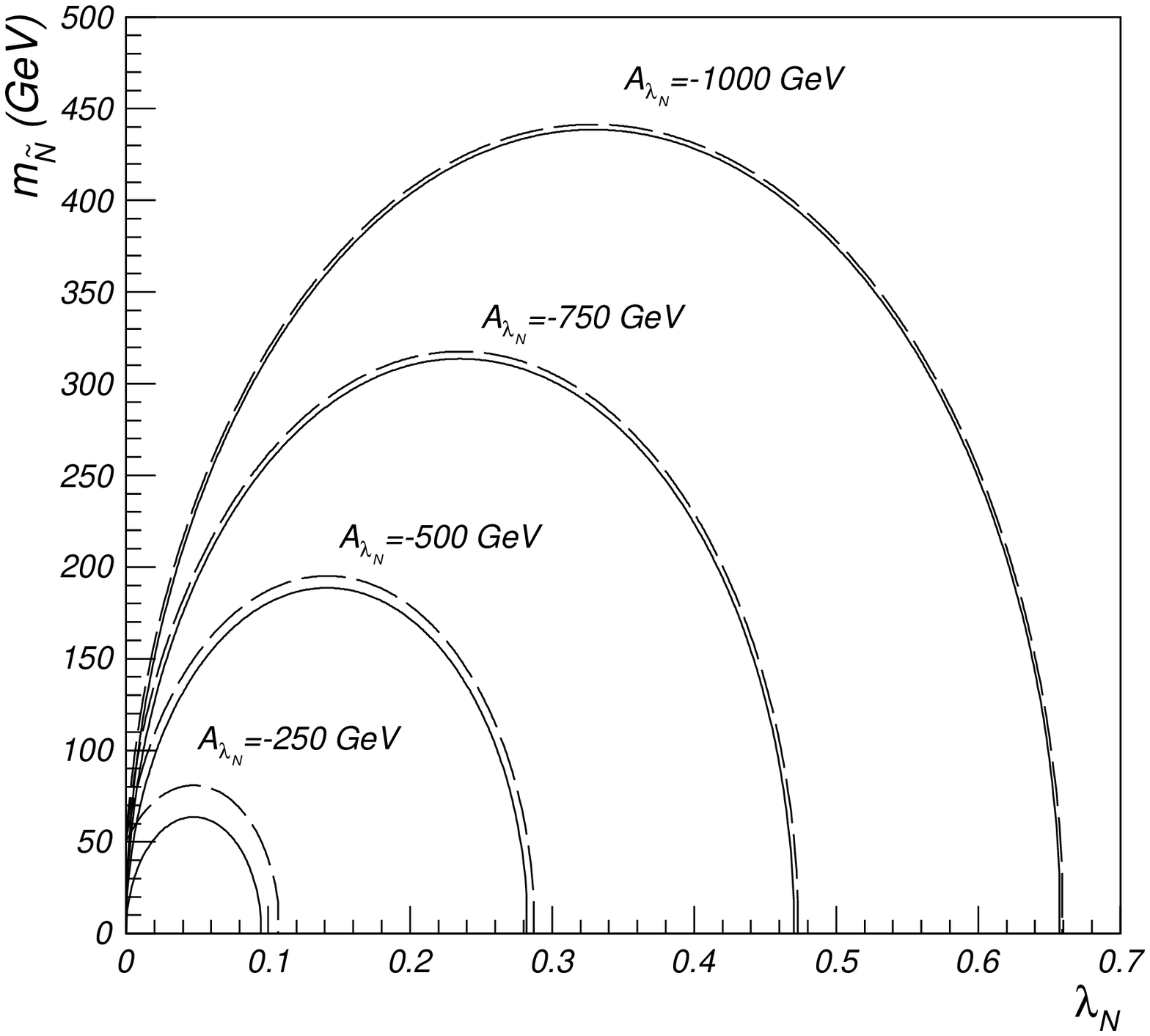,width=10.cm}
	\end{center}
  \captions{Trajectories in the $(\mn,\ln)$ plane with fixed RH sneutrino mass, given various values of $\aln$. For each choice of $\aln$ the dashed line represents the trajectory along which $\snmassr=50$~GeV and the solid one corresponds to $\snmassr=0$. We have used $\tan\beta=5$, $\l=0.3$, $\k=0.2$, and $\mu=200$~GeV.
  }
  \label{fig:spec}
\end{figure}

The flexibility of this construction stems from the fact that the new free parameters $\left\{\ln,\,m_{\tilde N},\,\aln\right\}$ can be chosen in such a way that they provide a wide range of RH sneutrino masses and couplings while on the other hand not affecting the rest of the NMSSM spectrum. 
This is illustrated in Fig.\,\ref{fig:spec}, where we show the trajectories for a fixed sneutrino mass in the $(\mn,\ln)$ plane for various choices of the trilinear parameter. 
We observe that very small RH sneutrino masses can be obtained for any choice of the soft mass parameter (for example, increasing  $|\aln|$ larger values of $\ln$ are possible). 
This is potentially interesting, since it suggests that no large non-universalities in the soft parameters are needed in order to have a very light RH sneutrino (contrary to what happens with very light neutralinos, where the bino mass parameter has to be significantly reduced). For concreteness, the soft masses for sleptons are fixed to $m_{L,R}=250$~GeV, squark masses are set to $m_{Q,U,D}=1$~TeV or 1.3~TeV, depending on the specific example, although our conclusions regarding dark matter are not sensitive to these choices. We also adopt the GUT relation for gaugino masses at low energy. 

In order for very light RH sneutrinos to be viable dark matter candidates they have to reproduce the correct value for their thermal relic abundance, $\snrelic$. 
In \cite{Cerdeno:2009dv} some of us showed that there were three distinct scenarios in which sneutrinos with masses below $~10$~GeV could be in agreement with the constraint set (at the $2\,\sigma$ level) by the WMAP satellite $0.1008\le\snrelic\le0.1232$ \cite{Komatsu:2010fb}.
The three cases correspond to RH sneutrinos annihilating preferentially either in fermions ($\tilde N \tilde N \to f\bar f$, mainly into $b\bar b$), or in a pair of very light pseudoscalar Higgses ($\tilde N \tilde N \to A^0_1 A^0_1$), or in RH neutrinos ($\tilde N \tilde N \to NN$).

In our calculations the analysis of the NMSSM phenomenology has been performed with the {\tt NMHDECAY\,2.3.7} code \cite{Ellwanger:2005dv,nmh2}, which minimises the scalar potential, dismissing the pressence of tachyons and/or false minima, and computes the Higgs boson masses including 1- and 2-loop radiative corrections, as well as the rest of the supersymmetric masses. It also implements the different collider constraints that apply to the Higgs sector. Based on this code, we have built a set of routines which numerically calculate the RH sneutrino spectrum and relic density, which is then compared to the observational bound extracted from the WMAP results. For details of our calculation we refer the reader to Ref.\,\cite{Cerdeno:2009dv}, where the amplitudes for RH sneutrino annihilation are explicitly computed for each possible channel.
Experimental constraints on low energy observables are also taken into account. In particular we incorporate the recent bounds on rare decays ($\bsg$ and $\bmumu$) and the muon anomalous magnetic moment. 

\begin{table}
  \begin{center}
    \vspace*{-1cm}
    \begin{tabular}{|c|r|r|r|r|c|c|}
      \hline
      					& ff1) 		& ff2)		& aa1)		&aa2)  	&nn1)	&nn2)\\
      \hline
      $\tan\beta$ 			& 5			& 5  		& 5		& 9		& 3   & 4.8\\
      $A_{\lambda}$		& 550 		& 500 		& 400		& 400		& 589     &645\\ 
      $A_{\kappa}$		&-200 		& 0 		& 0		& 0		& -30     &-86\\  
      $\mu$				& 130  		& 120 		& 200		& 200		& 204     &168\\
      $\lambda$	       		& 0.2 	   	& 0.3		& 0.04	    	& 0.115 	    	&
      0.41 -- 0.54& 0.43 -- 0.51\\
      $\kappa$         		& 0.1		& 0.2  		& 0.03		& 0.05		& 0.031   &0.15\\    
      \hline
      $M_1$			& 200  		& 150		& 150 		& 150 		& 365     &178\\
      $m_{L,E}$   		&  250  	& 250    	& 250		& 250		& 250     &250\\
      $m_{Q,U,D}$ 		& 1000 		& 1000  	& 1000 		& 1000 		& 1300    &1300\\
      $A_{E}$     		& -2500  	& -2500 	& -2500		& -2500		& -2500   &-2500\\
      $A_{U,D}$     		&  1500  	& 1000  	& 1000		& 1500		& 1000    &1000\\
      \hline
      \hline
      $\hmassl$   		&  62.4  	& 115.9 	& 114.4		& 115.0 &
      52.5 -- 60.7 & 72.9 -- 90.9\\      
      $\hmassm$   		& 119.4 	& 158.5 	& 300.0		& 178.9&
      116.6 -- 124.8 & 123.2 -- 126.2\\      
      $\hmassh$   		& 634.2 	& 592.6 	& 740.8		& 919.6 &
      637.5 -- 638.9 & 756.9 -- 758.6\\
      $\phmassl$  		& 199.6  	& 51.2 	   & 6.64		&
      11.7& 46.9 -- 47.5 & 126.3 -- 133.2\\      
      $\phmassh$  		& 632.5  	& 589.6 	& 739.7		& 919.2 &
      637.9 -- 638.3 & 754.5 -- 758.6\\
      \hline
      \hline
      $\hcompld$    	    	&  0.05  	& 0.21 		& 0.20		& 0.10
      &-0.004 -- 0.011 & 0.010 -- 0.011\\
      $\hcomplu$        	&  -0.003  	& 0.98 		& 0.98		& 0.95
      & -0.34 -- -0.23 & -0.44 -- -0.41\\
      $\hcompls$        	&  0.999  	& 0.05 		& -0.01		& -0.30
      &  0.94 -- 0.97 & 0.90 -- 0.91\\
      $S_{H_2^0}^1$        	&  0.21  	& 0.08 		& 0.01		& 0.05
      & 0.34 & 0.23\\
      $S_{H_2^0}^2$             &  0.98  	& -0.07 	& 0.01		&
      0.30   & 0.88 -- 0.92 & 0.88 -- 0.89 \\
      $S_{H_2^0}^3$        	&  -0.008       & 0.99 		& 0.9999
      & 0.95   &  0.21 -- 0.32 & 0.40 -- 0.43\\
      $S_{H_3^0}^1$        	&  0.98  	& 0.97 		& 0.97		& 0.99
      & 0.94 & 0.97\\
      $S_{H_3^0}^2$        	&  -0.21  	& -0.20 	& -0.20		&
      -0.11   & -0.32 & -0.21\\
      $S_{H_3^0}^3$        	&  -0.05  	& -0.09 	& -0.009	&
      -0.01   & -0.11 -- -0.09 & -0.10 -- -0.09\\
      \hline
      \hline
      BR($\bsg$) $\times 10^4$&  4.15 & 4.20 & 3.97 & 3.82 & 4.13 & 3.95\\ 
      $\asusy \times 10^{10}$ &4.08 & 4.49 & 8.49 & 3.59 & 0.62 -- 0.84 & 1.77 -- 1.94\\   
      BR($\bmumu) \times 10^{9}$& 3.11 & 3.11 & 3.11 & 3.15 & 3.10 & 3.10\\         
      \hline
    \end{tabular}
  \end{center}
  \captions{
    Sets of inputs corresponding to the examples used in the analysis. The resulting masses of the CP-even and CP-odd Higgses are indicated, together with the composition of the CP-even Higgses. All the masses are in GeV. The corresponding values for low energy observables are also indicated. In scans nn1) and nn2) the value 
    of $\lambda$ is determined by requiring a specific RH neutrino mass as a function of the rest of the parameters, $\rhnmass=8$~GeV and $\rhnmass=15$~GeV in examples nn1) and nn2), respectively.  See the 
    text for more details.} 
  \label{tab:cases}
\end{table}

In particular, we impose the experimental
bound on the branching ratio of the rare $\bsg$ decay,
$2.85\times10^{-4}\le\,{\rm BR}(\bsg)\le 4.25\times10^{-4}$ at
$2\sigma$ level, obtained
from the experimental world average 
reported by the Heavy Flavour Averaging Group \cite{Asner:2010qj},
and the theoretical calculation in the Standard Model
\cite{bsg-misiak,bsg-misiak2}, 
with errors combined in quadrature.  
We also take into account the recent upper constraint on
the $\bmumu$ branching ratio obtained by the CMS collaboration,
BR$(\bmumu)<1.9\times10^{-8}$ at $95\%$ c.l. \cite{Chatrchyan:2011kr}, which improves the previous results from the LHCb and CDF collaborations.

Concerning the muon anomalous magnetic moment, a constraint on the
supersymmetric contribution to this observable, $\asusy$, can be
extracted by comparing the experimental result
\cite{g-2}, with the theoretical evaluations of the
Standard Model contributions \cite{Davier:2009zi}. 
When $e^+e^-$ data are used the experimental excess in
$a_\mu\equiv(g_\mu-2)/2$ would constrain a possible supersymmetric
contribution to be $3.3\times10^{-10}\le\asusy\le47.9\times10^{-10}$ at the $2\sigma$ level, where
theoretical and experimental errors have been combined in
quadrature. However, when tau data are used a smaller discrepancy
with the experimental measurement is found \cite{Davier:2009ag}. Due to this reason, in our
analysis we will not impose this constraint, but only indicate the resulting prediction in our benchmark examples.

Finally, the presence of an new unrealistic vacuum in the NMSSM field space which is not included in the \nmh\ code has been pointed out in Ref.\,\cite{Kanehata:2011ei}.
It was found that the direction where $\langle H_1 \rangle = \langle H_2 \rangle \neq 0$ and $\langle S \rangle \neq 0$, with vanishing D-terms, is equivalent to the so-called MSSM unbounded from below (UFB)-1 direction \cite{Casas:1995pd}, but 
lifted up by the singlet Higgs $S$ in the NMSSM.
This means that the UFB-1 in the MSSM becomes just another minimum in the NMSSM.
Imposing that the realistic minimum is deeper than this new one leads to constraints that can be very stringent for large values of the $\al$ and $\ak$ parameters. In our analysis we have implemented these constraints too. 

Let us now explain in more detail how the correct annihilation cross section can be obtained for very light RH sneutrinos.
In Table\,\ref{tab:cases} we detail the inputs of a set of benchmark scenarios that will be used throughout the text. We also include information about the resulting Higgs spectrum.

\subsection{$\tilde N \tilde N \to f\bar f$}
\label{sec:relicnn}

\begin{figure}[t!]
  \begin{center}
  	\epsfig{file=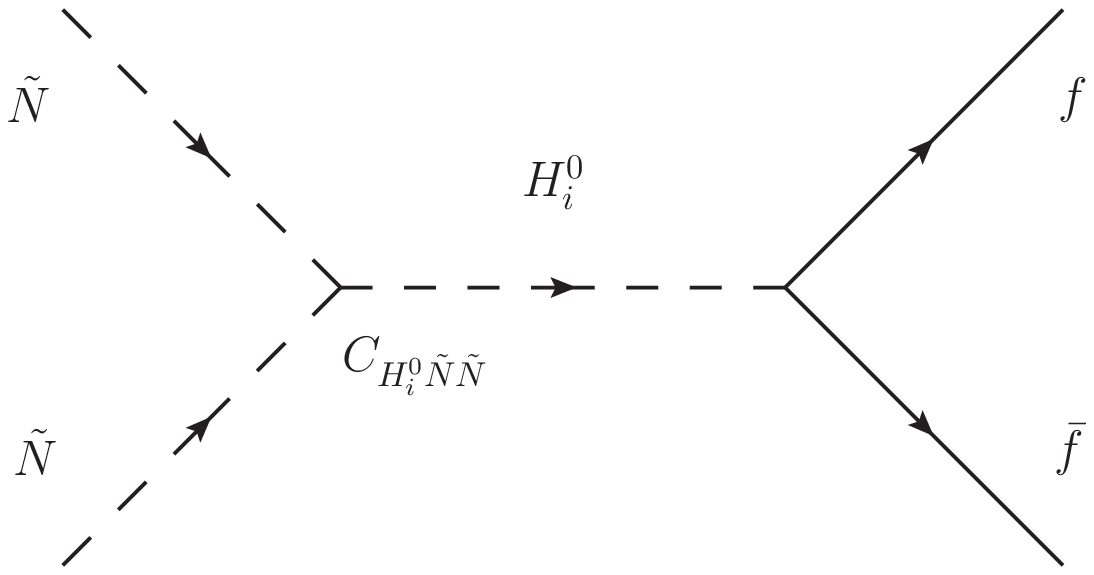,height=3.5cm}
  \end{center}
  \captions{Diagram contributing to the annihilation of RH sneutrinos
    into $f\bar f$.
    \label{nnbb}}
\end{figure}

This case is very simple to analyse. There is only one Feynman diagram that contributes, namely the exchange of a CP-even Higgs, $\higgsi$ along the $s$-channel depicted in Fig.\,\ref{nnbb} (the $s$-channel annihilation mediated by the $Z$ boson vanishes for a pure RH sneutrino). 
For light sneutrinos the main annihilation product is a $b\bar b$ pair. Annihilation into $c\bar c$ can also be significant, as we discuss later, depending on the composition of the lightest Higgs. 

Under these circumstances it is easy to derive an analytical approximate expression for the thermally averaged annihilation cross section by means of a partial wave expansion $\langle \sigma v_{Mol}\rangle\approx a+b x$, where $x=T/m$ is proportional to the WIMP velocity-square.
This approximation holds when one is far enough from resonances and thresholds for new final states\footnote{Although for very light sneutrinos resonances can generally be avoided (as they would require very light CP-even Higgses), the threshold for annihilation into $b\bar b$ happens around $4$~GeV and therefore we should expect deviations from this approximation for sneutrinos lighter than $6$~GeV.}.

For this specific diagram the integral of the matrix element describing each annihilation process, $\snr \snr \rightarrow X_1 X_2$, which we define in terms of the scattering angle in the CM frame, $\theta_{CM}$, as 
\begin{eqnarray}
  \widetilde{w}_{X_1 X_2}(s)\equiv \frac{1}{2}\int_{-1}^{+1} \!d \cos
  \theta_{CM}  
  \,|{\cal A}(\snr \snr \rightarrow X_1 X_2)|^2, \label{wtildedef:eq}
\end{eqnarray}
can be written as
\begin{eqnarray}
	\tilde w_{b\bar b}&=&\left(\frac{g\,m_b}{2\mw\cos\beta}\right)^2
	\sum_{i,j=1}^3 
	\frac{\chisnsn\chjsnsn\hcompid\hcompjd}{\Delta_{ij}}
	\,(2s-8m_b^2)\, ,\\
	\tilde w_{c \bar c}&=&\left(\frac{g\,m_c}{2\mw\sin\beta}\right)^2
	\sum_{i,j=1}^3 
	\frac{\chisnsn\chjsnsn\hcompiu\hcompju}{\Delta_{ij}}
	\,(2s-8m_c^2)\, ,
\end{eqnarray}
where $\chisnsn$ is the RH sneutrino coupling to the Higgs $H_i^0$, and $\Delta_{ij}$ is the square denominator of the Higgs propagator, both defined in Appendix A of
Ref.\,\cite{Cerdeno:2009dv}. 
The following convention is used to express the composition of the CP-even Higgs mass eigenstates, $\higgsi=\hcompid\,H_d +\hcompiu\,H_u +\hcompis\,S$.

In terms of the quantities $\tilde w_{X_1 X_2}$ the annihilation cross section can be calculated numerically as detailed in Appendix B of Ref.\,\cite{Cerdeno:2009dv}. Moreover, the coefficients of the partial wave expansion of the thermally averaged annihilation cross section can also be calculated analytically \cite{swo88}. Following the prescription of Ref.\,\cite{Nihei:2001qs}, the expressions for the velocity-independent contribution to the annihilation cross section into a pair of $b\bar b$ or $c\bar c$ then read
\begin{eqnarray}
  a_{b\bar b}&=& 
  \frac{3}{4\pi}
  \left(\frac{g\,m_b}{2\mw\cos\beta}\right)^2
  \frac{(\snmassrsq-m_b^2)^{3/2}}{\snmassrqb} 
  \, {\cal D}^2\,,\nonumber\\
  a_{c\bar c}&=& 
  \frac{3}{4\pi}
  \left(\frac{g\,m_c}{2\mw\sin\beta}\right)^2
  \frac{(\snmassrsq-m_c^2)^{3/2}}{\snmassrqb} 
  \, {\cal U}^2\,,
  \label{abbacc}
\end{eqnarray}
where we have defined
\begin{equation}
  {\cal D}\equiv\sum_{i=1}^3
  \frac{\chisnsn\hcompid}{4\snmassrsq-\hmassisq}
  \quad;\quad
  {\cal U}\equiv\sum_{i=1}^3
  \frac{\chisnsn\hcompiu}{4\snmassrsq-\hmassisq}\,.
  \label{defcal}
\end{equation}
In most cases the lightest Higgs contribution will dominate in the expressions above (especially if its mass is small), so that we can define
\begin{equation}
  {\cal D}\approx{\cal D}_1\equiv
  \frac{\chlsnsn\hcompld}{4\snmassrsq-\hmasslsq}
  \quad;\quad
  {\cal U}\approx{\cal U}_1\equiv
  \frac{\chlsnsn\hcomplu}{4\snmassrsq-\hmasslsq}\,.
    \label{defcal1}
\end{equation}

The contribution from the $b$ parameter in the $\langle\sigma
v_{Mol}\rangle$ expansion is suppressed (since $x_f\approx1/20$) and will be neglected in the following discussion. 

Notice that 
\begin{equation}
	a_{c\bar c}=
	\left(\frac{\snmassr^2-m_{c}^2}{\snmassr^2-m_{b}^2}\right)^{3/2}
	\frac{m_c^2}{m_b^2\tan^2\beta}
	\frac{{\cal U}^2}{{\cal D}^2}\, a_{b\bar b}
	\approx
	\frac{0.18}{\tan\beta^2}\,
	\frac{{\cal U}^2}{{\cal D}^2}\, a_{b\bar b}
\end{equation}
for RH sneutrinos with a mass $\snmassr\approx7-9$~GeV. Thus, annihilation into $c\bar c$ can only be dominant for small values of $\tan\beta$ and depending on the properties of the Higgs sector. Since it is the lightest Higgs the one that contributes the most to both ${\cal D}$ and ${\cal U}$, the above condition could happen if its $H_u$ component is much larger than its $H_d$ component. 
Although this is not a very common situation in our scans, we will keep this possibility open and explicitly consider both cases, where annihilation into either $b\bar b$ or $c\bar c$ dominates\footnote{It should be noted that the contribution from annihilation into $\tau\bar\tau$ can be larger than that corresponding to $c\bar c$. However, it has the same dependence on the lightest Higgs composition than the contribution from $b$ quarks (i.e., $a_{\tau\bar \tau}$ is proportional to ${\cal D}^2$) and because of the different Yukawa couplings  $a_{\tau\bar \tau}\ll a_{b\bar b}$ is always satisfied. Thus, we only have two possible regimes, where either $b\bar b$ or $c\bar c$ is the leading contribution.}.

For moderately heavy WIMPs the sneutrino relic density is usually  approximated as 
\begin{equation}
	\Omega h^2 \approx\frac{1}{x_f\,\sqrt{g^*(x_f)}}
	\frac{1.07\times10^{9}\,{\rm GeV}^{-1}}{M_P\,
	(a+\frac{b}{2}\,x_f)
	}\,,
	\label{relic}
\end{equation}
where $M_P=1.22\times10^{19}$ GeV is the Planck mass and $g^*(x_f)$ is the number of relativistic degrees of freedom at the decoupling temperature.
However, very light dark matter with mass smaller or of order $10$~GeV would have decoupled when the temperature of the Universe was around $400$~MeV, precisely when quarks confined into hadrons. This hadronization implies that the number of relativistic degrees of freedom drops dramatically around the decoupling temperature of very light WIMPs, producing an enhancement of their relic abundance \cite{Gondolo:1990dk}.
Furthermore, this enhancement is sensitive to the phase transition model that is considered (in particular, it depends on the deconfinement temperature, $T_c$). 
In our analysis we have taken $T_c=400$~MeV.

For a quick analytical estimate let us momentarily assume that the QCD transition is close to first order and that the relativistic number of degrees of freedom suddenly varies from $\sqrt{g^*}\approx9$ to $~3.7$, considering also the uncertainty in $T_c$.
Notice that this implies essentially an increase of a factor 2.5 in the approximate evaluation of $\snrelic$ of Eq.\,(\ref{relic}).
In order to reproduce the WMAP result, this results in a condition on the annihilation cross section that can be written as
\begin{eqnarray}
	\langle\sigma v\rangle\approx a_{b\bar b}\approx0.77-1.9\,{\rm pb}
	&\quad {\rm if}\quad& 
	{\cal D} \gg \frac{0.4}{\tan\beta}\,{\cal U}\,,\label{omegaconu}\\
	\langle\sigma v\rangle\approx a_{c\bar c}\approx0.77-1.9\,{\rm pb}
	&\quad {\rm if}\quad& 
	{\cal D} \ll \frac{0.4}{\tan\beta}\, {\cal U}\,,
	\label{omegacond}
\end{eqnarray}
where the lower (upper) value applies to a RH sneutrino which is heavier
(lighter) than $\snmassr\sim8$~GeV and therefore decouples above (below) $T_c$.
These values are generic for a dark matter particle in which $s$-wave annihilation dominates \cite{Birkedal:2004xn}.

So far we have used an analytical approach with several approximations so that the correlation between the sneutrino annihilation cross section and its scattering cross section off nuclei (which we calculate in the next section) is manifest. 
However, we stress that in our results we use the full calculation of $\snrelic$ following the same numerical method that we detail in Ref.\,\cite{Cerdeno:2009dv}. Also, the QCD transition is taken into account through a parametrization of $g^*(x_f)$ according to \cite{Olive:1980dy} with $T_c=400$~MeV. 

The question is then whether or not it is possible to find a sufficiently large annihilation cross section in this model and which is the choice of input parameters to achieve this. 
The quantities $a_{b\bar b}$  and $a_{c\bar c}$ are very sensitive to the structure of the Higgs sector and the new couplings $\chisnsn$, and our model provides much flexibility in this sense. 
For example, both $a_{b\bar b}$  and $a_{c\bar c}$ increase as the mass of the lightest Higgs (which is the leading term) decreases, but they also depend on the composition of this lightest Higgs. 
In the NMSSM it is possible to have a very light Higgs without violating the current experimental limits as long as its singlet composition is large enough. 

Thus one possible scenario in which a sufficiently large annihilation cross section can be achieved involves a light singlet-like Higgs. The $H_u$ or $H_d$ components, though small, would determine whether predominant annihilation occurs via $b\bar b$ or $c\bar c$, according to the expressions above. This kind of scenarios is interesting since the Higgs structure is completely different to that of the MSSM for very light neutralinos (in which Higgses are in the so-called intense coupling regime for which their masses are very similar and of order 100~GeV). On the other hand, as already commented in the Introduction, a light singlet-like Higgs is also one of the possible solutions for the very light neutralinos in the NMSSM.

\begin{figure}[t!]
  \hspace*{-0.5cm}
  \epsfig{file=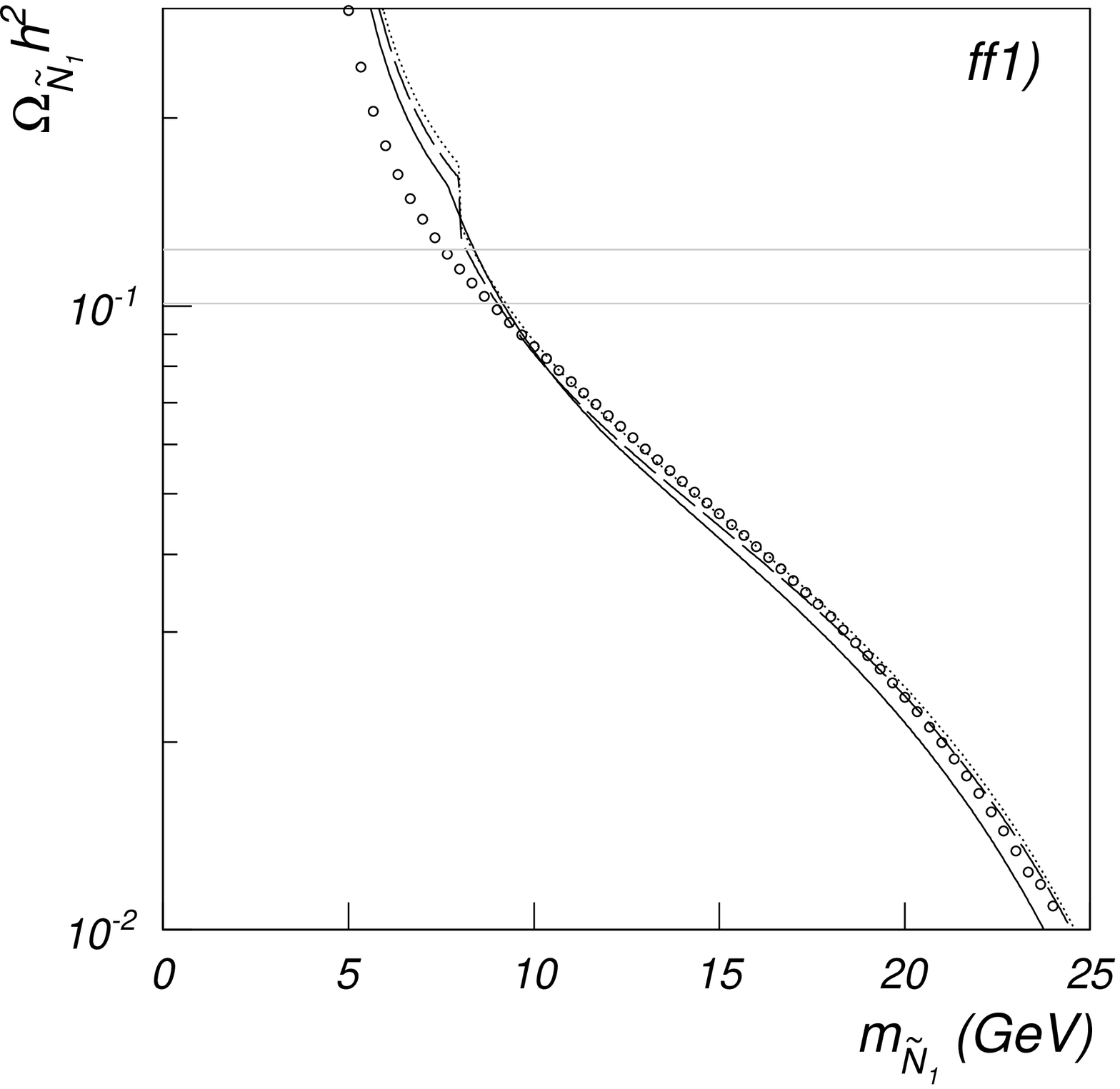,width=8.cm}
  \epsfig{file=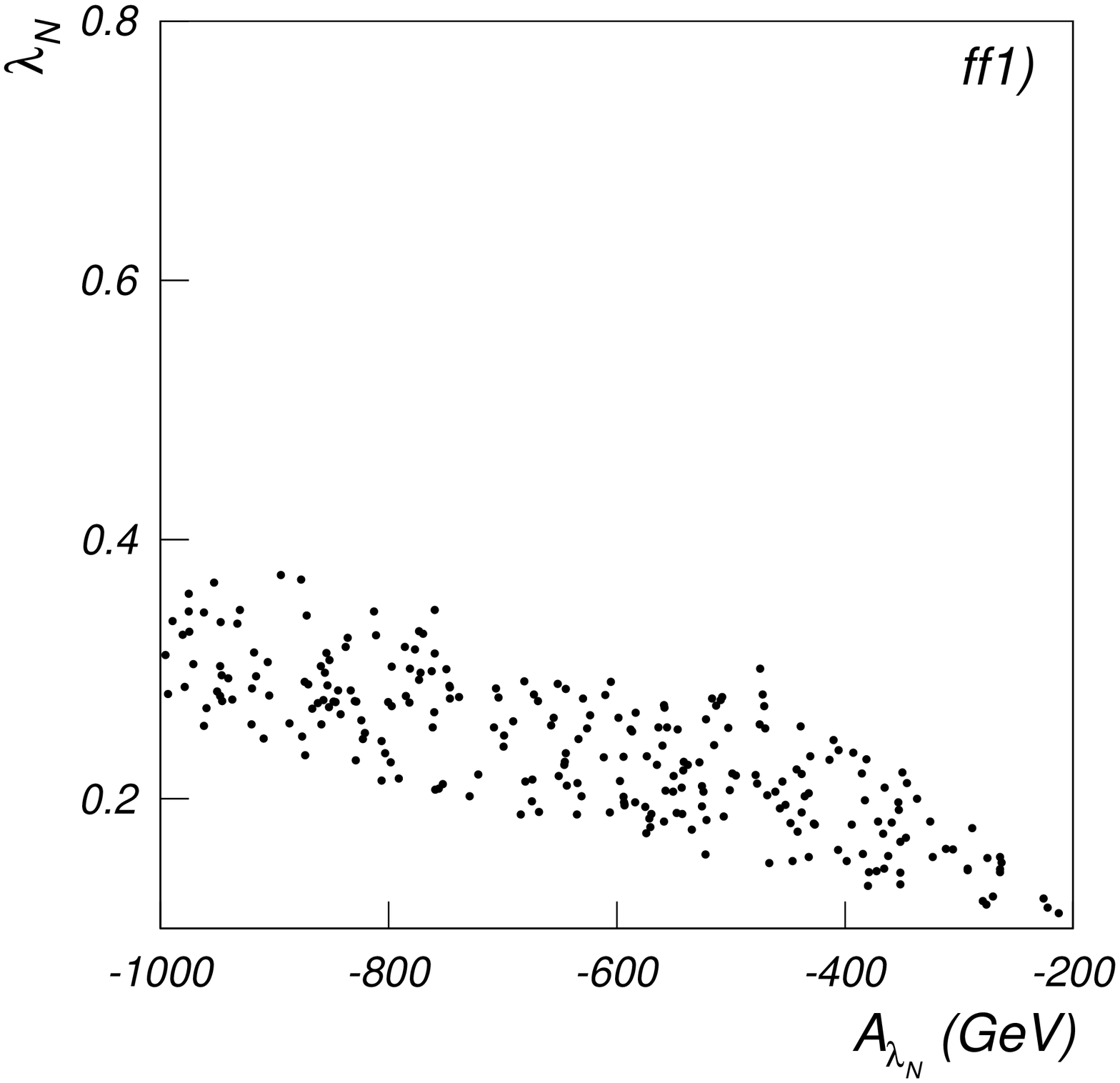,width=8.cm}
  \captions{Results for case ff1) of Table\,\ref{tab:cases}. 
Left) Sneutrino relic density as a function of the sneutrino mass for $\ln=0.25$ and $\aln=-500$~GeV. 
  The solid line corresponds to a calculation using the partial wave expansion of $\langle \sigma v\rangle$, the dashed line uses the approximation  $\langle \sigma v\rangle\approx a_{b\bar b}$ and the dotted line represents the contribution coming from only the lightest Higgs to $a_{b\bar b}$.
  The circles correspond to the results using a numerical evaluation of the relic abundance with a parametrization of $g^*$ in which $T_c = 400$~MeV.
  Right) Values of the parameters $\ln$ and $\aln$ for which the WMAP relic density is reproduced for sneutrinos in the mass range $\snmassr=5-25$~GeV.  }
  \label{fig:relicff1}
\end{figure}

In Table\,\ref{tab:cases} we display a specific example with these properties, namely case ff1). The lightest Higgs, with a mass of $62.4$~GeV is mostly singlet, whereas the second lightest Higgs is Standard Model (SM)-like with a predominant $H_u$ composition ($S_{H_2^0}^2=0.98$) and a mass of $119.4$~GeV.
The sneutrino relic abundance is represented on the left-hand side of
Fig.\,\ref{fig:relicff1} as a function of the sneutrino mass for $\ln=0.25$
and $\aln=-500$~GeV.
The solid line corresponds to the estimation using the partial wave expansion of $\langle \sigma v\rangle$. The dashed line corresponds to the approximation  $\langle \sigma v\rangle\approx a_{b\bar b}$ and the dotted line represents the contribution coming from only the lightest Higgs to $a_{b\bar b}$. In extracting these approximations we have assumed a sharp QCD transition with $T_c=400$~MeV.
As we can see, this is one of the examples for which the contribution of the lightest Higgs provides a good approximation to the total result.
The predictions from our numerical code are represented by circles. The rapid decrease of the relic density towards larger masses is due to the resonance with the lightest Higgs which takes place for $\snmassr\approx\hmassl/2\approx31$~GeV.

Scanning in the $\ln$ and $\aln$ parameters we can vary the $\chisnsn$ coupling in such a way that the correct relic density is obtained for a range of RH sneutrino masses. 
In order to illustrate this, on the right-hand side of Fig.\,\ref{fig:relicff1} we represent the distribution of points in the $(\ln,\aln)$ plane which reproduce the correct relic abundance for RH sneutrinos in the mass range $\snmassr=5-25$~GeV.

Another possible scenario consists of having a lightest Higgs with a mass of order $114$-$120$~GeV and which is mostly $H_u$ (as in the MSSM) and then increasing $\chisnsn$ by taking larger values of the parameter $\ln$ (as we showed in Fig.\,\ref{fig:spec}, a larger $|\aln|$ may then be needed in order to keep the small values of the RH sneutrino mass). 

\begin{figure}
  \hspace*{-0.5cm}
  \epsfig{file=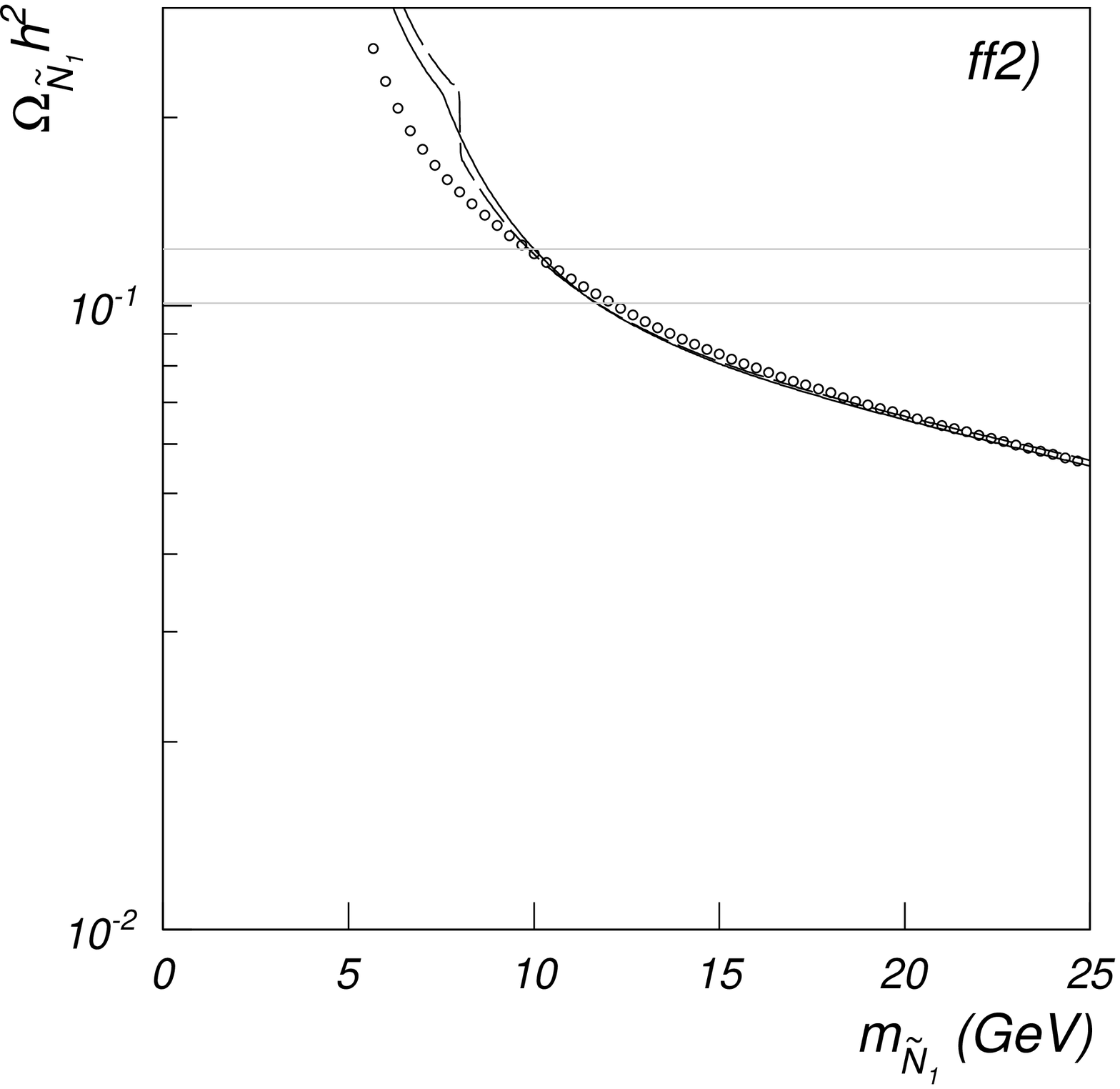,width=8.cm}
  \epsfig{file=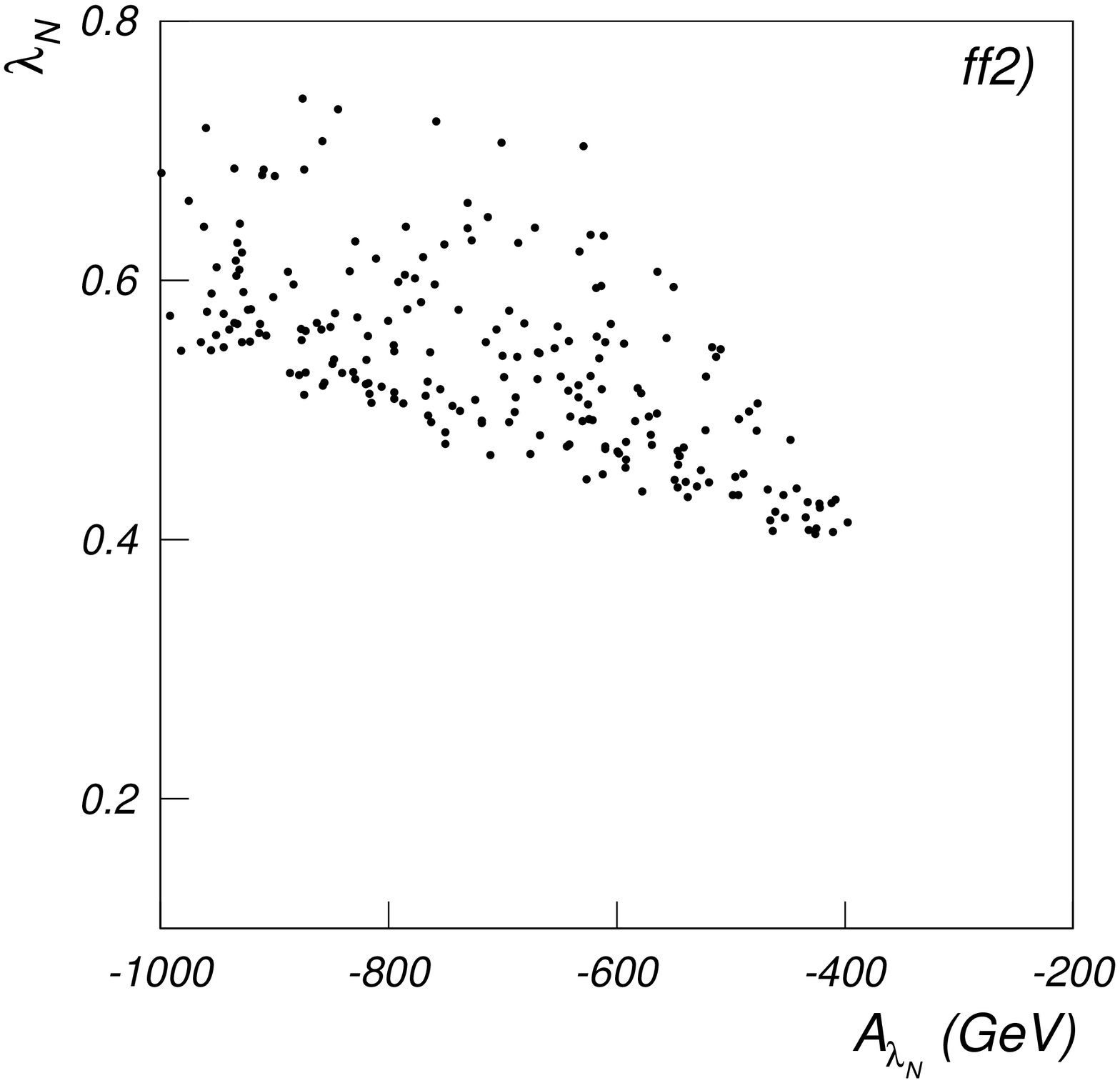,width=8.cm}
  \captions{The same as in Fig.\,\ref{fig:relicff1} but for case ff2) of Table\,\ref{tab:cases} with $\ln=0.5$ and $\aln=-500$~GeV.
  }
  \label{fig:relicff2}
\end{figure}

These features are clearly seen in Fig.\,\ref{fig:relicff2}, where we consider the example ff2) in Table\,\ref{tab:cases}, which falls into this category. 
The relic density is plotted as a function of the sneutrino mass on the left-hand side of the figure.
In this particular case, the masses of the lightest Higgs ($116$~GeV) and second lightest Higgs ($159$~GeV) are relatively close to each other, and both have to be taken into account when computing the relic density. Thus, whereas the approximation of considering only the term $a_{b\bar b}$ (dashed line) remains valid, the contribution from including only the lightest Higgs deviates significantly from the correct result (the line is not represented in the plot, as it leads to a relic density which is one order of magnitude above the exact value).
Notice also, that although in this scenario the $H_u$ component of the lightest Higgs is not small, it is not sufficiently larger than the $H_d$ component for the annihilation into $c\bar c$ to become comparable to the $b\bar b$ channel. This would only be possible for smaller values of $\tan\beta$ and a much more careful choice of the parameters entering the Higgs mass matrix. 

Regarding the allowed regions in the $(\ln,\,\aln)$ plane, these are represented on the right-hand side of the figure for sneutrino masses in the range from $5$ to $25$~GeV and, as mentioned above, feature larger values of the $\ln$ parameter in order to compensate for the larger Higgs mass. 
In general it is still possible to reproduce the correct relic abundance for very light sneutrinos for this range of sneutrino masses and a wide area of the parameter space. 

Summarising, the conditions under which very light RH sneutrinos can have the correct relic abundance when they annihilate into a fermion-antifermion pair are relatively easy to fulfil in our model. 
Of course, this is due to the flexibility of the new input parameters $\ln$, $\mn$, and $\aln$, which modify the RH sneutrino mass and couplings without altering any other feature of the NMSSM spectrum. 
Thus significant constraint appears in the new parameters. The low-energy constraints can be fixed only with an adequate choice of the NMSSM inputs and the RH sneutrino sector freely modified after that.
Something that should be emphasized is that since $\ln$ can be chosen to be rather large, the Higgs coupling to $b$ quarks needs not be too large in order to reproduce the correct relic abundance, in other words, the value of $\tan\beta$ can be kept small. This is an important difference with respect to the case of the neutralino.

\subsection{$\snr\snr\to\phiggsl\phiggsl$}

The second possibility for very light RH sneutrinos to have the correct relic abundance is that they annihilate predominantly into a pair of very light pseudoscalar Higgses. This channel is possible provided the pseudoscalar is mostly singlet, thereby evading present experimental constraints and requires some tuning of the NMSSM parameters. 

In Ref.\,\cite{Gunion:2005rw} it was already shown that a very light pseudoscalar Higgs could be viable in the NMSSM and that this made it possible for the lightest
neutralino to have the correct relic density in this extension of the MSSM. In particular, the neutralino annihilation cross section is enhanced through a resonance
with the pseudoscalar Higgs when $2\,\neutmass \approx \phmassl$ or because the annihilation channel into $\phiggsl\phiggsl$ becomes kinematically allowed.

Obtaining very light pseudoscalars in the NMSSM requires a careful tuning of some of the parameters so that either the $U(1)_R$ or $U(1)_{PQ}$ symmetries of the model are quasi-restored and the CP-odd Higgs corresponds to the pseudo-Goldstone boson of the broken symmetry~\cite{Gunion:2005rw}. For example, this can be achieved by taking $\k\to0$ or the trilinear terms $\al,\ak\to0$.
Very light pseudoscalars are very constrained by collider searches and its composition has to be mostly singlet-like in order to avoid these. 
In particular, they can lead to observable signals in radiative $\Upsilon$ decays. The most recent data from the CLEO collaboration \cite{:2008hs} set stringent upper bounds whose effect in the NMSSM parameter space was investigated in Refs.\,\cite{Dermisek:2006py,Gunion:2008dg} for the case $\phmassl\lsim9.2$~GeV and later analysed in more detail for pseudoscalar masses between $9.2$ and $10.5$~GeV in \cite{Domingo:2008rr}.
The constraints coming from the analysis of Ref.\,\cite{Dermisek:2006py} are incorporated in the code {\tt NMHDECAY\,2.3.7}, which we use in our analysis. The bounds from Ref.\,\cite{Domingo:2008rr} are easy to implement in our analysis. In all cases, these are avoided if the lightest pseudoscalar is singlet-like.

All the above can be applied to our model, since the inclusion of RH sneutrinos has no influence on any aspect of the NMSSM spectrum (or the NMSSM vacuum \cite{Cerdeno:2009dv}). 
Notice however that the RH sneutrino has no coupling with the CP-odd Higgs and therefore there is no $s$-channel annihilation mediated by this particle (and therefore no resonant enhancement as in the case of the neutralino).
It is possible, nevertheless, that a RH sneutrino particle annihilates preferentially into a pair $\phiggsl\phiggsl$, and this is the case we study here.

\begin{figure}[t!]
  \begin{center}
	\epsfig{file=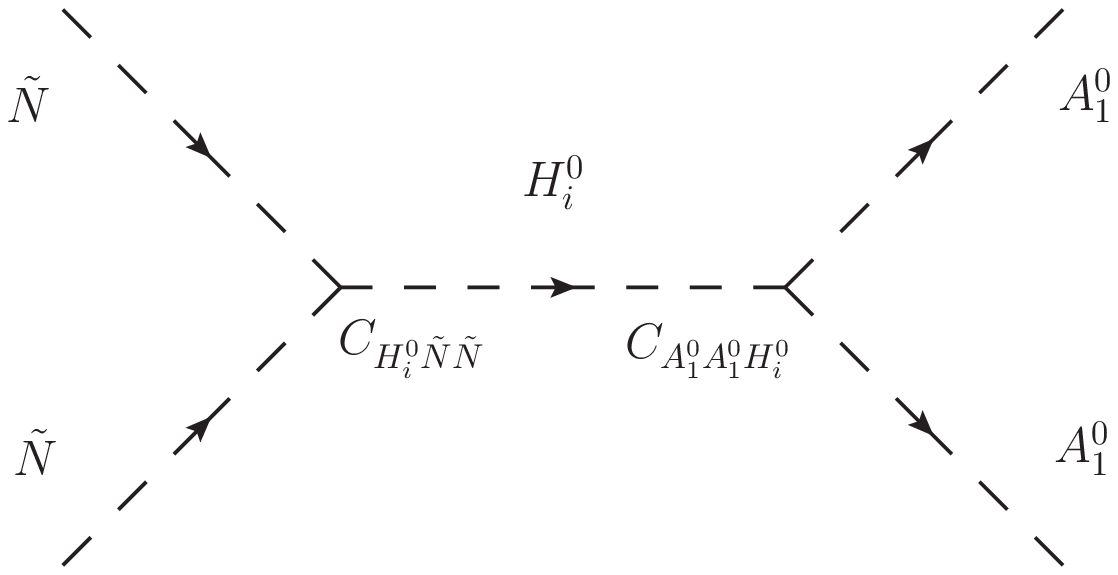,height=3.cm}
	\epsfig{file=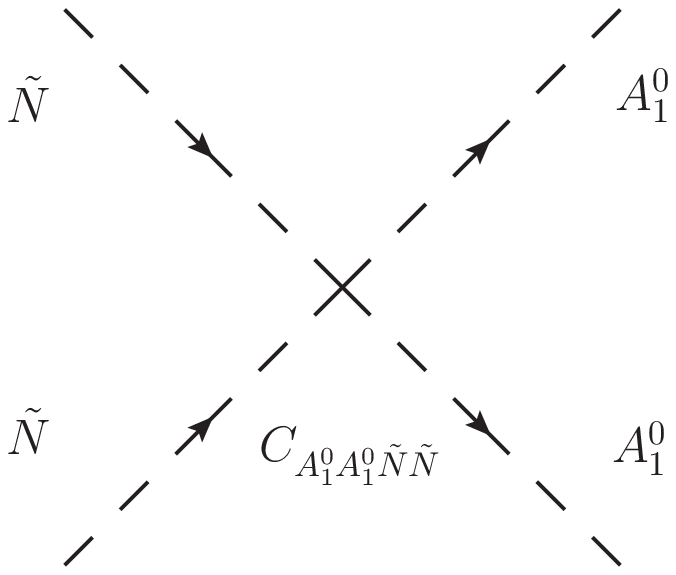,height=3.cm}	
  \end{center}
  \captions{Diagrams contributing to the annihilation of RH sneutrinos
    into $\phiggsl\phiggsl$.
    \label{nnaa}}
\end{figure}

The Feynman diagrams which are involved in this process are represented in Fig.\,\ref{nnaa} and consist of a quartic coupling and $s$-channel mediated by CP-even Higgses.
The explicit expression for $\tilde w_{\phiggsl\phiggsl}$ can be found in Ref.\,\cite{Cerdeno:2009dv}.  It can easily be seen that in the partial wave expansion of the annihilation cross section the $a_{\phiggsl\phiggsl}$ term is non-vanishing and is generally a good approximation. 
The relevant couplings are now the quartic coupling $\caasnsn$ and the CP-even coupling to a pair of CP-odd Higgses $\chaa$. 
For singlet-like pseudoscalars, $\caasnsn\approx\ln
(\k-2\ln)$, whereas  $\chaa$ is much more sensitive to the CP-even Higgs composition but is independent on the $\ln$ parameter.

The contribution from the $s$-channel is generally sizable since it involves
the VEVs of the scalar Higgses. For example, in the case of a pure singlet pseudoscalar and a lightest Higgs which is mostly $H_u$ one has $\chaa\approx i(\l\k v_1+\l^2 v_2)$. This channel, if open, is easily more important than the $\snr\snr\to f\bar f$ channel discussed in the previous section.
Furthermore, since this coupling is more effective, the value of $\ln$ (which now only appears through the sneutrino-sneutrino-Higgs vertex) that is necessary in order to reproduce the correct relic density is typically smaller than in the former section. The exact value for $\ln$ is now very dependent on the rest of the parameters. 

\begin{figure}
  \hspace*{-0.5cm}
  \epsfig{file=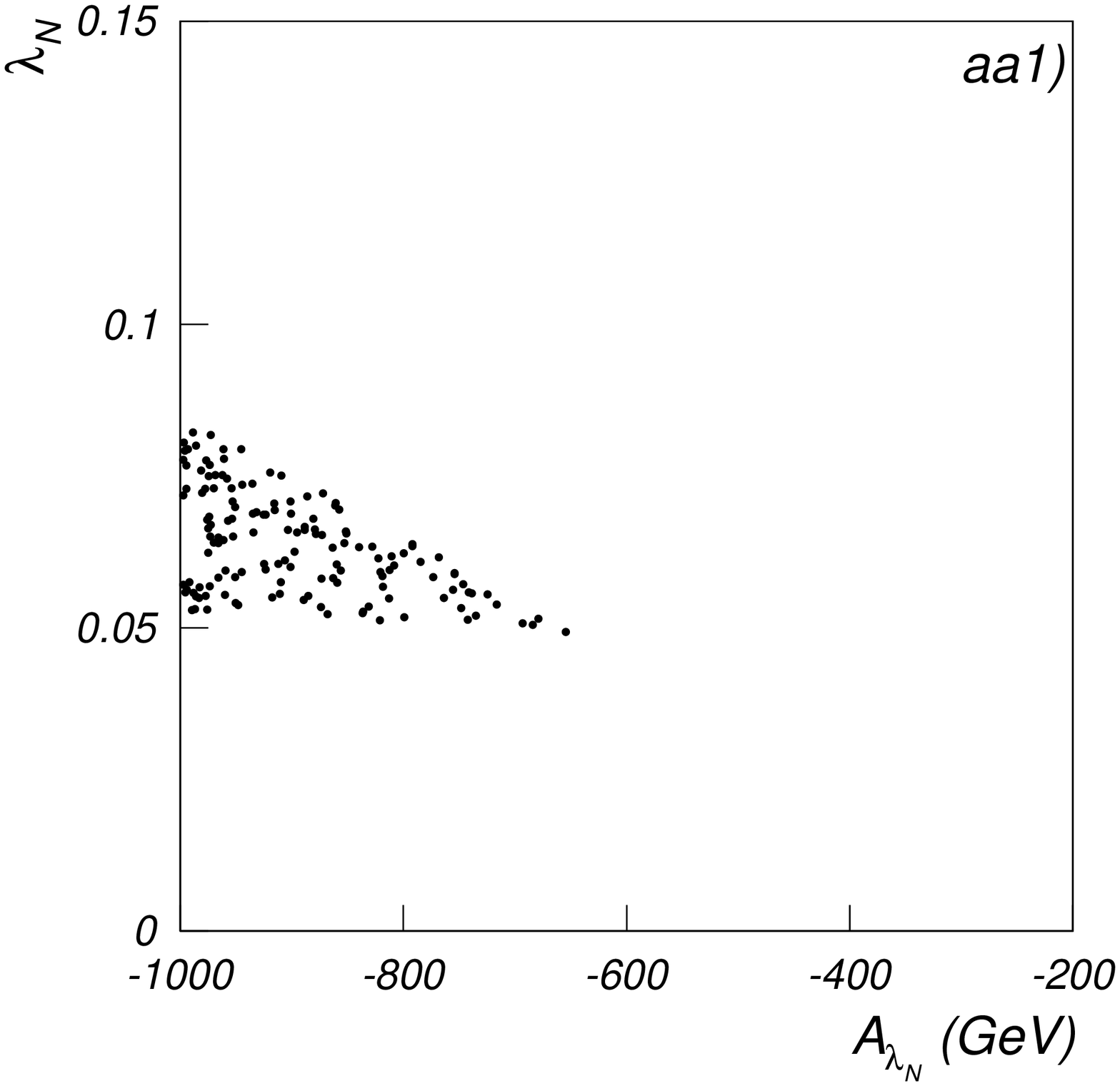,width=8.cm}
  \epsfig{file=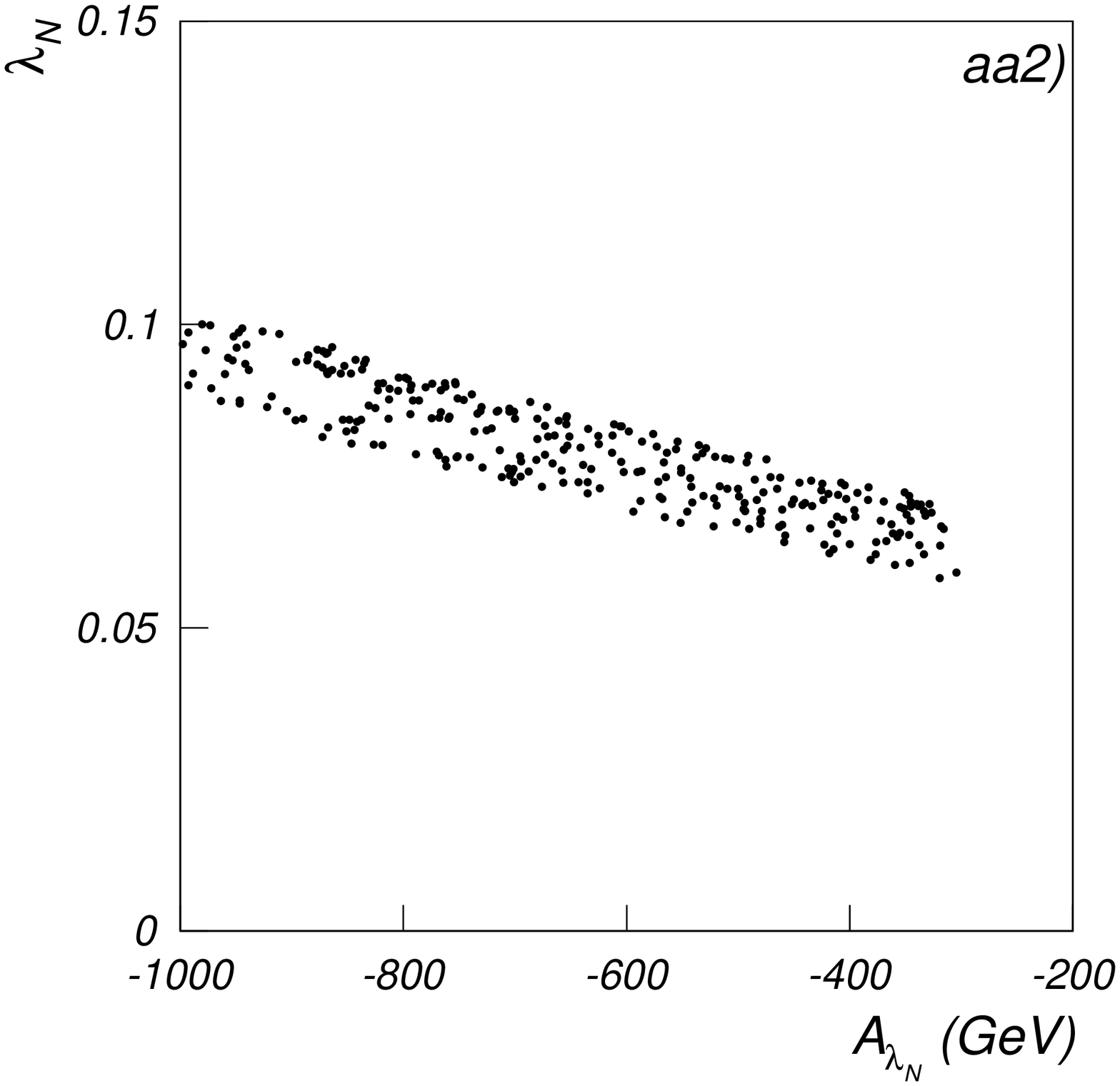,width=8.cm}
  \captions{The same as in the right-hand side of Fig.\,\ref{fig:relicff1} but for cases aa1) and aa2) of Table\,\ref{tab:cases} on the left and right, respectively.
  }
  \label{fig:lnalnaa}
\end{figure}

The input parameters for two explicit examples of this kind are given in
Table\,\ref{tab:cases} and labelled aa1) and aa2). We have chosen two cases
with a pseudoscalar mass of $\phmassl\approx7$ and $12$~GeV, respectively.
The values of $\ln$ and $\aln$ for which the correct relic density can be obtained are represented in Fig.\,\ref{fig:lnalnaa} for the range of RH sneutrino masses $\snmassr=6-45$~GeV. 
We clearly see the above mentioned decrease in $\ln$ when comparing these plots with the examples of the previous section.
Regarding the CP-even Higgs spectrum, the lightest Higgs is SM-like in both examples, with a predominant $H_u$ composition and a mass of approximately $114$~GeV.

\subsection{$\tilde N \tilde N \to \rhn\rhn$}
\label{sec:verylightnn}

\begin{figure}[t!]
  \begin{center}
  	\raisebox{1ex}{\epsfig{file=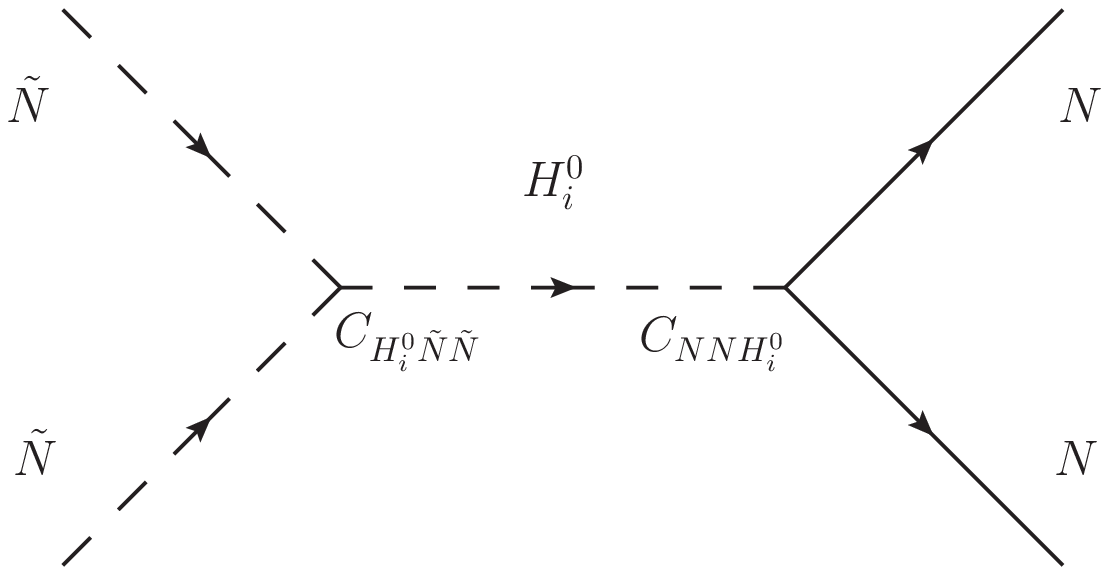,height=3.cm}}
  	\epsfig{file=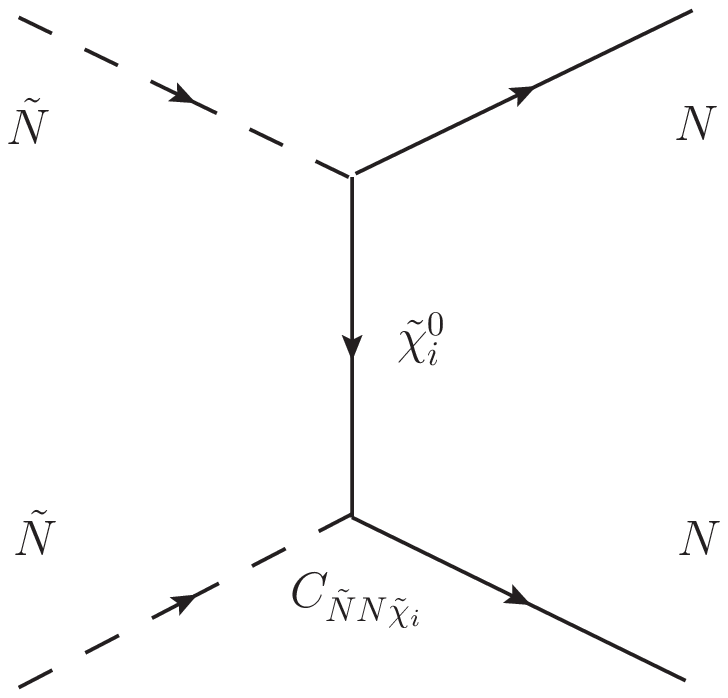,height=3.5cm}
 \end{center}
  \captions{Diagrams contributing to the annihilation of RH sneutrinos
    into $\rhn\rhn$.
    \label{nnnn}}
\end{figure}

Let us finally address a last possibility that is characteristic of this model, namely annihilation into a pair of RH neutrinos, $\snr\snr\to\rhn\rhn$. 
Remember in this sense that the RH neutrino mass in this model is a consequence of radiative Electroweak symmetry-breaking and is thus related to the VEV of the singlet field as
\begin{equation}
	\rhnmass=2\ln v_s=2\frac{\mu\,\ln}{\l}\ .
	\label{rhnmass}
\end{equation}
It is therefore possible to obtain a small value of the RH neutrino mass independently of the sneutrino mass (\ref{eq:snmass}), for which we still have two more free parameters to play with, namely the soft mass and the trilinear parameter $\aln$.

Since there is a lower bound in the value of the $\mu$ parameter in order to satisfy the experimental constraint on the chargino mass, $\mu\gsim105$~GeV, and an upper constraint $\l\lsim0.6$ in order to avoid Landau poles in the RGE of the NMSSM, a reduction in $\rhnmass$ necessarily implies a small value of $\ln$. For example, adopting the above constraints, one sees that $\ln\lsim0.018\,(0.11)$ for $\rhnmass=7\,(40)$~GeV.

The annihilation into a pair of RH neutrinos can proceed through $s$-channel
Higgs exchange or $t,u$-channel neutralino exchange, as illustrated in
Fig.\,\ref{nnnn}. Notice that both the Higgs-neutrino-neutrino coupling and
the sneutrino-neutrino-neutralino coupling are proportional to $\ln$ ($\cnnhi=-i/\sqrt{2}\ln\hcompls$ and $\csnnneui=-i\sqrt{2}\ln\hcompls\neutcomps$). 
Therefore the upper bound on this quantity that we derived in the previous paragraph is a serious handicap in order to obtain a sufficiently large annihilation cross section. 
One possibility is to consider a resonant enhancement of the $s$-channel when $2\snmassr\approx \hmassi$. 
However, if we want this channel to dominate over $\snr\snr\to f\bar f$, the corresponding Higgs exchange for the latter channel (which is obviously also enhanced by the resonance) has to be suppressed. 
The ratio of the contribution from both $s$-channels in the vicinity of the resonance with the lightest Higgs can be expressed as the fraction of the corresponding couplings, 
\begin{equation}
	R_{NN/b\bar b}\equiv\frac{\langle\sigma v\rangle_{\snr\snr\to\rhn\rhn}}{\langle\sigma v\rangle_{\snr\snr\to b\bar b}}=
	\frac{\cnnhl^2}{6Y_b^2 (\hcompld)^2}=
	\frac{1}{3}\left(\frac{\mw\ln\cos\beta}{g\,m_b}\right)^2\left(\frac{\hcompls}{\hcompld}\right)^2\,,
\end{equation}
where the factor 6 takes into account the color factor of the $b\bar b$ channel and a factor $1/2$ for identical particles in the final state of the $NN$ diagram. 
Thus if we demand the above ratio to be larger than one, the lightest Higgs needs to be almost a pure singlino and its $H_d$ composition extremely small. For example, for $\tan\beta=3$ and $\rhnmass=7$~GeV it implies $|{\hcompls}/{\hcompld}|\gsim10\,(30)$ for $R_{NN/b\bar b}=1\,(10)$.

This can be considered as a condition on the parameters entering the CP-even Higgs mass matrix (see e.g., expression (2.8) in \cite{Cerdeno:2004xw}). In particular, the NMSSM input parameters can be chosen in such a way that the ${\cal M}^2_{S,23}$ element of the Higgs mass matrix is larger than ${\cal M}^2_{S,13}$. In this sense, a small value of $\al$ is welcome but also a careful choice of the parameter $\k$.

The condition above clearly favours the use of small values of $\tan\beta$. 
This can be problematic since the supersymmetric corrections to BR($\bsg$) become sizable and it easily exceeds the experimental bound.  Once more, avoiding this constraint further limits the choice of initial parameters. Similarly, the resulting value of $\asusy$ is very small and difficult to reconcile with experimental results from $e^+e^-$ data.
Since the upper bound on $\ln$ is relaxed for larger $\rhnmass$, the amount of fine-tuning which is needed in order to obtain predominant annihilation into a pair of RH neutrinos decreases as $\rhnmass$ increases. 
We have found numerical solutions for $\rhnmass\gsim7$~GeV with the correct relic density and satisfying the condition $R_{NN/b\bar b}\ge10$.

Finally, the $t,u$-channel can also be enhanced if the neutralino mass is decreased, however, this is generally not sufficient in order to recover the correct relic abundance for very light sneutrinos. 

\begin{figure}[!t]
  \hspace*{-0.5cm}
  \epsfig{file=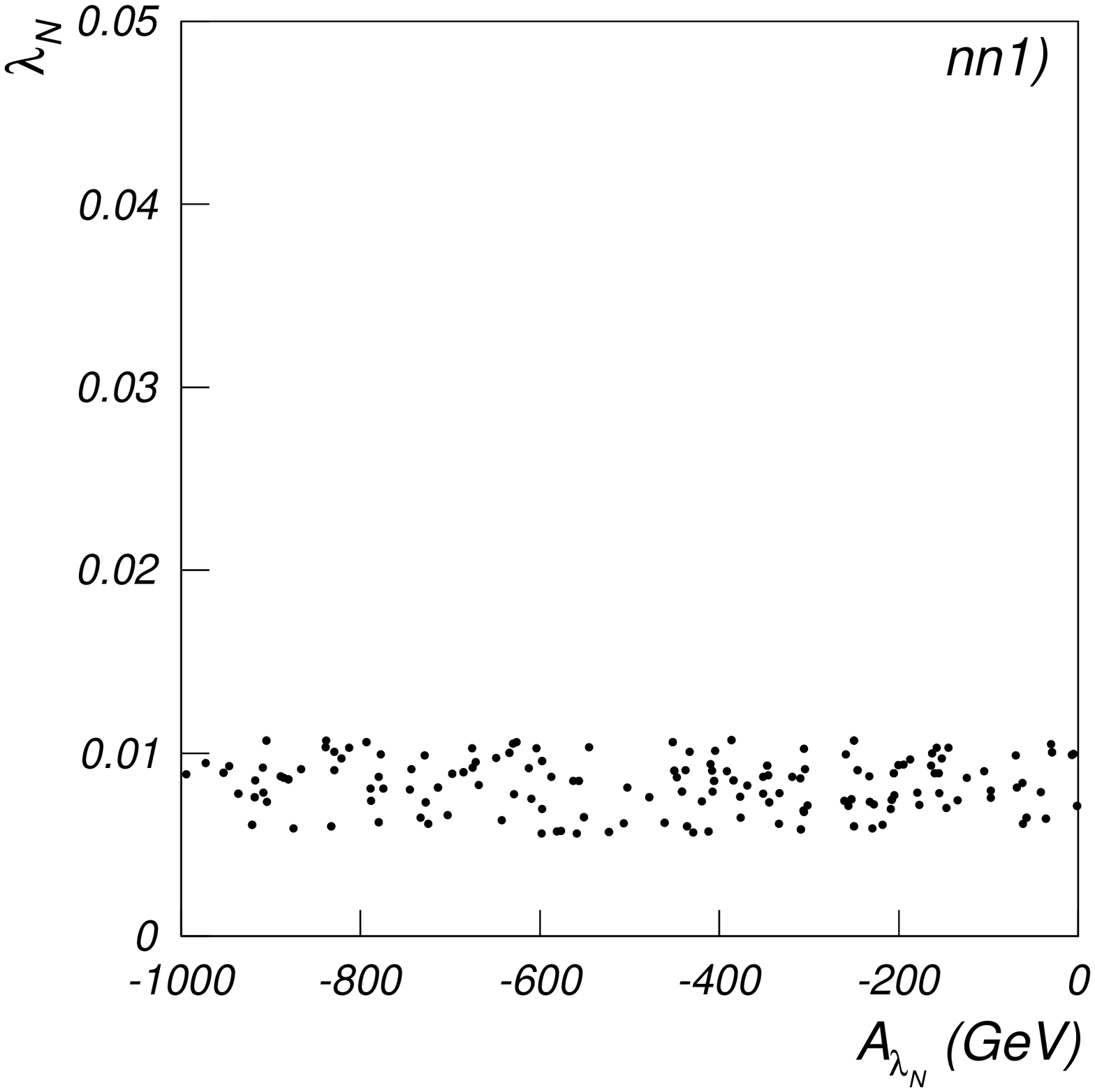,width=8.cm}
  \epsfig{file=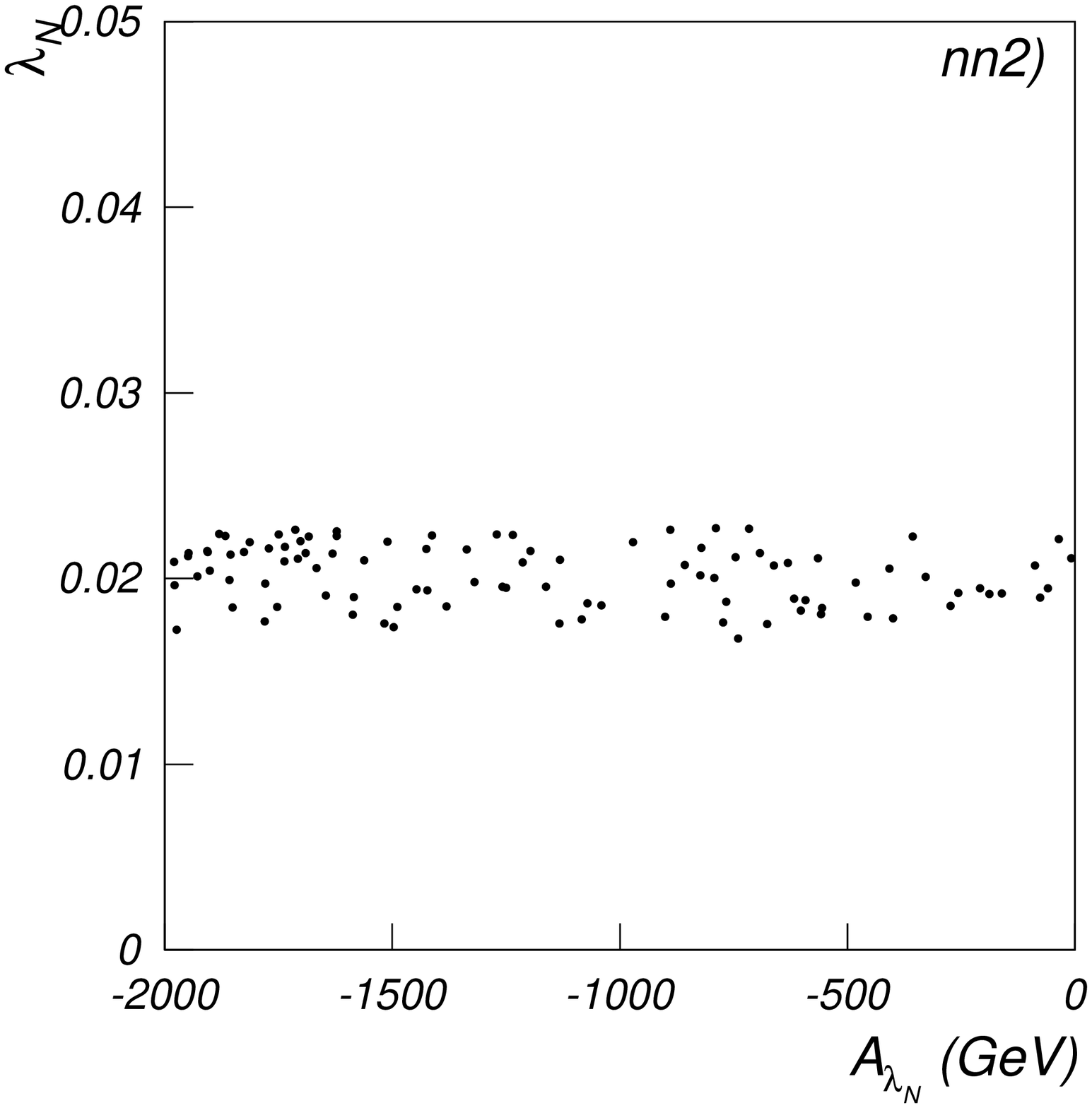,width=8.cm}
  \captions{The same as in the right-hand side of Fig.\,\ref{fig:relicff1} but for cases nn1) and nn2) of Table\,\ref{tab:cases} on the left and right, respectively.
  }
  \label{fig:lnalnnn}
\end{figure}

Two explicit examples are given in Table\,\ref{tab:cases} which satisfy the conditions described above, and are labelled as nn1) and nn2) in Table\,\ref{tab:cases}. 
For each of them, a scan is performed in the parameters $\l$, $\ln$, $\aln$ and $\mn$ in order to account for all possible RH sneutrino masses and couplings. 
The value of the $\l$ and $\ln$ parameters is set by relation (\ref{rhnmass}), which we use to fix the RH neutrino masses to $\rhnmass=8$~GeV in nn1) and $15$~GeV in nn2).
As it was already explained, the value of $\tan\beta$ is chosen to be as small as possible and the NMSSM parameters $\al$, $\ak$ and $\k$ have also been fixed to values which lead to a small $H_d$ component for the lightest Higgs.
In particular, the lightest Higgs in case nn1) has a mass of approximately
$50-60$~GeV and $|\hcompld|\sim0.004-0.01$, $|\hcomplu|\sim 0.2-0.4$. Similarly, the Higgs
in case nn2) has a mass in the range $70-90$~GeV and $|\hcompld|\sim0.01$,
$|\hcomplu|\sim-0.4$ (due to the variation in $\lambda$ the mass of the Higgs also varies and these conditions can be fulfilled by a small range of masses).
Since the correct relic abundance is only obtained when the resonant condition with the lightest Higgs is satisfied, this implies  a value of the RH sneutrino masses of $\snmassr\approx25-30$~GeV and $30-45$~GeV, in cases nn1) and nn2), respectively. 
  
As explained above, if we consider lighter RH neutrino masses, the value of $\ln$ has to be decreased. In our scans we have found that the correct RH sneutrino relic abundance could be obtained for $\ln\sim0.01$ in case nn1) and $\ln\sim0.02$ in case nn2). The viable points are represented in Fig.\,\ref{fig:lnalnnn}. 
Notice that these values of the coupling constant are considerably smaller than those obtained for the $\snr\snr\to f\bar f$ case.


\section{Direct detection}
\label{sec:direct}

\begin{figure}[!t]
  \begin{center}
  	\epsfig{file=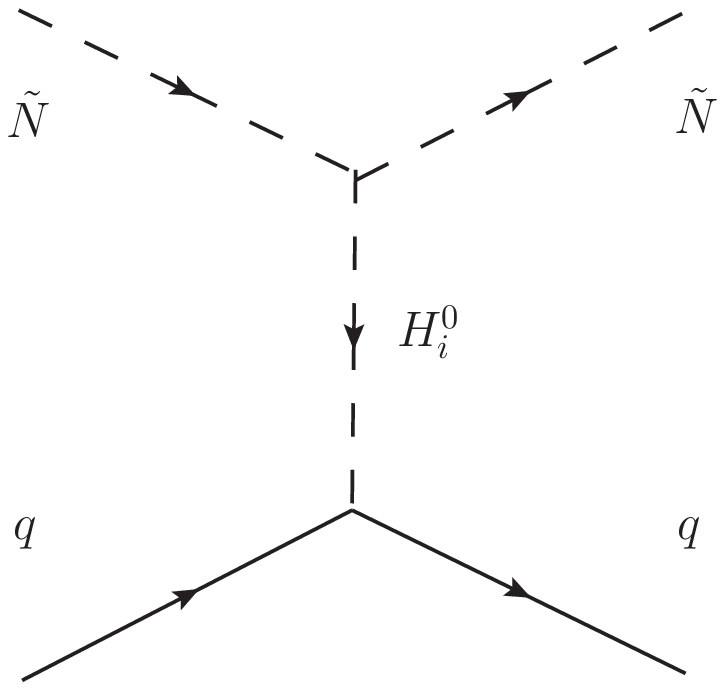,height=3.5cm}
  \end{center}
  \captions{Diagram contributing to the spin-independent elastic scattering of RH sneutrino off quarks.
  }
  \label{fig:cross-diag}
\end{figure}

Let us now address the detectability of these particles in direct detection experiments.
In general, WIMPs could be observed through their elastic scattering off nuclei (see Ref.\,\cite{Cerdeno:2010jj} for a recent review),  their interaction with quarks being described by an effective Lagrangian that is valid in the non-relativistic regime where the collision takes place.  
In the case of RH sneutrinos there is only one Feynman diagram contributing at tree level to this process, namely, the $t$-channel exchange of neutral Higgses shown in Fig.\,\ref{fig:cross-diag}. This leads to a Lagrangian describing the four-field interaction which only contains a scalar coupling,
\begin{equation} 
  {\cal L}_{eff}\supset \alpha_{q_i} \tilde N\tilde N
  \bar q_i q_i
\end{equation}
with
\begin{equation}
  \alpha_{q_i}\equiv\sum_{j=1}^3\frac{\chsnsn Y_{q_i}}{m_{H_j^o}^2}
  \label{alphaq}
\end{equation}
where $\chsnsn$ is the sneutrino-sneutrino-Higgs coupling, $Y_{q_i}$ is the
corresponding quark Yukawa coupling, and $i$ labels up-type quarks ($i=1$) and down-type quarks ($i=2$). 
The spin-independent part of the sneutrino-nucleon elastic scattering cross section thus reads
\begin{equation}
  \crosssec = \frac1\pi\frac{m_p^2}{(m_p+\snmassr)^2}\,f_p^2\,,
\end{equation}
where $m_p$ is the proton mass and 
\begin{equation}
  \frac{f_p}{m_p}=
  \sum_{q_i=u,d,s}f_{Tq_i}^p\frac{\alpha_{q_i}}{m_{q_i}}+ 
  \frac{2}{27}\ f_{TG}^p\sum_{q_i=c,b,t}\frac{\alpha_{q_i}}{m_{q_i}}\ .
  \label{fpsneutrino}
\end{equation}
The hadronic matrix elements, $f_{Tq}^p(=f_{Tq}^n=f_{Tq})$ and
$f_{TG}^p(=f_{TG}^n=f_{TG})$, are
defined as $\langle p |m_q \bar q q|p\rangle=m_pf_{Tq}^p$
and $f_{Tq}^p=1-\sum_{q=u,d,s}f_{Tq}^p$, and determined experimentally
as $f_{Tu}^p=0.020$, $f_{Td}^p=0.026$ and $f_{Tu}^p=0.229$.
Being a scalar field, the effective Lagrangian contains no axial-vector coupling and thus the spin-dependent cross section vanishes.
Using the explicit expressions of the quark Yukawa couplings, 
this quantity can be expressed as
\begin{equation}
  \frac{f_p}{m_p}=
  \frac{g}{2\mw\cos\beta}
  \sum_{j=1}^3\frac{\chjsnsn}{m_{H_j^o}^2}
  \left(\hcompjd\left(f_{Td}+f_{Ts}+\frac{2f_{TGP}}{27}\right)
  +\frac{\hcompju}{\tan\beta}\left(f_{Tu}+\frac{4f_{TGP}}{27}\right)
  \right)\,,
  \label{fprelic}
\end{equation}
where the term proportional to $\hcompjd$ corresponds to the interaction with the down-type quarks (the dominant contribution is due to the quark $s$) and the term proportional to $\hcompju$ corresponds to up-type quarks.

If the Higgs spectrum features a lightest Higgs with SM-like properties, i.e., with a mass of order $114-120$~GeV, then the approximation $4\snmassrsq\ll\hmasslsq$ holds for very light sneutrinos and the above equation can be approximated as
\begin{equation}
  \frac{f_p}{m_p}
  \approx
  \frac{0.31\,g}{2\mw\cos\beta}
  \left({\cal D}
  +\frac{0.42}{\tan\beta}\,{\cal U}
  \right)\,.
  \label{fprelicsm}
\end{equation}
If, on the other hand, the lightest Higgs is lighter than the SM-like one (and necessarily featuring a larger singlet composition), the contribution from this lightest Higgs generally dominates and expression (\ref{fprelic}) can be approximated as 
\begin{equation}  
  \frac{f_p}{m_p}
  \approx
  \frac{0.31\,g}{2\mw\cos\beta}
  \left({\cal D}_1
  +\frac{0.42}{\tan\beta}\,{\cal U}_1
  \right)\,\left(\frac{4\snmassrsq-\hmasslsq}{\hmasslsq}\right)\,.
  \label{fprelic1}
\end{equation}
where we have used the quantities ${\cal D}_1$ and ${\cal U}_1$ defined in Eq.(\ref{defcal1}). Notice that the approximation  $4\snmassrsq\ll\hmasslsq$ is not necessarily good now, since the Higgs can be very light, and this leads to the inclusion of the last factor. 

For moderate and large values of $\tan\beta$ the contribution from the $s$ quark is the leading one to the spin-independent cross section. 
However, the second term can become sizable and even dominate for small $\tan\beta$ and if the lightest Higgs is mostly $H_u^0$. Both in the case of the relic density, as well as in the scattering cross section, the condition that determines when the coupling from down-type quarks dominates over the coupling from up-type quarks is approximately the same, ${\cal D}_1\gg(0.4/\tan\beta)\,{\cal U}_1$ (see equations (\ref{omegaconu}) and (\ref{omegacond})). It should be emphasized that this rarely happens, but we include this possibility here for completeness.

The resulting spin-independent contribution to the RH sneutrino elastic scattering cross section off nuclei is then approximated as 
\begin{equation}
  \crosssec \approx \frac1\pi\frac{m_p^4}{(m_p+\snmassr)^2}
  \left(\frac{0.31\,g}{2\mw\cos\beta}\right)^2
  \left({\cal D}
  +\frac{0.42}{\tan\beta}\,{\cal U}
  \right)^2\,.
  \label{crosssm}
\end{equation}
In those cases where the lightest Higgs is lighter than the SM one, a better approximation (that incorporates the effect of the resonance) is 
\begin{equation}
  \crosssec \approx \frac1\pi\frac{m_p^4}{(m_p+\snmassr)^2}
  \left(\frac{0.31\,g}{2\mw\cos\beta}\right)^2
  \left({\cal D}_1
  +\frac{0.42}{\tan\beta}\,{\cal U}_1
  \right)^2\left(\frac{4\snmassrsq-\hmasslsq}{\hmasslsq}\right)^2\,.
  \label{crossud}
\end{equation}
We will use these expressions to extract some analytical predictions for the
detectability of sneutrinos.
Once more, in our numerical calculations the full expression for $\crosssec$ has been included, without using any numerical approximations.

\subsection{$\tilde N \tilde N \to f\bar f$}

Let us address first the case in which sneutrino annihilation into a pair of fermions is dominant. 
Using equations (\ref{abbacc}) and (\ref{defcal1}) the quantities ${\cal D}$ and ${\cal U}$ can be determined as a function of the lightest Higgs mass. Inserting these in Eq.\,(\ref{crosssm}) we obtain the following predictions for $\crosssec$,
\begin{equation}
  \crosssec \approx
  \frac{0.13\,\snmassrqb\,m_p^4\,a_{b\bar b}}
  {m_c^2(m_p+\snmassr)^2(\snmassrsq-m^2_{b})^{3/2}}\,
  ,\quad {\rm if}\quad
	{\cal D} \gg \frac{0.42}{\tan\beta}\,{\cal U}\,,
	\label{siabbsm}
\end{equation}
and
\begin{equation}
  \crosssec \approx
  \frac{0.13\,\snmassrqb\,m_p^4\,a_{c\bar c}}
  {m_b^2(m_p+\snmassr)^2(\snmassrsq-m^2_{c})^{3/2}}\,
  ,\quad {\rm if}\quad
	{\cal D} \ll \frac{0.42}{\tan\beta}\,{\cal U}\,.
	\label{siaccsm}
\end{equation}
In those cases where the lightest Higgs mass is small we can use expression (\ref{crossud}) and substitute the quantities ${\cal D}_1$ and ${\cal U}_1$ to obtain
\begin{equation}
  \crosssec \approx
  \frac{0.13\,\snmassrqb\,m_p^4\,a_{b\bar b}}
  {m_c^2(m_p+\snmassr)^2(\snmassrsq-m^2_{b})^{3/2}}\,\left(\frac{4\snmassrsq-\hmasslsq}{\hmasslsq}\right)^2
  ,\quad {\rm if}\quad
	{\cal D}_1 \gg \frac{0.42}{\tan\beta}\,{\cal U}_1\,,
	\label{siabb}
\end{equation}
and
\begin{equation}
  \crosssec \approx
  \frac{0.13\,\snmassrqb\,m_p^4\,a_{c\bar c}}
  {m_b^2(m_p+\snmassr)^2(\snmassrsq-m^2_{c})^{3/2}}\,\left(\frac{4\snmassrsq-\hmasslsq}{\hmasslsq}\right)^2
  ,\quad {\rm if}\quad
	{\cal D}_1 \ll \frac{0.42}{\tan\beta}\,{\cal U}_1\,.
	\label{siacc}
\end{equation}
In both cases  there exists a correlation between the scattering cross section and the annihilation cross section but with a slightly different proportionality factor.
Interestingly, imposing the correct relic density, and therefore the results for $a_{b\bar b}$ and $a_{c\bar c}$ of equations (\ref{omegaconu}) and (\ref{omegacond}), in the expressions above leads to a prediction of the cross section of order $10^{-4}$~pb for masses of order $\snmassr\sim8$~GeV, which can be compatible with the CoGeNT result.

\begin{figure}[t!]
  \hspace*{-0.5cm}
  \epsfig{file=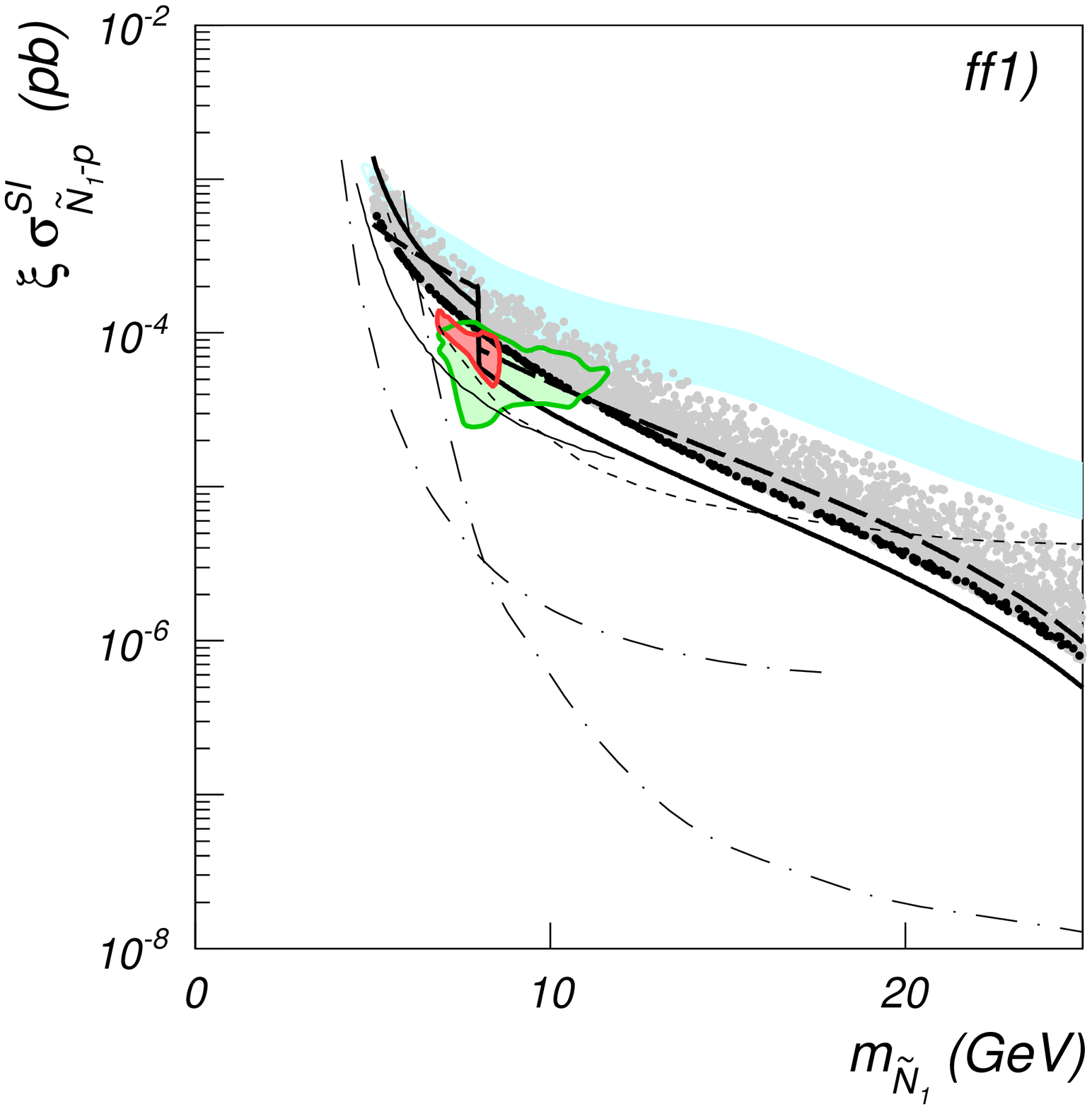,width=8.cm}
   \epsfig{file=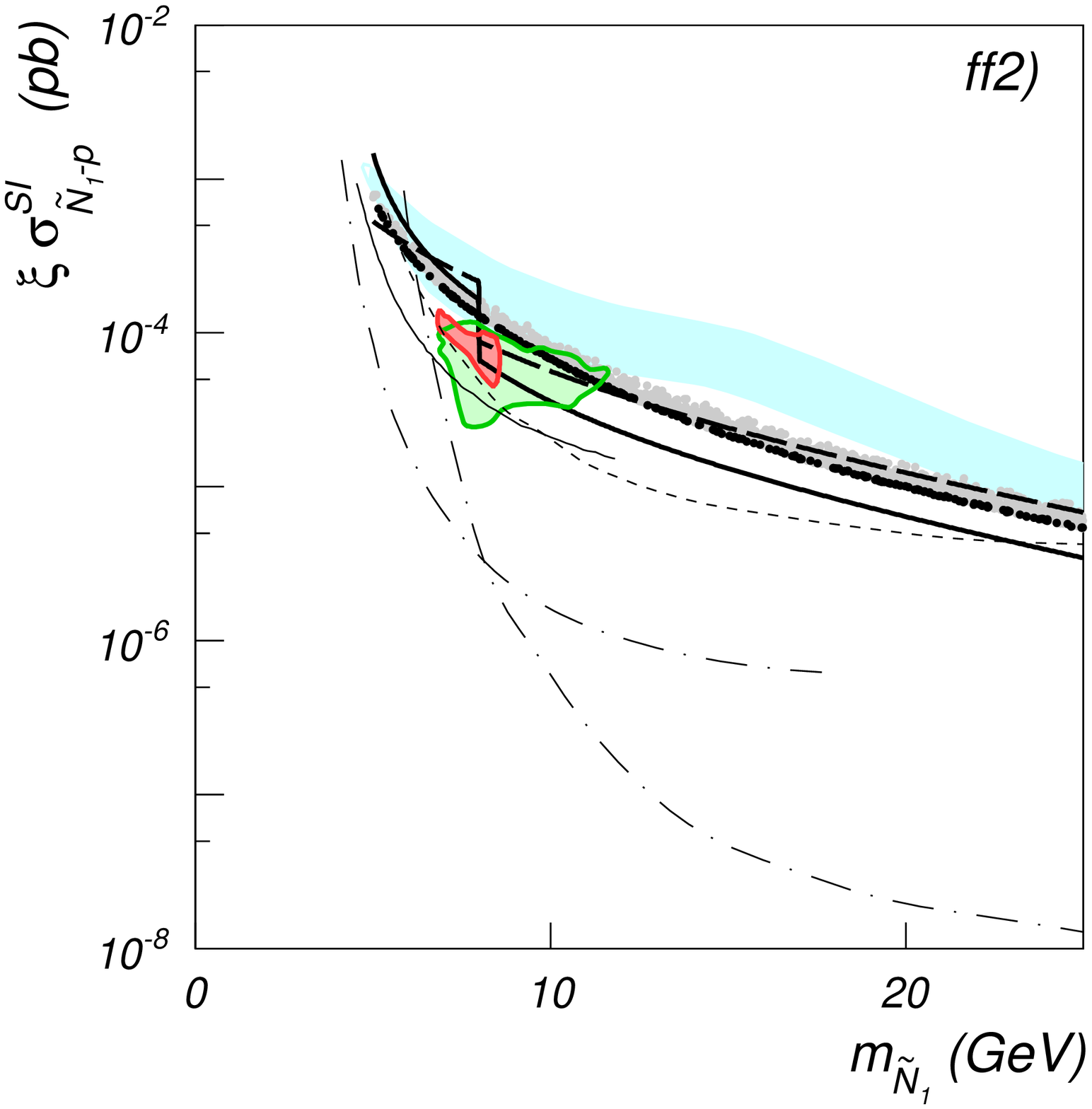,width=8.cm}
  \captions{Spin-independent part of the RH sneutrino-quark elastic cross section as a function of the RH sneutrino mass for cases ff1) and ff2) and a scan in $\mn$, $\ln$ and $\aln$. The black points correspond to those with the correct relic abundance, whereas the gray ones are those in which $\snrelic$ is smaller than the WMAP constraint. The regions compatible with the CoGeNT observation of an irreducible excess and annual modulation are shown as the large (green) and small (red) areas bounded by a solid line. The shaded (blue) area is consistent with the DAMA/LIBRA experiment if no channelling effects are considered. The dashed, solid and dot-dashed lines correspond to the exclusion regions from the SIMPLE, CDMS and XENON experiments.}
    \label{fig:crossff}
\end{figure}

The theoretical predictions for $\crosssec$ are plotted as a function of the RH sneutrino mass in Fig.\,\ref{fig:crossff} for cases ff1) and ff2) of Table\,\ref{tab:cases} for which we have performed a random scan in the $\mn$, $\ln$ and $\aln$ parameters, retaining only those points for which the RH sneutrino relic abundance is in agreement with the WMAP constraint (black dots) or smaller (gray dots).
In case ff1) we have also represented the analytical approximation of Eq.\,(\ref{siabb}) as a thick solid line, which turns out to be a qualitatively good approximation. Deviations happen because the numerical computation of the relic density is more precise and also due to the contribution of the second-lightest Higgs to $\crosssec$. 
For illustrative purposes we also plot with a thick dashed line the theoretical predictions for the cross section if the term $a_{c\bar c}$ had been the dominant contribution to the annihilation cross section, as computed in Eq.\,(\ref{siacc}). 
Similarly, in case ff2) the thick and solid lines correspond to the approximations  Eqs.\,(\ref{siabbsm}) and (\ref{siaccsm}) obtained when annihilation into $b\bar b$ or into $c\bar c$ is dominant and including the contribution from the three CP-even Higgs bosons. 
The green area in both figures corresponds to the region consistent with the first CoGeNT results and the narrower red area is compatible with their latest ones. The cyan region is compatible with DAMA/LIBRA. Finally, the dashed, solid and dot-dashed lines correspond to the exclusion regions from the SIMPLE, CDMS and XENON experiments, respectively. 

Uncertainties in the hadronic matrix elements have not been included in the plot, but their effect is very easy to understand. 
The largest effect is due to the indetermination in the strange quark content of the quark. This propagates into the theoretical predictions for $\crosssec$ and can be responsible for a variation of about an order of magnitude \cite{Bottino:2001dj,Bottino:2011b,eoss-uncertainties,Ellis:2008hf}.  

From our results we conclude that the predicted $\crosssec$ can be in agreement with the CoGeNT region without having demanded any further constraint, and solely as a consequence of the correlation between the diagrams that contribute to the RH sneutrino annihilation cross section and those for direct detection.
This is however not only true for these particles, as it also happens with some other well-motivated WIMPs. In particular, this has already been pointed out for very light neutralinos both in the MSSM \cite{Bottino:2002ry,Bottino:2003cz,Cerdeno:2004zj,Bottino:2008mf} (although they are very constrained by low-energy observables \cite{Feldman:2010ke}) as well as in the NMSSM \cite{Aalseth:2008rx}.
A question worth investigating is then whether or not very light sneutrino dark matter can be distinguishable from other possible WIMPs.
In this sense, including information from other sources of dark matter detection, such as indirect and collider searches, can shed some light on the specific model.

\subsection{$\snr\snr \to \phiggsl\phiggsl$}

Let us describe now the situation when annihilation into a pair of very light pseudoscalars is dominant.
Following from the discussion about the relic abundance in Section\,\ref{sec:verylight} we see how in this case there is no correlation between the diagram contributing to the direct detection (which is mostly dependent on the $\ln$ parameter) and those for annihilation cross section (which are now dependent on a combination of various parameters). 

The implications for direct detection are easy to understand. If the annihilation cross section is dominated by the channel $\snr\snr\to\phiggsl\phiggsl$, in the regions with the correct relic abundance the contribution from $\snr\snr\to f\bar f$ is necessarily smaller than the one determined in the previous section.
In other words, $a_{b\bar b}$ and $a_{c\bar c}$ in expressions (\ref{omegaconu}) and (\ref{omegacond}) have now a lower value since the corresponding term $a_{\phiggsl\phiggsl}$ would provide the leading contribution to the calculation of the relic abundance. This entails a decrease in the predicted $\crosssec$, according to Eqs. (\ref{siabb}) and (\ref{siacc}).

\begin{figure}[t!]
  \hspace*{-0.5cm}
  \epsfig{file=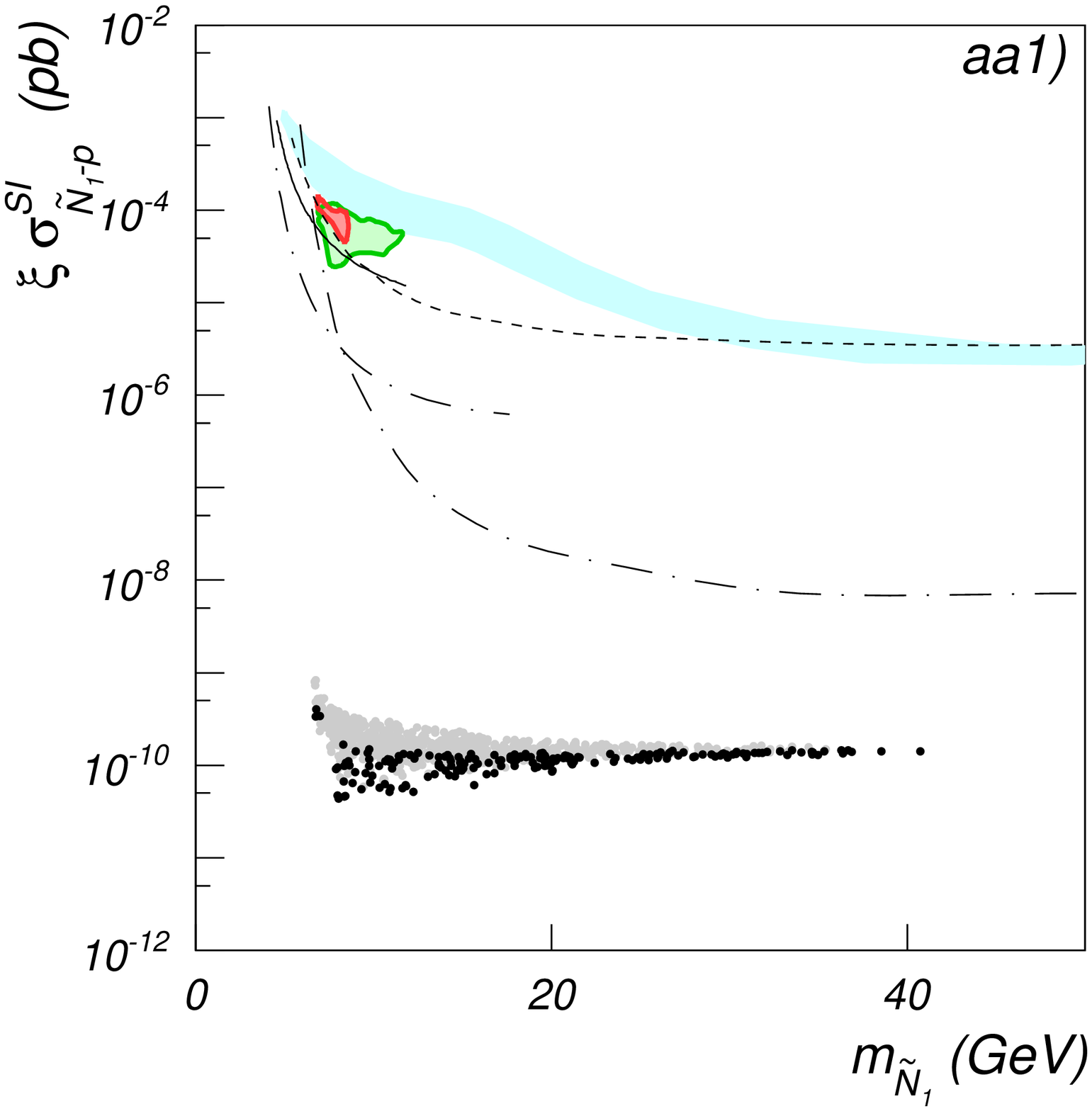,width=8.cm}
   \epsfig{file=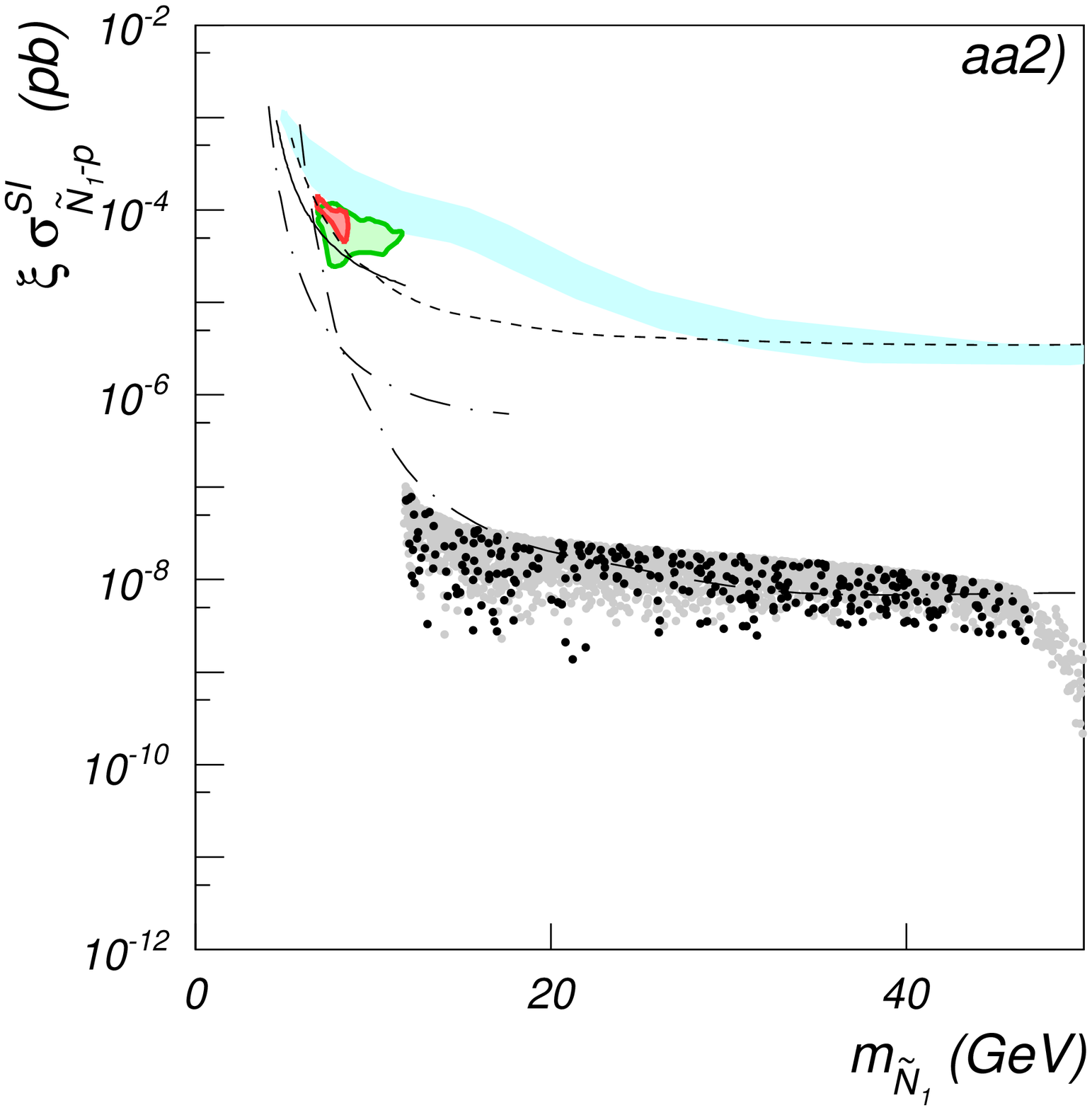,width=8.cm}
  \captions{The same as in Fig.\,\ref{fig:crossff} but for cases aa1) and aa2) in Table\,\ref{tab:cases}.}
    \label{fig:crossaa}
\end{figure}

As a consequence, points obtained in this scenario do not reproduce the CoGeNT results. Notice however that if the exclusion regions set by other experiments are taken at face value, these RH sneutrinos would still be allowed and survive as very light WIMP dark matter. We illustrate this in Fig.\,\ref{fig:crossaa}, where the theoretical predictions for $\crosssec$ as a function of the RH sneutrino mass are represented for cases aa1) and aa2). 
As we can see, these very light WIMPs can have a scattering cross section which is not in tension with current exclusion regions. The predictions can be very low as in case aa1) or within the reach of future experiments such as SuperCDMS. 

A similar phenomenon happens with very light neutralinos in the NMSSM when the resonant annihilation through a light pseudoscalar Higgs is invoked in order to account for the correct relic abundance. In that case the predictions for sneutrino direct detection can be as low as $\neutcrosssec\sim10^{-7}-10^{-10}$~pb \cite{Aalseth:2008rx,Cao:2011re,Cumberbatch:2011jp}.
Thus, although very light RH sneutrinos with annihilation into a pseudoscalar pair are clearly distinguishable from MSSM neutralinos, they might still be confused with neutralino dark matter in the NMSSM. However, we will see in the next section that the signals in colliders might differ. 

\subsection{$\snr\snr \to \rhn\rhn$}

Let us finally address the scenario in which annihilation into a pair of RH neutrinos dominates.
As in the case of annihilation into a pseudoscalar pair, the smallness of the $\ln$ parameter in the regions with the correct relic density implies that the resulting spin-independent RH sneutrino-proton cross section is significantly suppressed. 
Once more, this happens because the Feynman diagrams that contribute to RH sneutrino annihilation are unrelated to those for direct detection. 
As a consequence, points of the parameter space where $\snr\snr\to\rhn\rhn$
is the main annihilation channel would not account for the experimental
results of the CoGeNT collaboration, if these are confirmed, but would still survive the bounds imposed by CDMS and XENON.

\begin{figure}[t!]
  \hspace*{-0.5cm}
  \epsfig{file=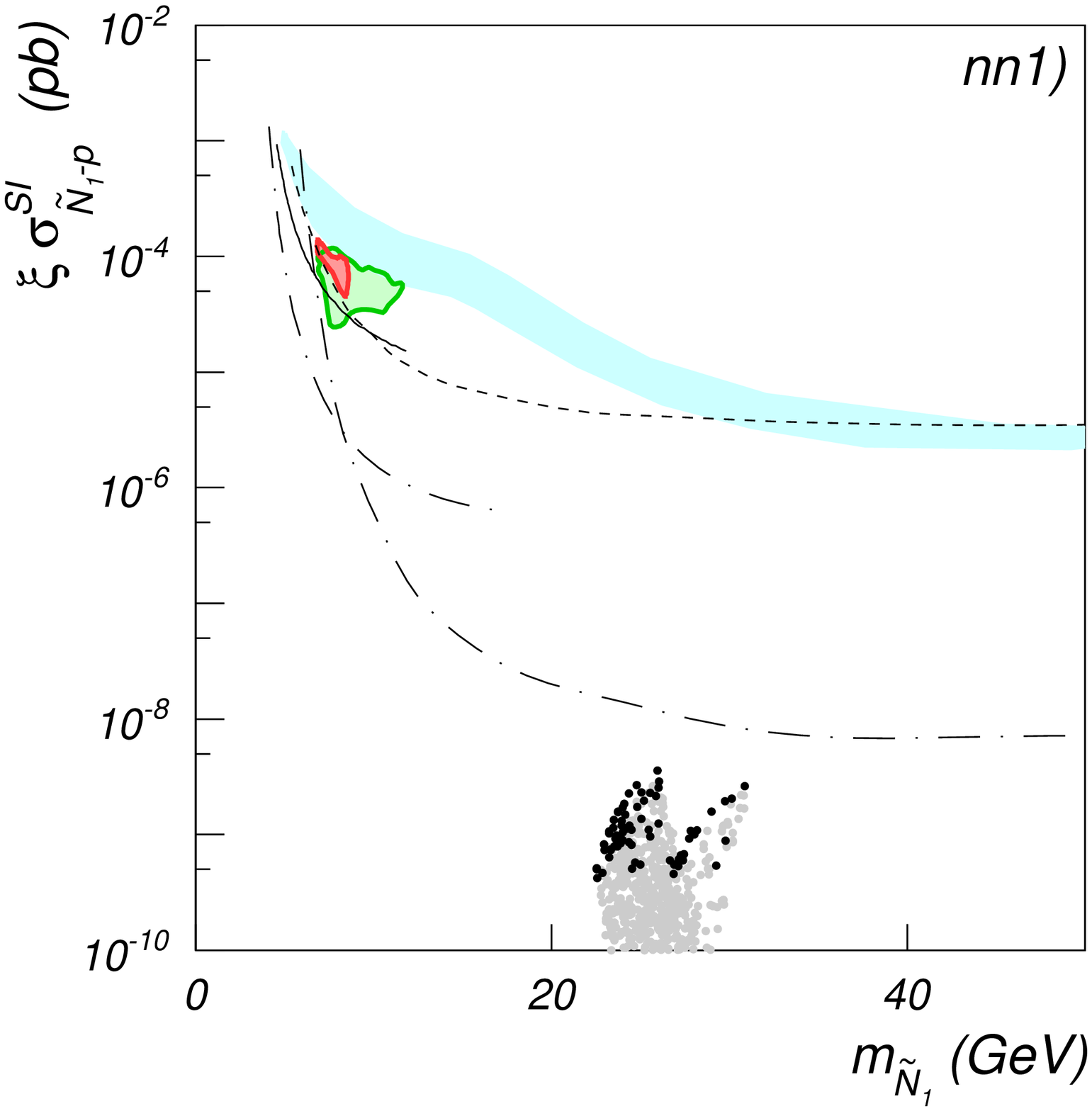,width=8.cm}
  \epsfig{file=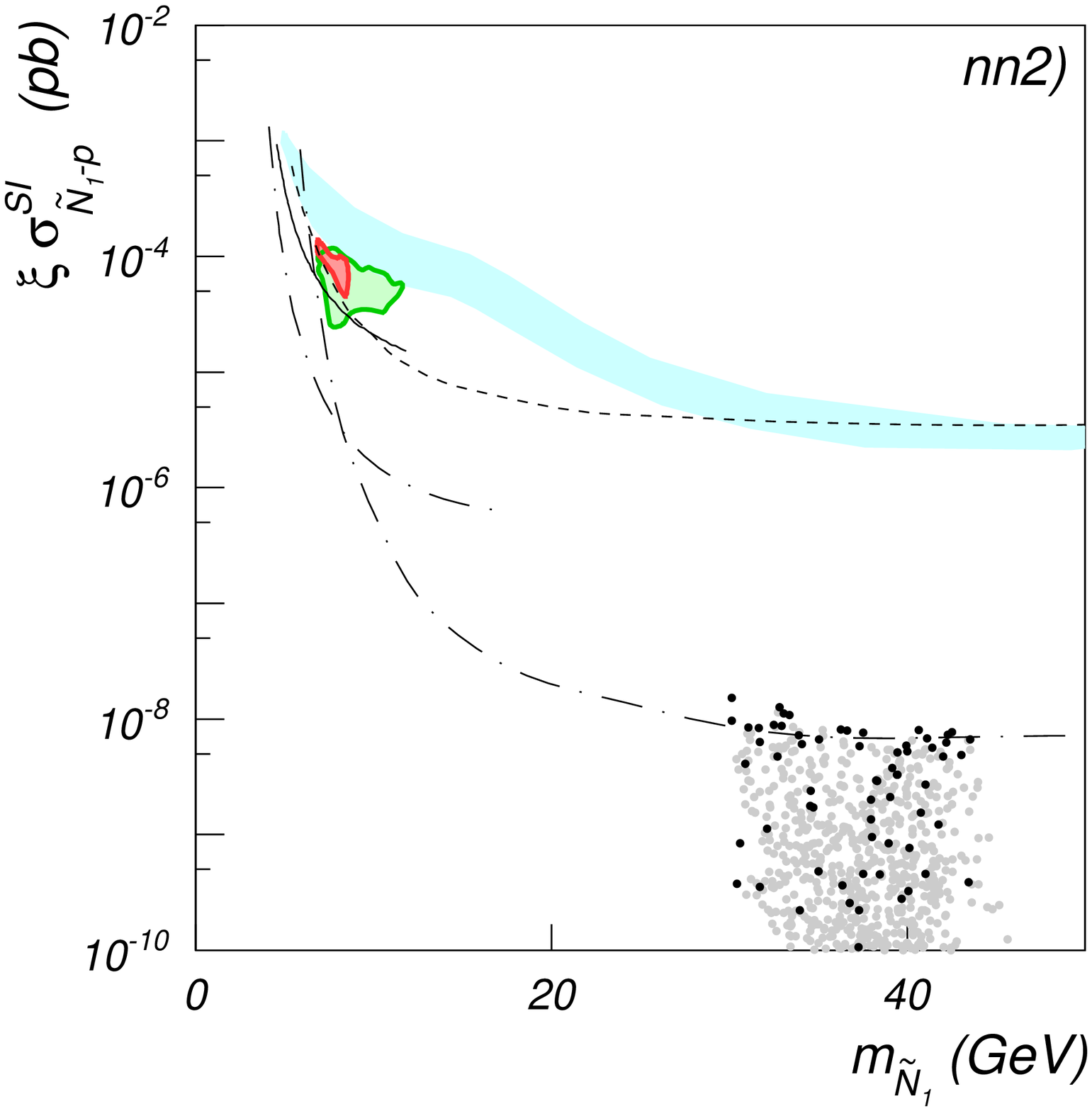,width=8.cm}
  \captions{The same as in Fig.\,\ref{fig:crossff} but for cases nn1) and nn2) in Table\,\ref{tab:cases}.}
    \label{fig:crossnn}
\end{figure}

This is clearly illustrated in Fig.\,\ref{fig:crossnn}, where the theoretical predictions for $\crosssec$ as a function of the sneutrino mass are represented for examples nn1) and nn2) of Table\,\ref{tab:cases}. 
The points with correct relic abundance accumulate in the narrow region where resonant annihilation is possible and can be as low as $\crosssec\sim10^{-8}-10^{-10}$~pb.

We have also found it very difficult to obtain viable RH sneutrinos with $\snmassr\lesssim 20$~GeV, as the parameters become extremely fine-tuned.
From the point of view of direct detection, this example looks very similar to the one described in the previous section, and therefore also to very light neutralinos in the NMSSM.

\section{Invisible Higgs decay}
\label{sec:colliders}

A common feature of most dark matter models featuring light WIMPs is the occurrence of a new invisible channel for the Higgs decay, namely the production of a dark matter pair. This also occurs in our scenario, since the light RH sneutrino has a sizable coupling to the CP-even Higgs bosons, and the resulting phenomenology can be very dependent on the specific scenario for sneutrino annihilation. 

For any of the CP-even Higgses, the decay width of the process $\higgsi\to\tilde N\tilde N$ can be expressed as
\begin{eqnarray}
\Gamma(H_i^0 \rightarrow \tilde{N}\tilde{N} )
 &=& \frac{1}{8 \pi m_{H_i}^2}
 \sqrt{ \frac{m_{H_i}^2}{4}-\snmassrsq }
 |{\cal M} |^2 \frac{1}{2}  \nonumber \\
 &=& \frac{1}{32 \pi m_{H_i^0}^2} 
 | \chisnsn |^2 
 \sqrt{m_{H_i^0}^2-4 \snmassrsq }\nonumber \\
 & \approx & \frac{1}{32 \pi m_{H_i^0} } | \chisnsn |^2
 \quad {\rm for } \quad m_{H_i^0} \gg 2\snmassr\ .
\label{HiIntoSnu}
\end{eqnarray}
As in other models for very light WIMPs, this decay mode can dominate for the lightest Higgs. This can be the case of the very light neutralino, both in the MSSM and NMSSM, but also for a generic light singlet scalar \cite{Burgess:2000yq,hm1,hm2,hm3,hm4,hm5,Kim:2009ke,Andreas:2010dz}. 
In our case the couplings $\chisnsn$ are very dependent on the properties of each of the annihilation scenarios that we presented in the previous Section, therefore we study each case separately.

\subsection{$\snr\snr \to f\bar f$}

In order to estimate the invisible Higgs branching ratio we have to compare its decay width into RH sneutrinos with the decay width into fermions, which normally account for the main visible decay channels\footnote{If the decay into a pair of very light pseudoscalars, $\higgsl\to\phiggsl\phiggsl$, is open, its contribution can also be sizable.}. 
The ratio between the lightest Higgs decay widths into RH sneutrinos and $b\bar b$ can be expressed as
\begin{equation}
	R_{\snr\snr/b\bar b}\approx \frac{|\chlsnsn|^2}{6\hmassl^2\,|\hcompld|^2}\left(\frac{2\mw\sin\beta}{g\,m_b}\right)^2\,,
\end{equation}
and we can define a similar quantity to compare with decays into $c\bar c$,
\begin{equation}
	R_{\snr\snr/c\bar c}\approx \frac{|\chlsnsn|^2}{6\hmassl^2\,|\hcomplu|^2}\left(\frac{2\mw\cos\beta}{g\,m_c}\right)^2\,,
\end{equation}
where we have used the approximation that $\hmassl^2\gg4\snmassrsq$. 
In Section.\,\ref{sec:verylight} we introduced two possible regimes in which the correct relic density could be obtained for a sneutrino annihilating into $f\bar f$. Depending on the lightest Higgs composition, the quantities ${\cal D}$ or ${\cal U}$, defined in Eq.(\ref{defcal}), provide the leading term for the annihilation into $b\bar b$ or $c\bar c$, respectively.

If we now impose that these sneutrinos reproduce the results from the CoGeNT
experiment $(\crosssec\sim10^{-4}\,{\rm pb}\sim2.6\times10^{-13}\,{\rm GeV}^{-2})$, the value of the coupling $|\chisnsn|$ can be determined through equation (\ref{crossud}), resulting in 
\begin{equation}
	R_{\snr\snr/b\bar b}\approx 2.8\times10^{-4}\,
	\frac{\cos^4\beta}{|\hcompld|^4}\,
	\frac{\hmassl^2\,\left(m_p+\snmassr\right)^2}{m_p^4}\,.
\end{equation}
It can easily be seen that this ratio is typically much larger than one for realistic examples. For example, if the Higgs mass is small then the Higgs mass has to be
mostly singlet, in which case $|\hcompld|$ becomes very small too and results in $R_{\snr\snr/b\bar b}\gg 1$. If the Higgs mass increases, the $H_d$ composition becomes larger (closer to 1) and it can be explicitly checked that in the limiting case of a SM-like Higgs with a mass of order 114~GeV this ratio is still large. Remember in this sense that $\tan\beta$ is small in our scenarios. 
This already implies that the lightest Higgs, irrespectively of its composition and mass, has a very large invisible decay width.

If we were in the regime where ${\cal U}\gg{\frac{0.31}{\tan\beta}\cal D}$ then Higgs decay into a $c\bar c$ pair would become larger than in $b\bar b$ (in fact, this coincides with condition under which sneutrino annihilation proceeds into $b\bar b$ or $c\bar c$). In such a case, the ratio of this decay mode with Higgs decay into sneutrinos can be written as
\begin{equation}
	R_{\snr\snr/c\bar c}\approx 1.7\times10^{-2}\,
	\frac{\sin^4\beta}{|\hcomplu|^4}\,
	\frac{\hmassl^2\,\left(m_p+\snmassr\right)^2}{m_p^4}\,,
\end{equation}
and the same considerations as above would apply. 

Thus, the lightest Higgs will always tend to decay mostly into invisible particles. Notice that these results are independent on whether ${\cal D}$ or ${\cal U}$ dominates in Eq.(\ref{defcal}).

\begin{figure}[t!]
  \epsfig{file=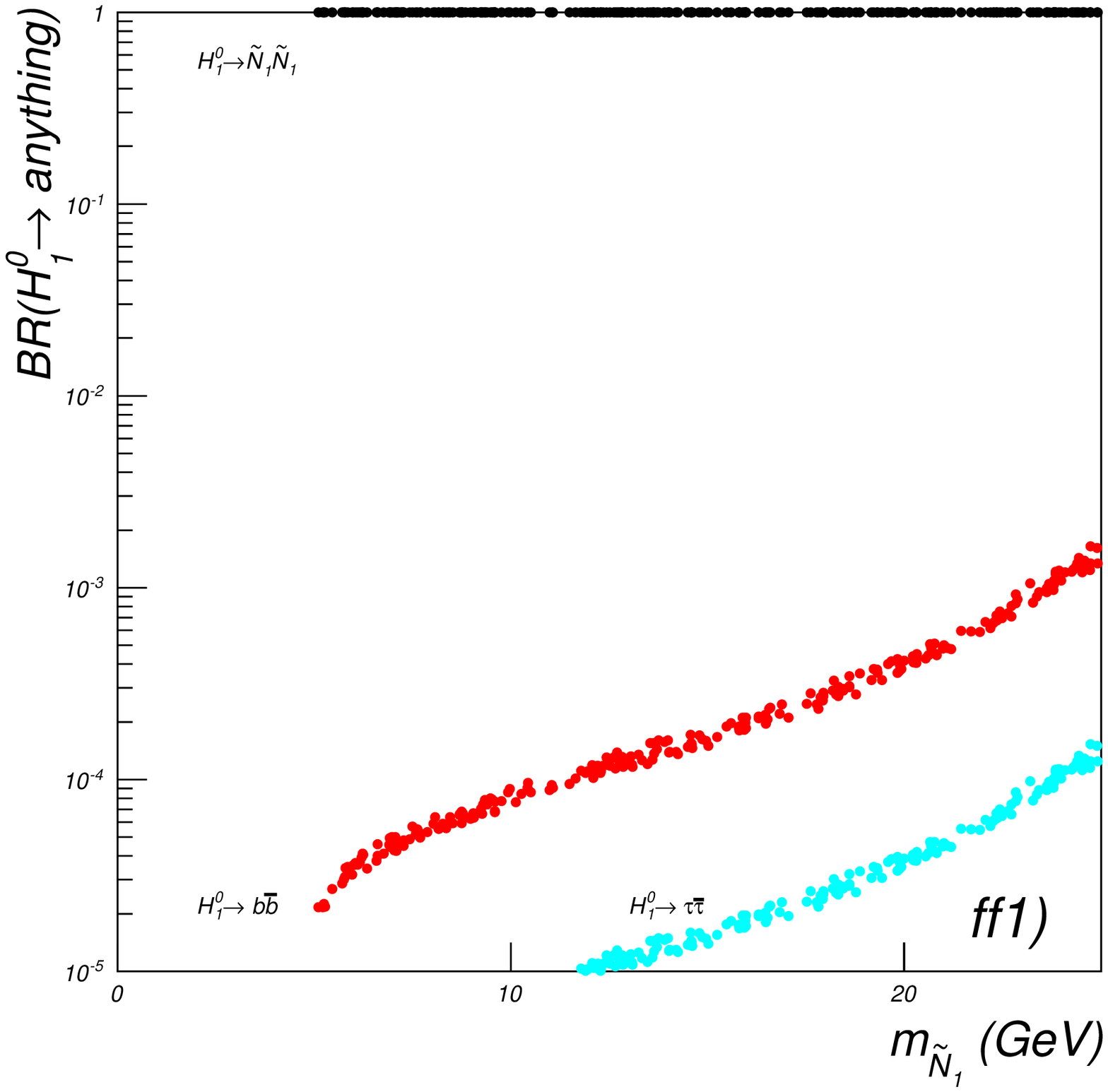,width=8.cm}  \epsfig{file=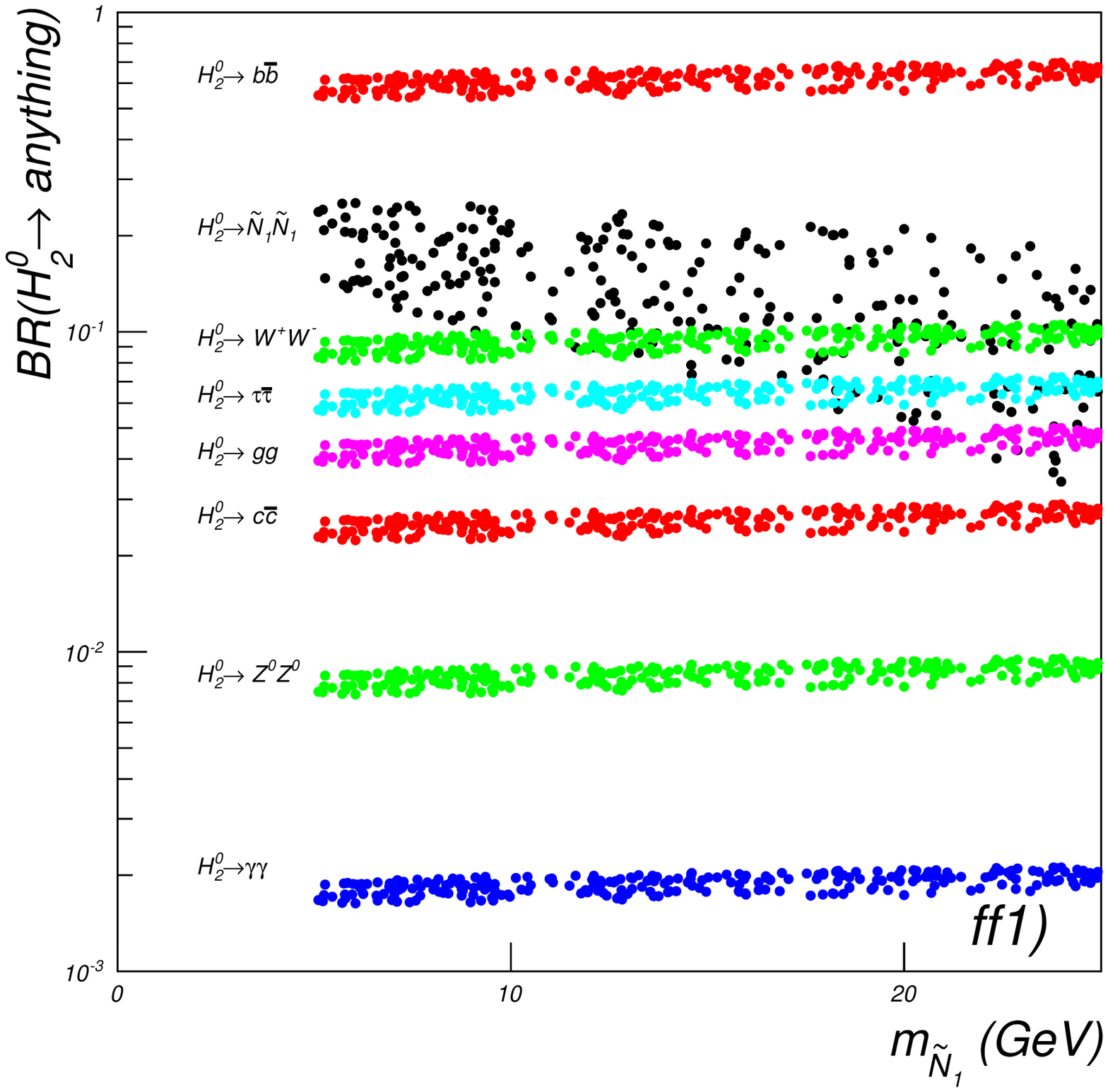,width=8.cm}\\
  \epsfig{file=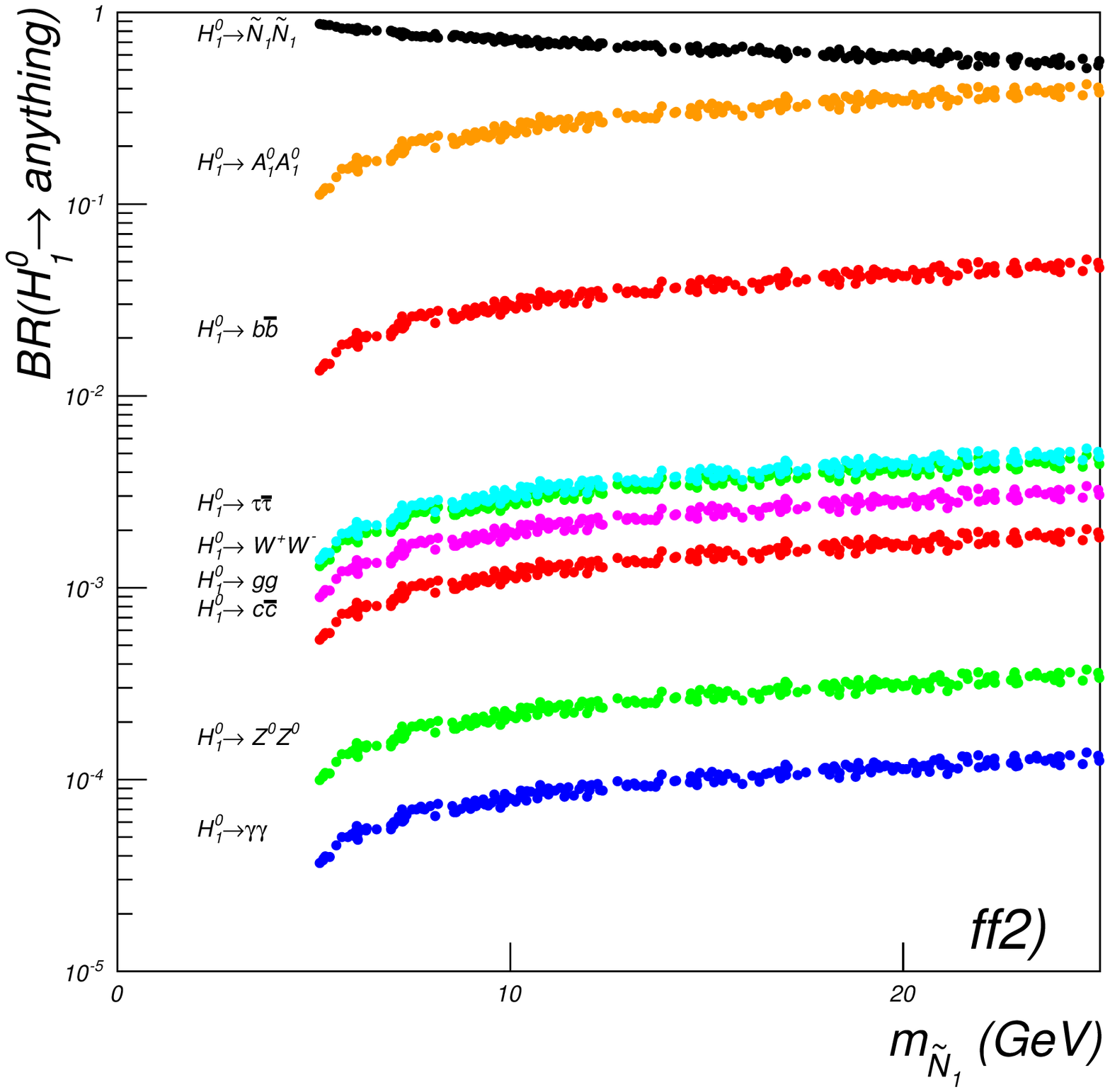,width=8.cm}  \epsfig{file=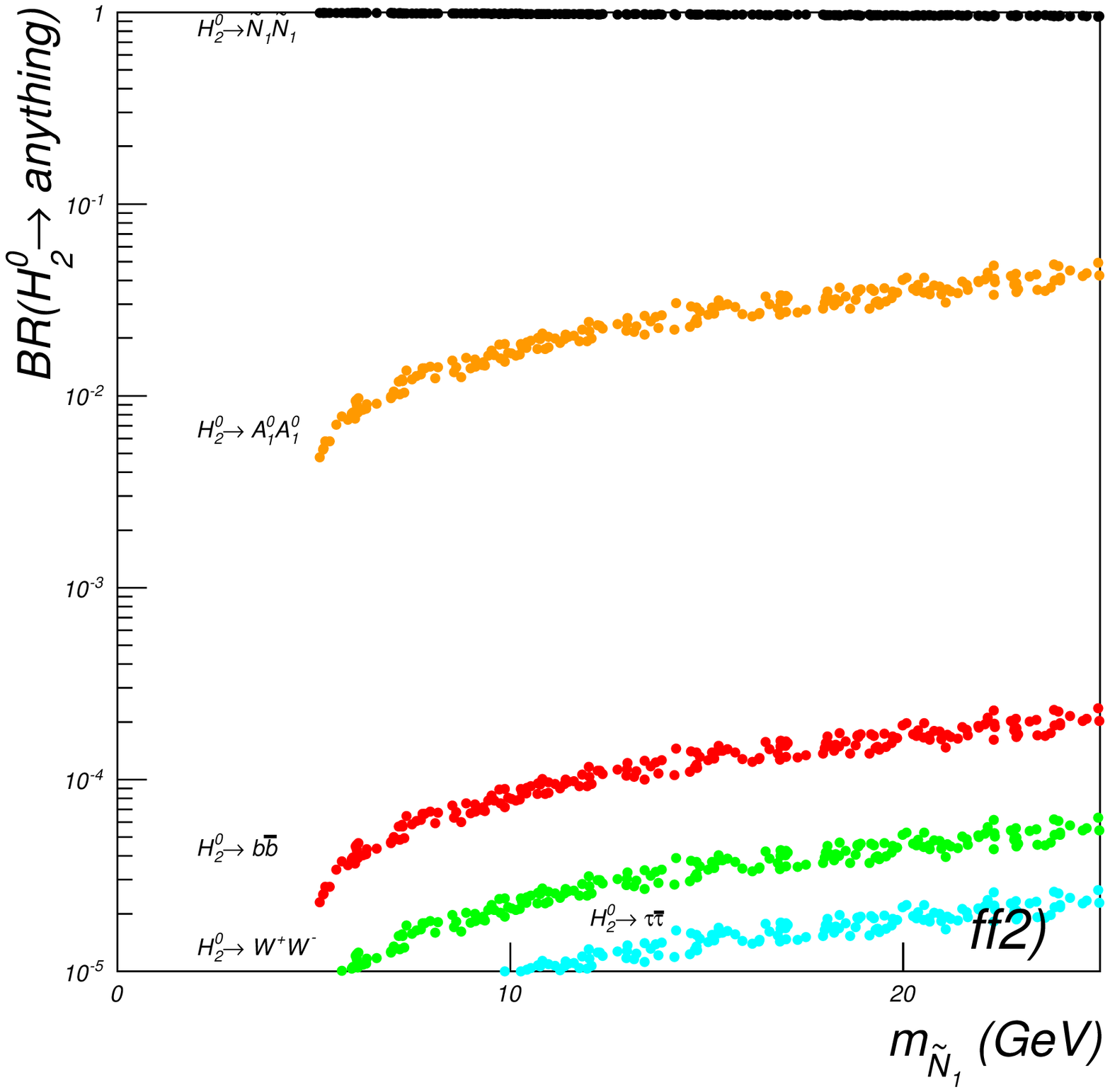,width=8.cm}
  \captions{Branching ratios of the decays of the lightest Higgs (left column) and second lightest Higgs (right column) in the possible different final states for cases ff1) and ff2) of Table\,\ref{tab:cases}. 	
  Only the points with correct relic abundance are plotted.}
  \label{fig:brnn_ff}
\end{figure}

Regarding the decay of the second lightest Higgs, the RH sneutrino coupling to $H_2^0$ is less constrained by the relic abundance condition and therefore we can have several possibilities. If the lightest Higgs is very light (e.g., below $100$~GeV) and therefore singlet-like, the most conventional situation is that the second lightest Higgs is similar to that of the SM, with a mass of order $110-120$~GeV, a large $H_u$ composition and a small but non-negligible $H_d$. Such a Higgs will decay mostly into a $b\bar b$ pair but it is likely that the presence of the very light sneutrino also induces a significant invisible decay width. 

This is actually the case of our example ff1). The branching ratios of the lightest and second-lightest Higgs are represented on the top row of Fig\,\ref{fig:brnn_ff}. Black dots correspond to the decays into a pair of RH sneutrinos and the rest of the decay products are indicated by different colours. As explained above  $H_1^0$ is invisible, decaying mostly into RH sneutrinos. On the other hand, $H_2^0$ (with a mass of $119$~GeV) could be observed through its decay into $b\bar b$. Notice however that this is significantly reduced with respect to an ordinary SM Higgs.

\begin{figure}[t!] 
	\begin{center}
	\epsfig{file=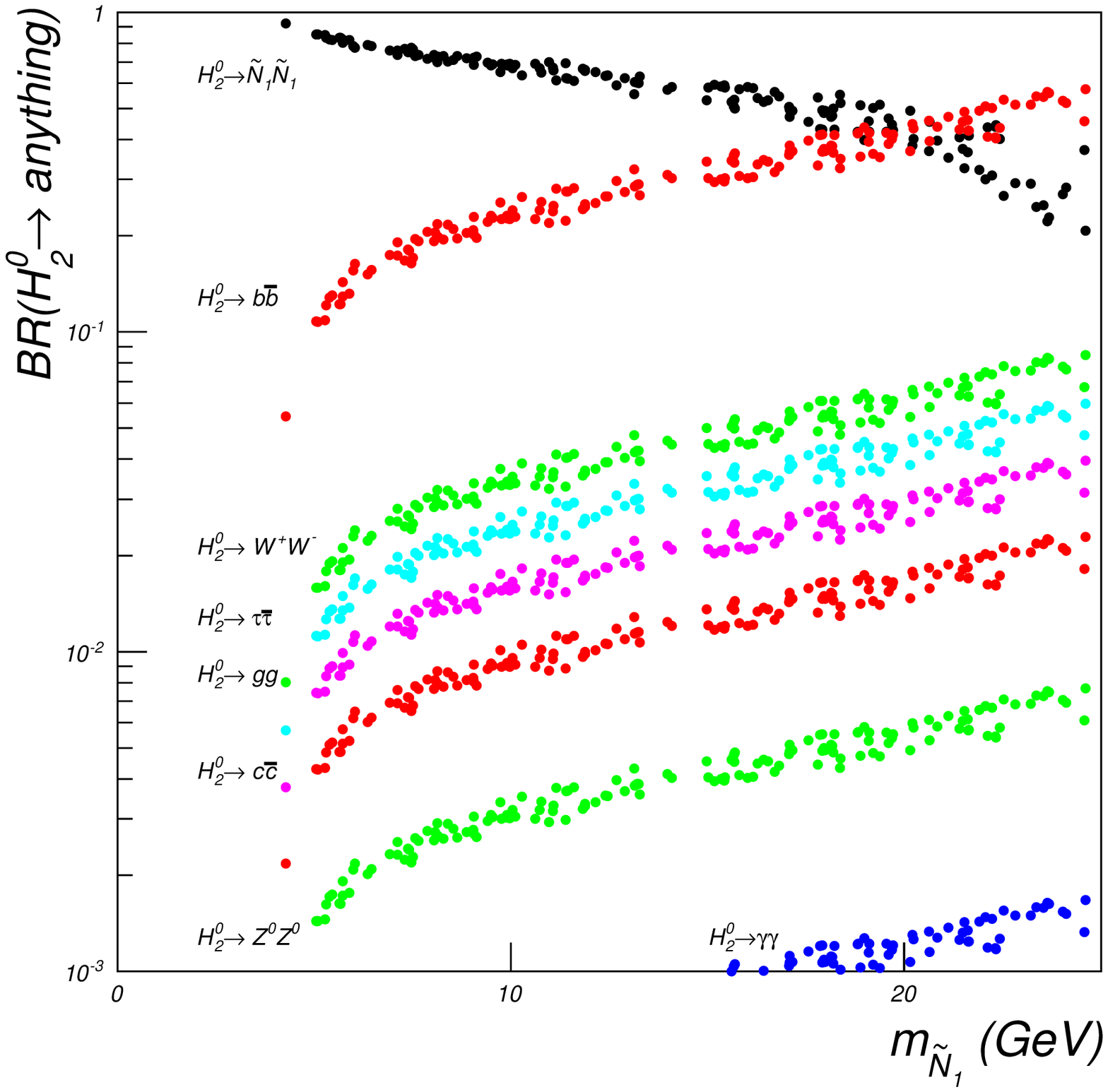,width=8.cm}
	\end{center}
	\captions{The same as in Fig.\,\ref{fig:brnn_ff} but with $\al=500$~GeV}
  \label{fig:brnn_ffb}
\end{figure}

A small change in the input parameters can alter significantly the
phenomenology of the second-lightest Higgs. For example, in case ff1) if we
take $\al=500$~GeV instead of $550$~GeV the invisible branching ratio
increases considerably and can even be dominant for light RH sneutrino masses. This is shown in Fig.\,\ref{fig:brnn_ffb}.

Another possibility for the second-lightest Higgs is that, if the lightest Higgs is SM-like and with a mass around $110-120$~GeV, $H_2^0$ can be mostly singlet. In such a case, not only the light SM-like Higgs is invisible (as already explained above), but also this second  Higgs. 
Case ff2) is one explicit example of this kind of scenarios. 
The branching ratios for $H_1^0$ and $H_2^0$ for this example are depicted at the bottom row of Fig\,\ref{fig:brnn_ff}.

Notice that in both cases the lightest pseudoscalar can also be relatively light, opening the decay channel $H_i^0\to \phiggsl\phiggsl$. Although this channel can be very efficient, it is nevertheless not sufficient to compete with decays into RH sneutrinos.

We therefore observe that, if the result of CoGeNT is imposed, the lightest Higgs is invisible, irrespectively of whether it is mostly singlino (and light) or the SM-like Higgs with a mass of around $110$ to $120$~GeV. 
Furthermore, in some scenarios also the second lightest Higgs can have a sizable invisible decay width.
This is a very interesting property of this scenario and provides some potential discrimination criterium to distinguish it from the case of very light neutralinos. 
In particular, in Ref.\,\cite{Cao:2011re} it was shown that in the case of very light NMSSM neutralinos the SM-like Higgs decay predominantly into a pair of very light Higgses.

\subsection{$\snr\snr \to \phiggsl\phiggsl$}

\begin{figure}[t!]
  \epsfig{file=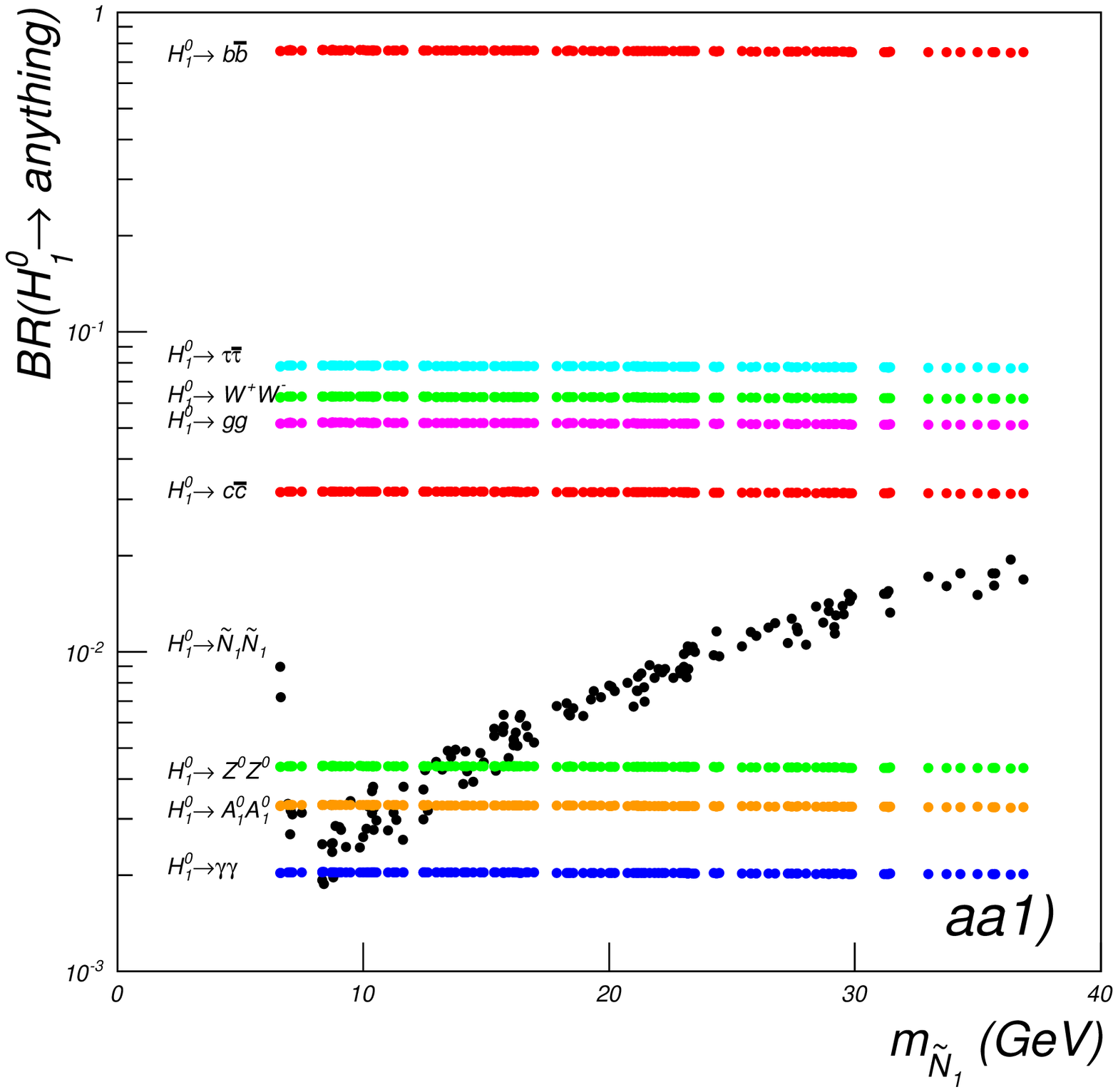,width=8.cm}  \epsfig{file=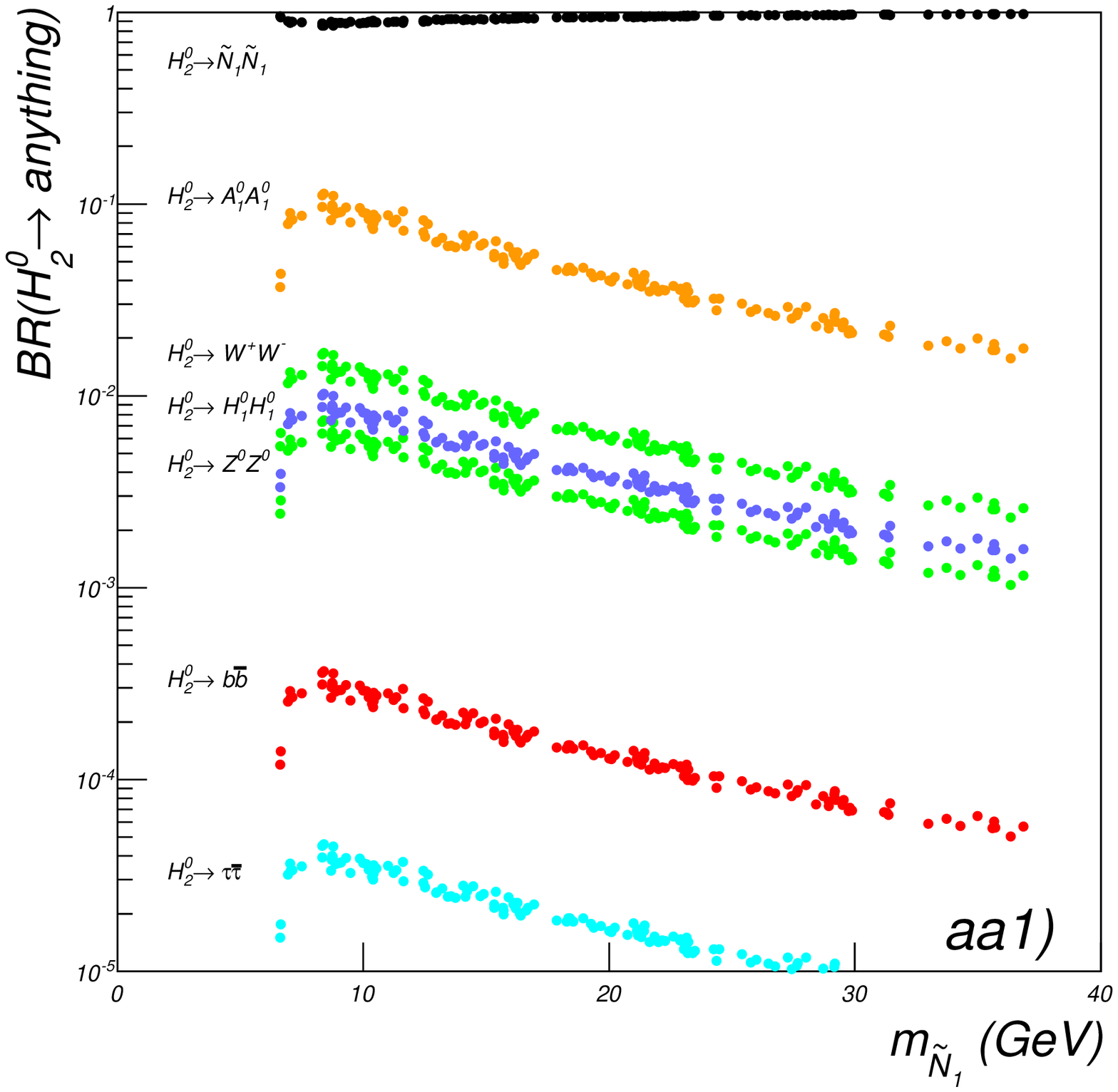,width=8.cm}\\
  \epsfig{file=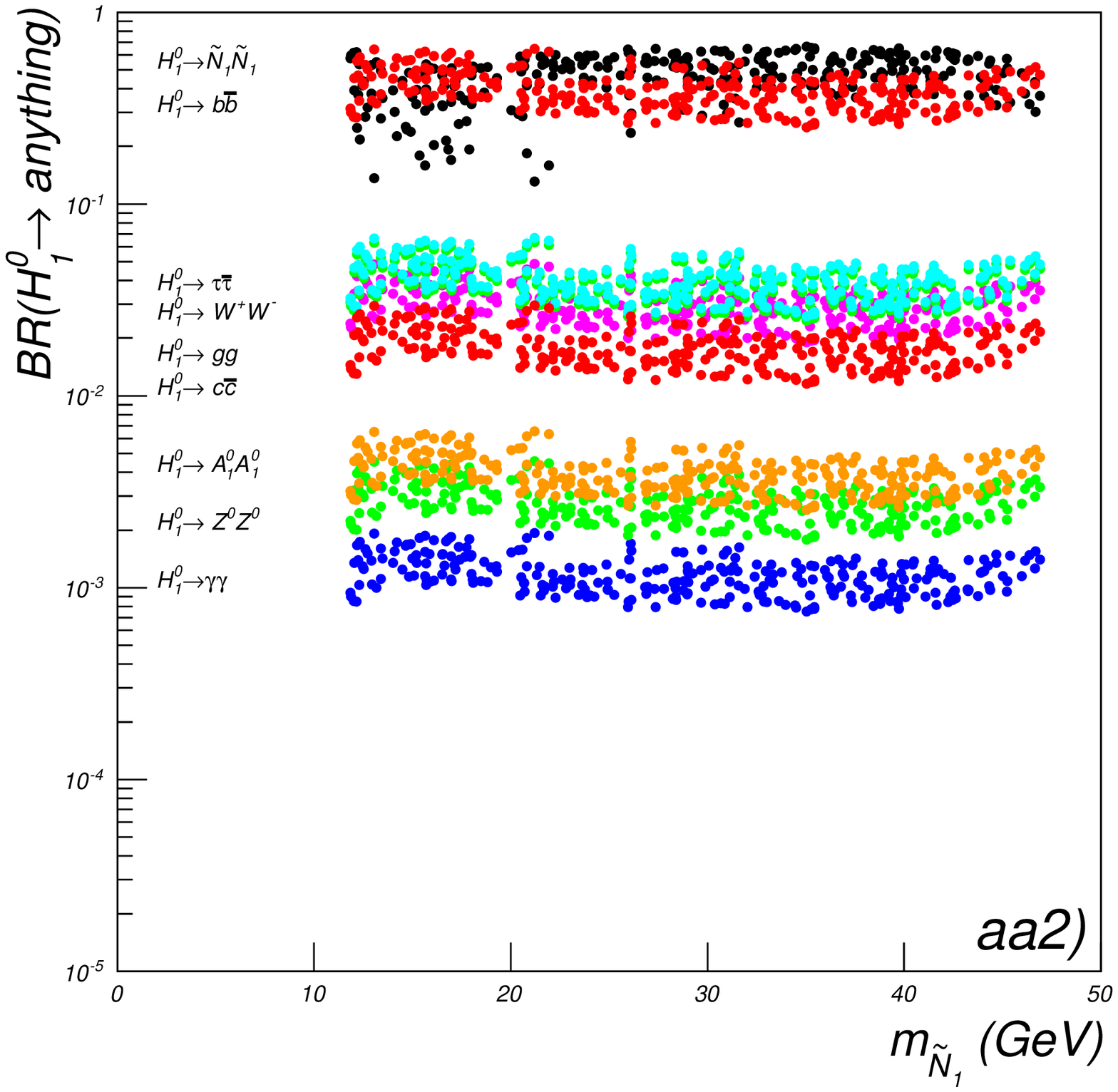,width=8.cm}  \epsfig{file=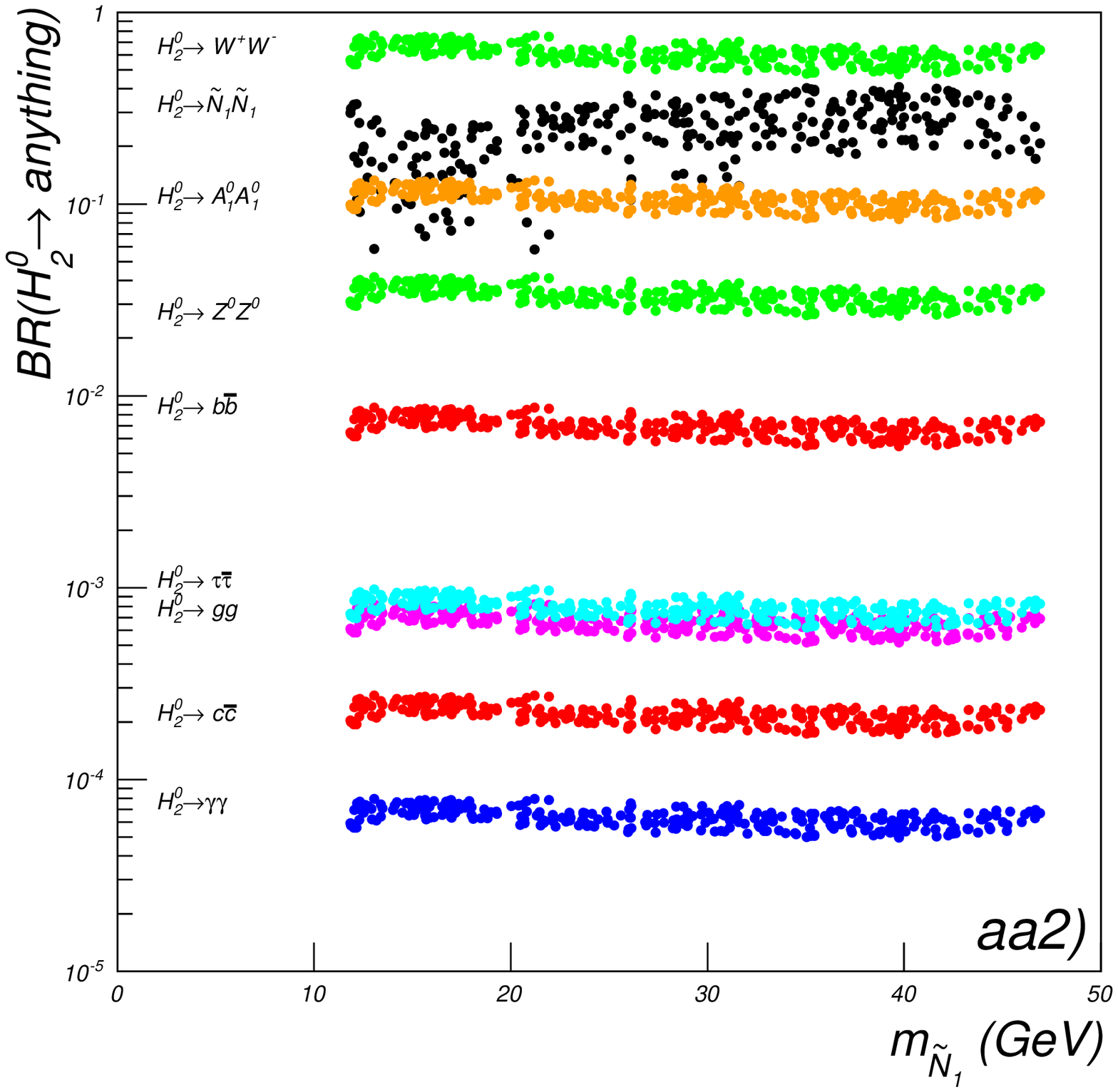,width=8.cm}
  \captions{The same as in Fig.\,\ref{fig:brnn_ff} but for cases aa1) and aa2) of Table\,\ref{tab:cases}.}
  \label{fig:brnn_aa}
\end{figure}

This example is qualitatively different from the one described in the previous Subsection. 
In particular, the lightest Higgs is SM-like and therefore the annihilation into $b\bar b$ is typically large. 
The coupling of the lightest Higgs to the RH sneutrino pair is very dependent on the parameter space and this has an impact on the predictions for the invisible branching ratio.
Now, since the coupling $\ln$ takes smaller values than in the previous section we may expect that the contribution to the invisible decay width of all the CP-even Higgses is suppressed.

This happens in case aa1), where the lightest Higgs has a large $H_u$ component. Annihilation into $b\bar b$ constitutes now the main decay channel, thereby making it look similar to a SM Higgs. The predictions for the different branching ratios are displayed on the upper left-hand side of Fig.\,\ref{fig:brnn_aa}, where we observe that $BR(\higgsl\to\snr\snr)$ is significantly smaller than in the previous section.
On the other hand, in case aa2) the first and second lightest Higgses are closer in mass and display a larger mixing. For example, despite having the same mass as in the previous example ($114$~GeV), the lightest Higgs has now a much larger singlet component. This is enough to enhance the predictions for the invisible branching ratio and, as we can observe in Fig.\,\ref{fig:brnn_aa}, $BR(\higgsl\to \snr\snr)\gsim BR(\higgsl\to b\bar b)$ for the whole range of RH sneutrino masses.

The second-lightest Higgs is normally singlet-like and as a consequence its annihilation into SM particles is already suppressed with respect to annihilation into RH sneutrinos and we should expect a sizable $BR(\higgsl\to \snr\snr)$ as well as  $BR(\higgsl\to \phiggsl\phiggsl)$ from the annihilation into a pair of very light (singlet-like) pseudoscalars. This is indeed what happens, as displayed on the right-hand side of Fig.\,\ref{fig:brnn_nn} for cases aa1) and aa2) on the top and bottom, respectively.
Again, the exact values are very dependent on the specific choice of the input parameters.

This situation differs from the case of very light neutralinos in the NMSSM. In the regions where the neutralino has the correct relic density due to the annihilation into a pair of very light pseudoscalars, the SM-like Higgs annihilates preferentially into a pair of pseudoscalars Ref.\,\cite{Cao:2011re}. However, in our case we observe that this is not necessarily the case and the SM-like Higgs (which in our examples coincides with $\higgsl$) either appears as a Higgs with SM-like decays (into $b\bar b$) or with a significant invisible decay. 

\subsection{$\snr\snr\to\rhn\rhn$}

\begin{figure}[t!]
  \epsfig{file=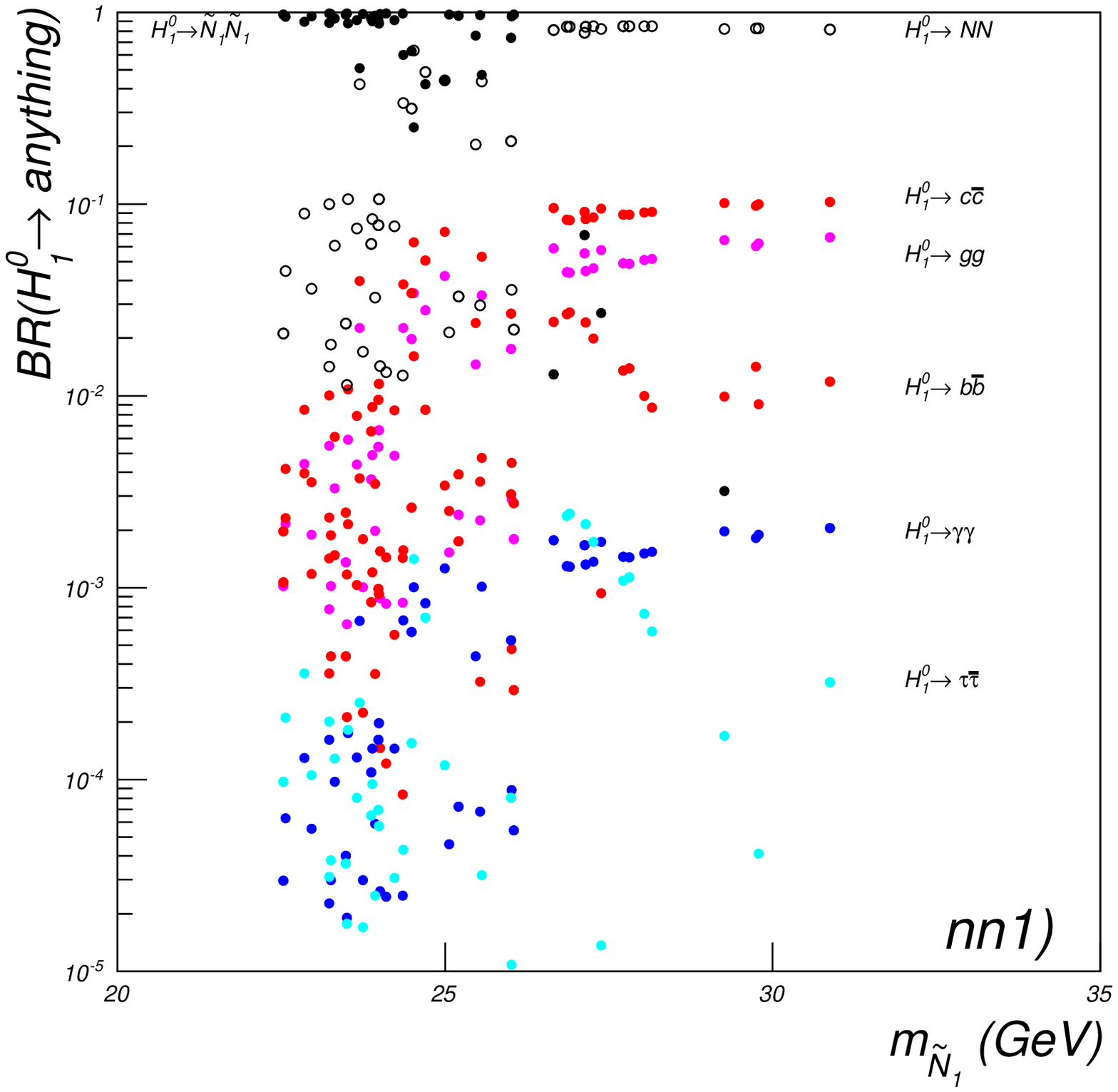,width=8.cm}  \epsfig{file=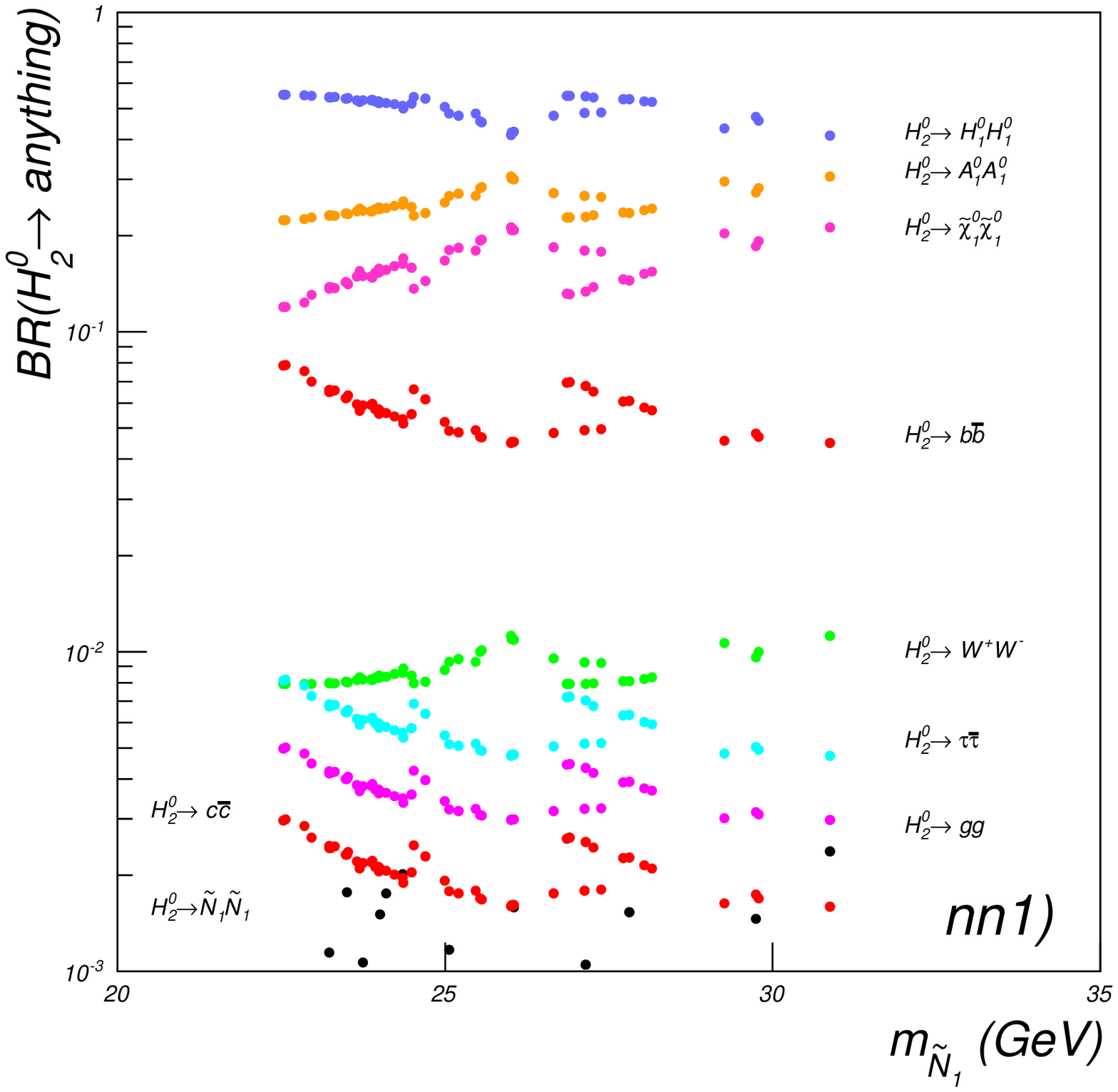,width=8.cm}\\
  \captions{The same as in Fig.\,\ref{fig:brnn_ff} but for case nn1) of Table\,\ref{tab:cases}.}
  \label{fig:brnn_nn}
\end{figure}

The Higgs spectrum in these examples features a light scalar Higgs, with a large singlet composition, and a second-lightest Higgs which is SM-like. 
In fact, as we stressed in Section\,\ref{sec:verylightnn}, the composition of the lightest Higgs has been carefully chosen in such a way that it couples preferentially to a pair of RH neutrinos (so that annihilation into these is the main channel). More specifically, we chose the input parameters in such a way that $\hcompls,\,\hcomplu\gg\hcompld$ is fulfilled. 
It is then clear that the main decay modes of the lightest Higgs are
$\higgsl\to\snr\snr$ (if 2$\snmassr<\hmassl$) and  $\higgsl\to\rhn\rhn$. Notice in this sense that while the decay into RH sneutrinos is an invisible channel, this needs not be the case for the decay into RH neutrinos.

In fact, RH neutrinos in our model are not stable particles. They couple to
a lepton and a $W^\pm$ boson through their mixing with LH neutrinos, and
thus can undergo a three body decay $\rhn\to l W^*\to ll\nu_L$, $\rhn\to l
W^*\to lq q$, $\rhn\to \nu_L Z^*\to \nu_L ll$ or $\rhn\to \nu_L Z^*\to \nu_L
qq$. Since the RH neutrino coupling to a lepton and a $W$ boson is suppressed by the small Yukawa coupling (which determines the neutrino LH-RH mixing) the lifetime of RH neutrinos can be either long enough for them to produce displaced vertices or too large, thereby counting also as an invisible component.

If Fig.\,\ref{fig:brnn_nn} we represent the resulting branching ratios for the different decay modes of the lightest and second-lightest Higgses in example nn1). Black dots correspond to BR$(\higgsl\to\snr\snr)$, whereas empty circles represent BR$(\higgsl\to\rhn\rhn)$. As commented above, these constitute the main decay modes for the lightest Higgs. 

Since the value of the coupling $\ln$ is very small in these examples, the
coupling of the RH neutrino and sneutrino to the heavier Higgses (which are,
respectively, mostly $H_u$ and $H_d$) are small. Thus there is no
contribution from these modes to the branching ratios of these. Still, since
we are dealing with the NMSSM, the SM-like Higgs can have exotic decays,
although these are more dependent on the particular choice of input
parameters. For example, in case nn1) the channel $H_2^0\to\higgsl\higgsl$
is open and provides the main contribution to the decay width of $H_2^0$.
Also, the neutralino is a relatively light singlino (with a mass of $40$ GeV) and also provides a prominent decay mode.
In sum, in this last case the second lightest Higgs (which is SM-like) can have a very similar phenomenology to the case of neutralino dark matter in the NMSSM. 

Summarising, this scenario could potentially be discriminated from other models with very light WIMPs through the study of the displaced vertices produced after the production of a pair of RH neutrinos and their subsequent decays. 
The signature of this decay mode at the LHC depends largely on properties of the RH neutrino which are intimately related to the details of the see-saw mechanism. This information is contained in the structure of the neutrino Yukawa coupling.
In particular, the RH neutrino lifetime, its decay products (especially whether it decays ``democratically" in electrons, muons or taus) and the effect of having three RH neutrinos (and sneutrinos) have to be carefully investigated.  This analysis is beyond the scope of the present work and will be the subject of a more specific future study.


\section{Indirect detection}
\label{sec:indirect}

As in most of the WIMP scenarios, annihilations of very
light RH sneutrinos
in the Galactic halo may contribute to the observed fluxes
of high energy $p/\bar p$, $e^+/e^-$, $\nu/\bar\nu$
and gamma rays. Among them, we will here address the possible signatures of our model
in the gamma ray flux from the Galactic Centre, a region which is currently
being observed by the Large Area Telescope on the Fermi satellite (Fermi-LAT) \cite{FermiDM}.
\footnote{Although we do not discuss here the possible signals in the high
energy cosmic ray (CR) like $e^+e^-$ or $p\bar p$, these may also 
constrain light dark matter candidates \cite{Cerdeno:2011tf,Lavalle:2010yw}.}

In the usual neutralino dark matter scenario, annihilation in the Galactic halo
tends to be suppressed due to the p-wave dominance of the annihilation cross
section because of the Majorana nature of the neutralino (see, e.g., the recent study of the gamma ray flux for neutralinos in the cMSSM in Ref.\cite{Ellis:2011du}). On the contrary, RH sneutrino annihilation is s-wave dominant, thus we expect
a larger gamma ray flux.

As explained in Section\, \ref{sec:verylight}, the dominant channel for sneutrino annihilation
can be $f\bar f$, $A^0_1A^0_1$ or $NN$, depending on the chosen parameters.
Among them, $NN$ mode dominant scenarios are potentially interesting because the gamma
ray spectrum produced in the subsequent decay of
the RH neutrino into three fermions through Electroweak interaction
($\rhn\to ll\nu_L$, $\rhn\to lq q$ or $\rhn\to qq\nu_L$) is distinctive from
the annihilation products of the $f\bar f$ channel. This possibility is not present in other WIMP models and thus is not generally addressed in the literature.

\begin{figure}[t!]
  \epsfig{file=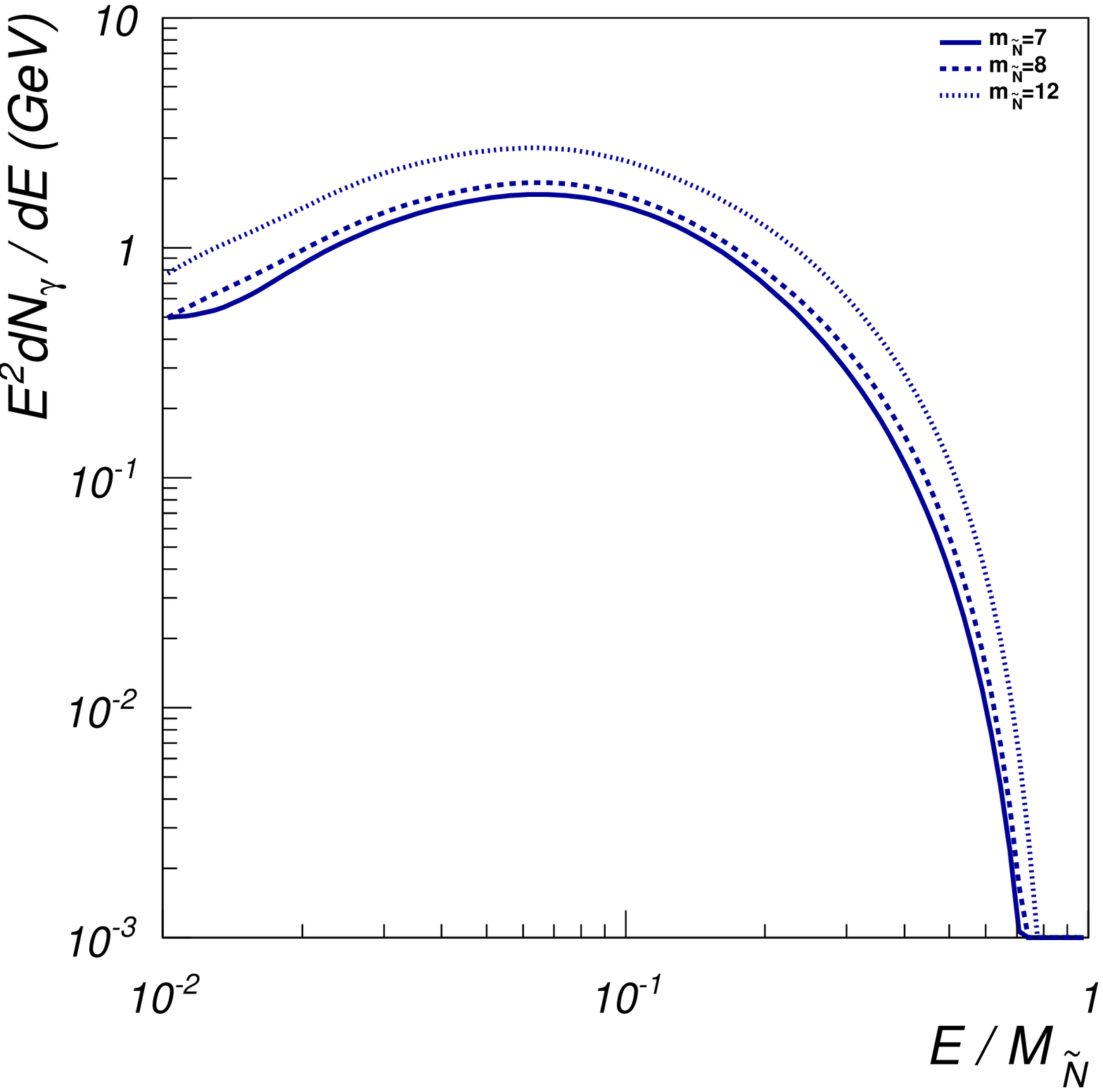,width=8.cm}
  \epsfig{file=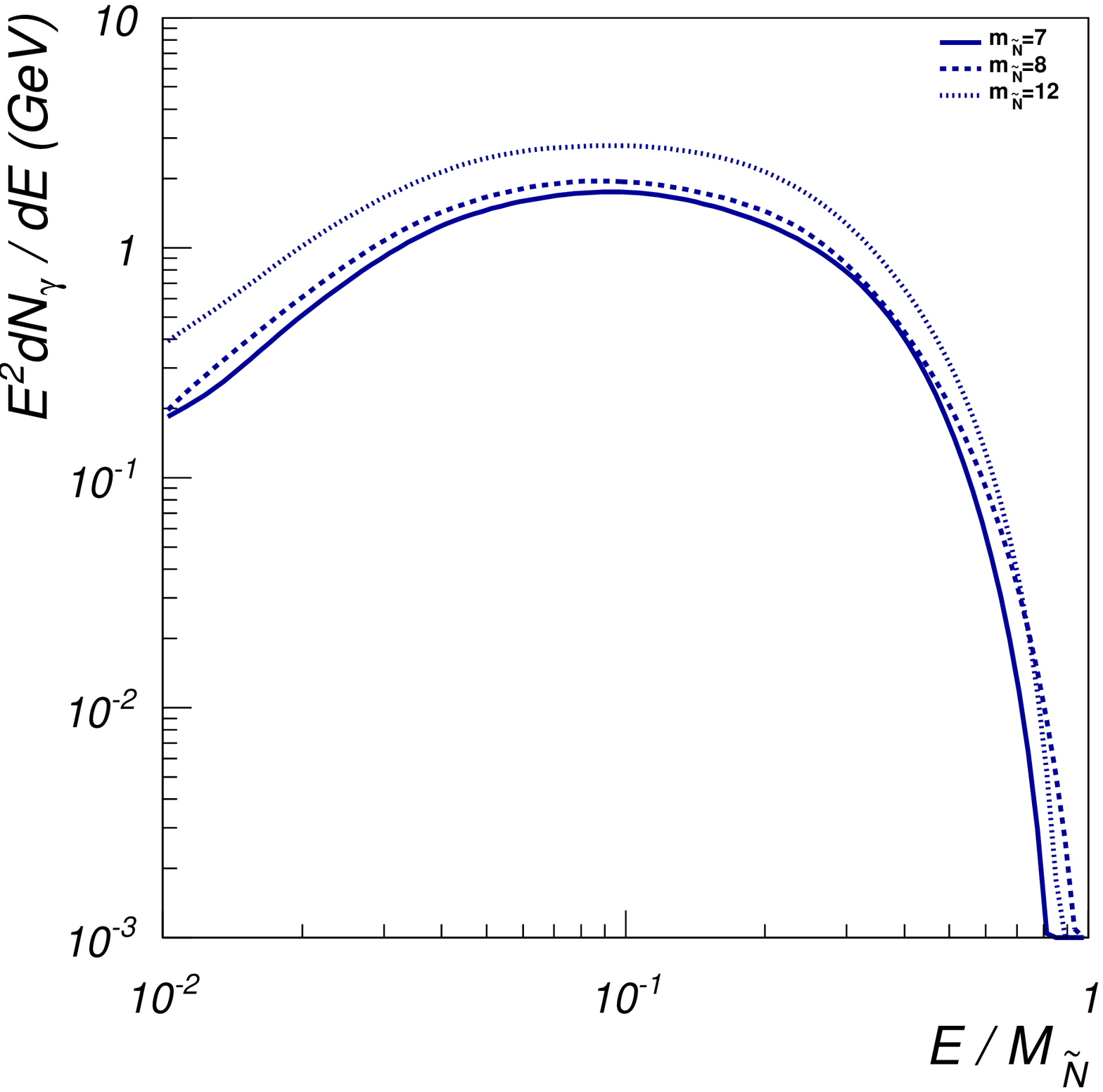,width=8.cm}
  \captions{Left: $dN_\gamma/dE$ from $\tilde N\tilde N\rightarrow b\bar b$. Right:
  $dN_\gamma/dE$ from $\tilde N\tilde N\rightarrow c\bar c$.}
  \label{fig:bcgamma}
\end{figure}

\begin{figure}[t!]
  \epsfig{file=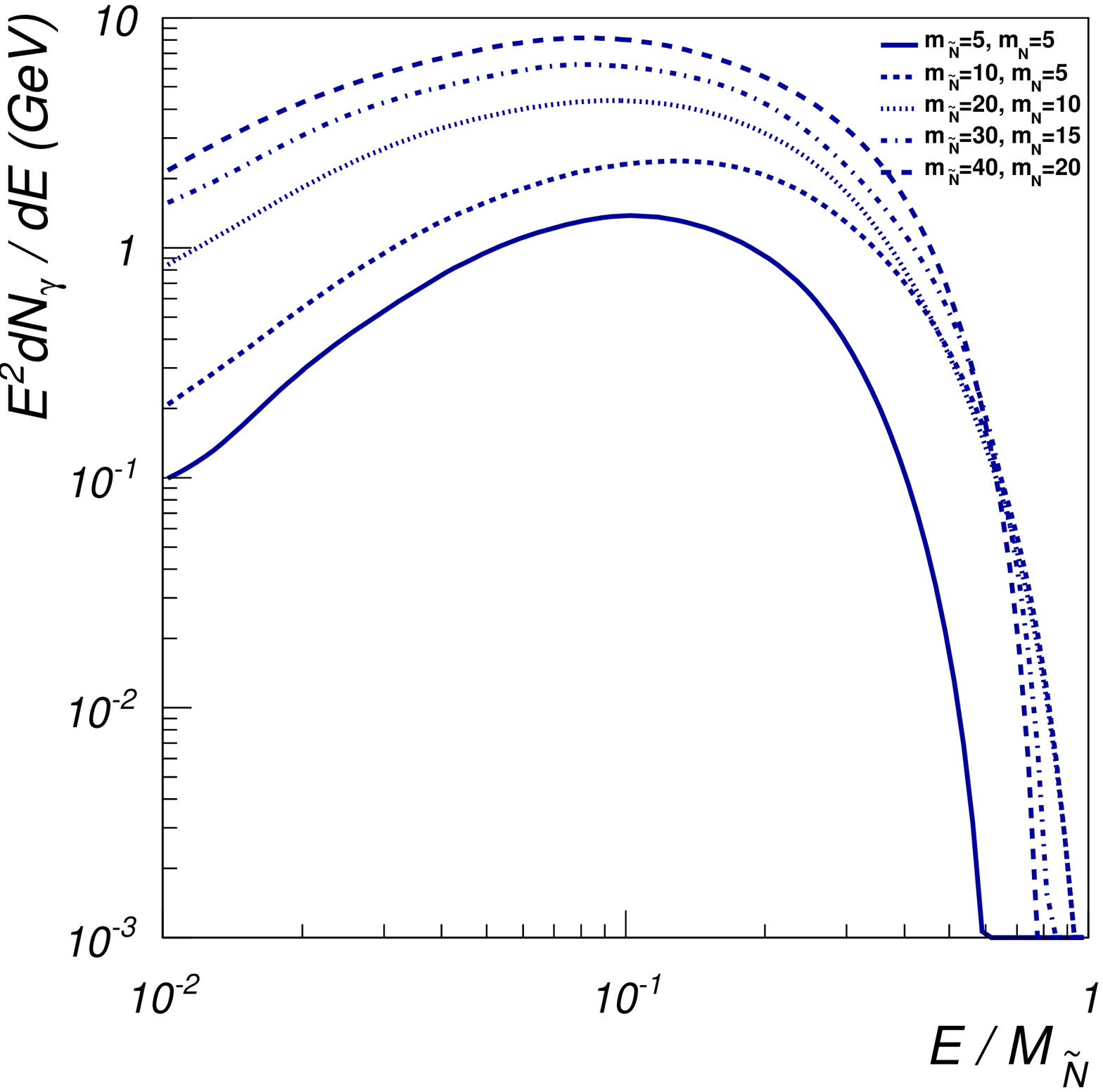,width=8.cm}
  \epsfig{file=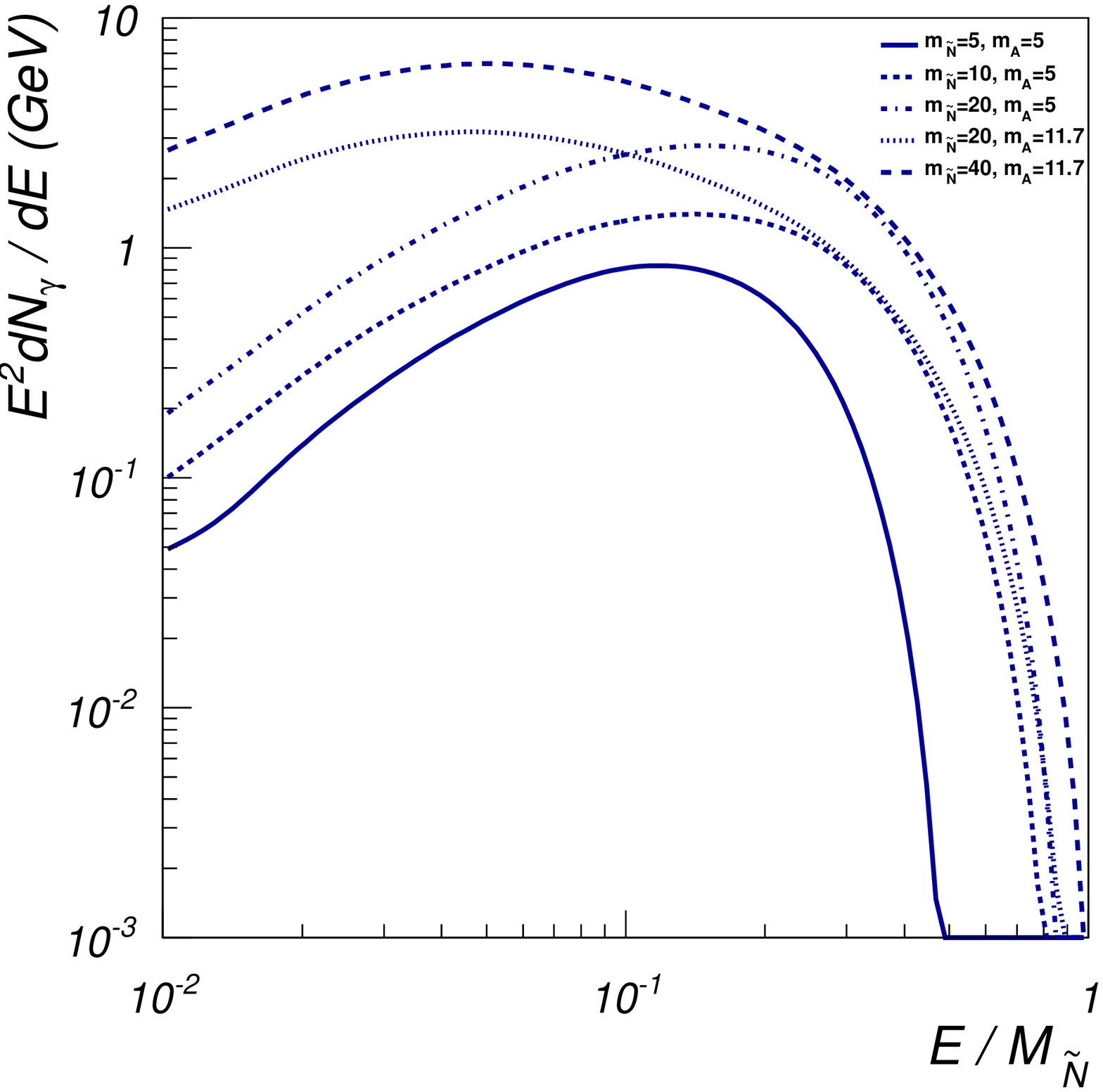,width=8.cm}
  \captions{Left: $dN_\gamma/dE$ from $\tilde N\tilde N\rightarrow NN$.
  Right: $dN_\gamma/dE$ from $\tilde N\tilde N\rightarrow \phiggsl\phiggsl$.}
  \label{fig:NAgamma}
\end{figure}

In general, the photon spectrum from WIMP annihilation or decay has three
components, the prompt, the inverse Compton scattering (ICS) and synchrotron
radiation. The prompt component involves both continuous and line emission. The
main sources of continuous spectra are $2\gamma$ decays of $\pi^0$ and
final state radiation. To calculate the expected spectrum
we have written a routine generating events of three body decays of $N$ with
an appropriate weight using the {\tt PY3ENT} routine in 
{\tt PYTHIA 6.4}
\cite{Sjostrand:2006za}.
For the $b\bar b$ and $A^0_1A^0_1$ modes, we also used {\tt PYTHIA 6.4}. In
Figs. \ref{fig:bcgamma} and \ref{fig:NAgamma}, we present the energy spectra of
gamma rays produced in the subsequent decays of final state particles
for each annihilation mode. Although
in various WIMP models the annihilation into $2\gamma$ appears at the
one-loop level, it is considered as a smoking gun channel of indirect dark matter
detection because it is distinguishable from any conceivable astrophysical
background. However, for very light RH sneutrino dark matter, the line emission
tends to be buried in the large astrophysical background after taking into account
the energy resolution
of the apparatus. Thus we can neglect its contribution in the analysis.

Other contributions are produced from the propagation of charged stable products
like $e^+/e^-$ and $p/\bar p$. Once they are produced, they
are diffused and loose their energies due to ICS with
ambient start light or cosmic microwave background radiation, and synchrotron
radiation by the Galactic magnetic field. Then the same energy loss mechanism
of charged
cosmic ray also generates photons. For heavy enough dark matter, these
contributions can exceed the prompt one in the relatively lower energy
region \cite{Cirelli:2010xx}. However, for light dark matter with mass $\sim{\cal O}(10{\rm GeV})$
the maximum energy of gamma rays from ICS by starlight with energy $\omega\sim 1{\rm
eV}$ is around $(E/m_e)^2\omega\sim 0.1{\rm GeV}$, which is near
the threshold energy of Fermi-LAT. Synchrotron
radiation on the other hand is
produced in the radio frequency range around $\nu_{\rm syn}\simeq
(1/3)~3eBE_e^2/4\pi m_e^3= 1.3~ {\rm kHz}~ (B/1{\rm
mG})~(E_e/m_e)^2\lesssim 10^2~
{\rm GHz} ~(B/1{\rm mG})$ \cite{Cirelli:2010xx,Bertone:2008xr}. Since the
constraints
arising from the observation of radio frequency are beyond the scope of this
article, in the present analysis we neglect the contributions
coming from both ICS and
synchrotron. Being able to ignore the ICS and the synchrotron radiation makes the analysis
much simpler because of the absence of complications stemming from uncertainties
in CR propagation model. 

Following the standard procedure
\cite{Bertone:2004pz} the 
gamma ray
flux can be written as
\begin{eqnarray}
  \frac{d\Phi_{\gamma}}{dE}\left(\Delta\Omega\right)
  &=&\sum_i\int_{\Omega<\Delta\Omega} dV~ 
  \frac{1}{2}\frac{\langle\sigma_i v\rangle n_{\rm DM}^2}{4\pi s^2}
  \frac{dN^i_{\gamma}}{dE}\nonumber\\
  &=&\sum_i\frac{1}{2}\int ds~d\Omega~
  \frac{\rho^2(r)}{4\pi\snmassr^2}\langle\sigma_i v\rangle
  \frac{dN^i_{\gamma}}{dE}\nonumber\\
  &=&\frac{1}{8\pi \snmassr^2}\sum_i \langle \sigma_i v\rangle
  \frac{dN^i_{\gamma}}{dE}\int ds~d\Omega~\rho^2(r)\nonumber\\
  &=&\frac{\rho_\odot^2 r_\odot }{8\pi\snmassr^2}\sum_i\langle
  \sigma_i v \rangle \frac{dN^i_{\gamma}}{dE}
  {\bar J}(\Delta\Omega)\Delta\Omega,
  \label{gammaflux}
\end{eqnarray}
where $n_{\rm DM}$ is the dark matter number density, 
$i$ represents all the possible annihilation channels
$\{e\bar e,\, \mu \bar\mu,\,\tau\bar\tau,\,u\bar u,\,d\bar d,\,c\bar c,\,b\bar b,\,\phiggsl\phiggsl,\,\rhn\rhn\}$
, $dN_\gamma^i/dE(E)$ is the expected number of photons in the energy range
of $(E,E+dE)$ produced from a given process $i$,
and the so-called halo factor $\bar J(\Delta\Omega)$ is defined as
\begin{equation}
  \bar J(\Omega)\Delta\Omega=\frac{1}{\rho_\odot r_\odot}
  \int_{\Delta\Omega}d\Omega~\int_{\rm line-of-sight} ds~
  \rho^2\left(r=\sqrt{r_\odot^2+s^2+2r_\odot s
  \cos\theta}\right).
  \label{Jomega}
\end{equation}
Here we have used the canonical value $\rho_\odot=\rho(r_\odot)=0.3$ GeV for the dark matter density around the Sun\footnote{More recent determinations in
Ref. \cite{Catena:2009mf,Strigari:2009zb,Pato:2010yq,Weber:2009pt,Salucci:2010qr}
indicate slightly larger values. In any case, the fluxes corresponding to
different values of the local density can be easily deduced from
the appropriate scaling.}
and a distance to the GC of $r_\odot=8.5$ kpc. The halo properties are thus factorized and
encoded in the single factor $\bar J(\Delta\Omega)$. 
In order to take into account the possible astrophysical uncertainties, we use two halo models which are supported by N-body simulations, namely NFW
\cite{Navarro:1996gj} and
Einasto \cite{Graham:2005xx} as well as the isothermal halo model
\cite{Begeman:1991iy,Bahcall:1980fb} as a reference. The corresponding density profiles
are parametrized as
\begin{equation}
  \rho(r)=\left\{
  \begin{array}{ll}
    \rho_s r_s/r(1+r/r_s)^2 &,{\rm NFW}\\
    \rho_s~\exp\left[-\frac{2}{\alpha}(r^\alpha-1)\right]&,{\rm Einasto}\\
    \rho_s/(1+(r/r_s)^2) &,{\rm Isothermal}\\ 
  \end{array}
  \right.
  \label{DMhalo}
\end{equation}
where $r_s=20$ kpc and $5$
kpc for NFW and isothermal model, respectively, and $\alpha = 0.17$. The value of $\rho_s$ is fixed to
reproduce the dark matter density around the solar system.
In Table \ref{tab:Jfactor},
we calculate $\bar J(\Delta\Omega)\Delta\Omega$ for each halo model considering a
region of interest (ROI) around the GC with radius $1^{\circ}$, $5^{\circ}$, $7^{\circ}$
and $10^{\circ}$.
\begin{table}
  \centering
\begin{tabular}{|c|cccc|}
  \hline
  $\theta_{\rm ROI}$  & $1^{\circ}$& $5^{\circ}$& $7^{\circ}$& $10^{\circ}$\\
  \hline
  $\Delta\Omega$ & $9.6\times10^{-4}$ & $2.4\times10^{-2}$ &
  $4.6\times10^{-2}$&$9.6\times10^{-2}$\\
  \hline
NFW& 1.35 & 5.95 & 7.91 & 10.5\\
Einasto& 2.10 & 11.6 & 15.2& 19.7\\
Isothermal& 0.0130 & 0.319 &0.615& 1.22\\
  \hline
\end{tabular}
\captions{Halo factors multiplied by given solid angles, $\bar
J(\Delta\Omega)\Delta\Omega$, for the different halo
models, NFW, Einasto and isothermal model.  ROI around GC
with radius $1^\circ$, $5^\circ$, $7^\circ$ and $10^\circ$ are used. }
  \label{tab:Jfactor}
\end{table}

To compare the predicted gamma ray flux from RH sneutrino
annihilation with the gamma ray flux currently observed by Fermi-LAT,
we have used data from the Fermi Science Support Center
(FSCC) archive \cite{FSSCurl} \footnote{ We extracted gamma ray data taken
from 4th of August in 2008 (15:43:37) to 4th of July in
2011 (14:08:11). We selected
signals classified as {\tt DIFFUSE} only, which are
appropriate to analyse the diffuse gamma ray emission using the {\tt gtselect}
tool in Fermi Science Tools
\cite{FSTurl}. In both processes of selecting data and
calculating the exposure map, we used
the specific instrument response function (IRF) 
{\tt P6\_V11\_DIFFUSE} to be consistent.}.
We have selected the ROI as a circular
region with radius $1^\circ$, $5^\circ$, $7^\circ$ and $10^\circ$
around the GC with RA$=266.46^\circ$ and
Dec$=-28.97^\circ$. Following the suggestion of the FSCC, we used
a zenith angle cut $105^\circ$ to reject photons coming from the
Earth.
We used {\tt gtbin} tool to make
20 bins which is equally spaced in logarithmic scale in energy
from $0.1$ GeV to $100$ GeV.

To obtain the flux map for gamma rays from the counts map of
photons actually detected in the experiment, we
divided the count map by
the exposure map in each position and energy bins. This
can be calculated from the spacecraft data and the instrument
response function using the {\tt gtltcube, gtexpcube2} tools.
We used the {\tt gtexpcube2} tool
to reflect the azimuthal dependence of the effective
area of the apparatus.

As described in \cite{Vitale:2009hr}, we have taken the systematic uncertainty
in the effective area of the apparatus as $10$ \% at $0.1$ GeV, decreasing
to $5$ \% at $0.56$ GeV and increasing to $20$ \% at $10$ GeV, and
interpolated between the points. We have extrapolated above $10$ GeV with
a constant value. Since the systematic uncertainty in the
background model is not fully understood yet, we used a
fixed background emission template provided by the Fermi Science Team
\cite{FSbkg}. 

There are three main components of background\footnote{Here we use the
terminology {\it background} although we include both
background and foreground in the analysis.},
the diffuse galactic emission (DGE), the resolved point sources (PS) and
the isotropic gamma ray background (IGB).
Here we used the DGE model map {\tt gll\_iem\_v02.fit} and IGB model
supplied by the Fermi Science Team.
For PS, we used the preliminary second Fermi-LAT catalog
found in \cite{2FGL}. PS in the catalog are fitted in
{\tt PowerLaw}, {\tt LogParabola} and {\tt PLExpCutoff} whose
form can be found in the accompanying draft. We modeled its
contribution by summing over the modeled fluxes of PS in a given ROI.

\begin{table}[!ht]
  \centering
\begin{tabular}{|c||c|c|c|c|c|}
  \hline
  name & $\langle \sigma v\rangle_{v=0}$ (cm$^3$/s) 
  & Mode
  & $m_{\tilde N_1}$ (GeV) & $m_{N}$  (GeV)  & $m_{A^0_1}$ (GeV) \\
  \hline
  \hline
  bb-8)& $3.47\times10^{-26}$ &$b\bar b$ & 8 & * & * \\
cc-8)& $3.47\times10^{-26}$ &$c\bar c$ & 8 & * & * \\
  \hline  aa1-10)& $4.52\times10^{-26}$ & $A^0_1A^0_1$ & 10 & * & 6.64 \\
aa1-20)& $3.19\times10^{-26}$ &$A^0_1A^0_1$ & 20 & * & 6.64 \\
aa1-40)& $2.90\times10^{-26}$ &$A^0_1A^0_1$ & 40 & * & 6.64 \\
  \hline
aa2-20)& $3.19\times10^{-26}$ &$A^0_1A^0_1$& 20 & * & 11.7 \\
aa2-40)& $2.90\times10^{-26}$ &$A^0_1A^0_1$& 40 & * & 11.7  \\
  \hline
nn-23)& $3\times10^{-26}$ &$NN$ & 23 & 8 & * \\
nn-30)& $3\times10^{-26}$ &$NN$ & 30 & 15 & * \\
nn-40)& $3\times10^{-26}$ &$NN$ & 40 & 15 & * \\
  \hline
nnB-22)& $3\times10^{-28}$ &$NN$ & 22 & 8 & * \\
nnB-25)& $3\times10^{-25}$ &$NN$ & 25 & 8 & * \\
  \hline
\end{tabular}
  \captions{Benchmark points chosen to calculate expected gamma ray fluxes.
  We assume sneutrino annihilates only into specified mode. 
  The parameters marked with * are
  not relevant in the analysis.}
  \label{tab:BenchID}
\end{table}

To illustrate the various
annihilation channels and mass spectra in our model
we have chosen 12 representative
benchmark points as in Table \ref{tab:BenchID}, inspired in the examples discussed in the previous sections. 
For simplicity, we assumed
that annihilations occur only through the given specific channels.

\subsection{$\snr\snr \to f\bar f$}

\begin{figure}[t!]
  \epsfig{file=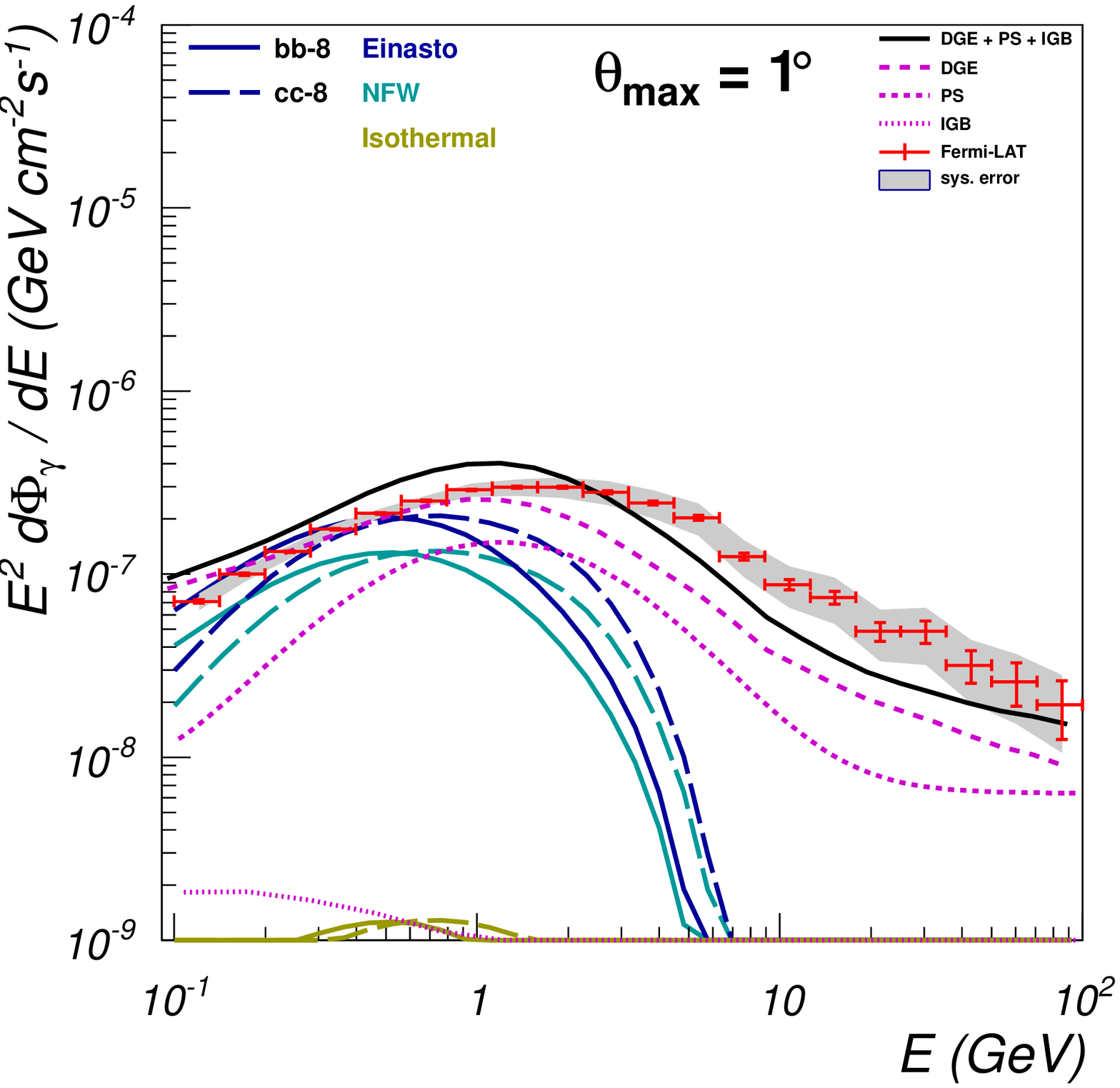,width=8.cm}
  \epsfig{file=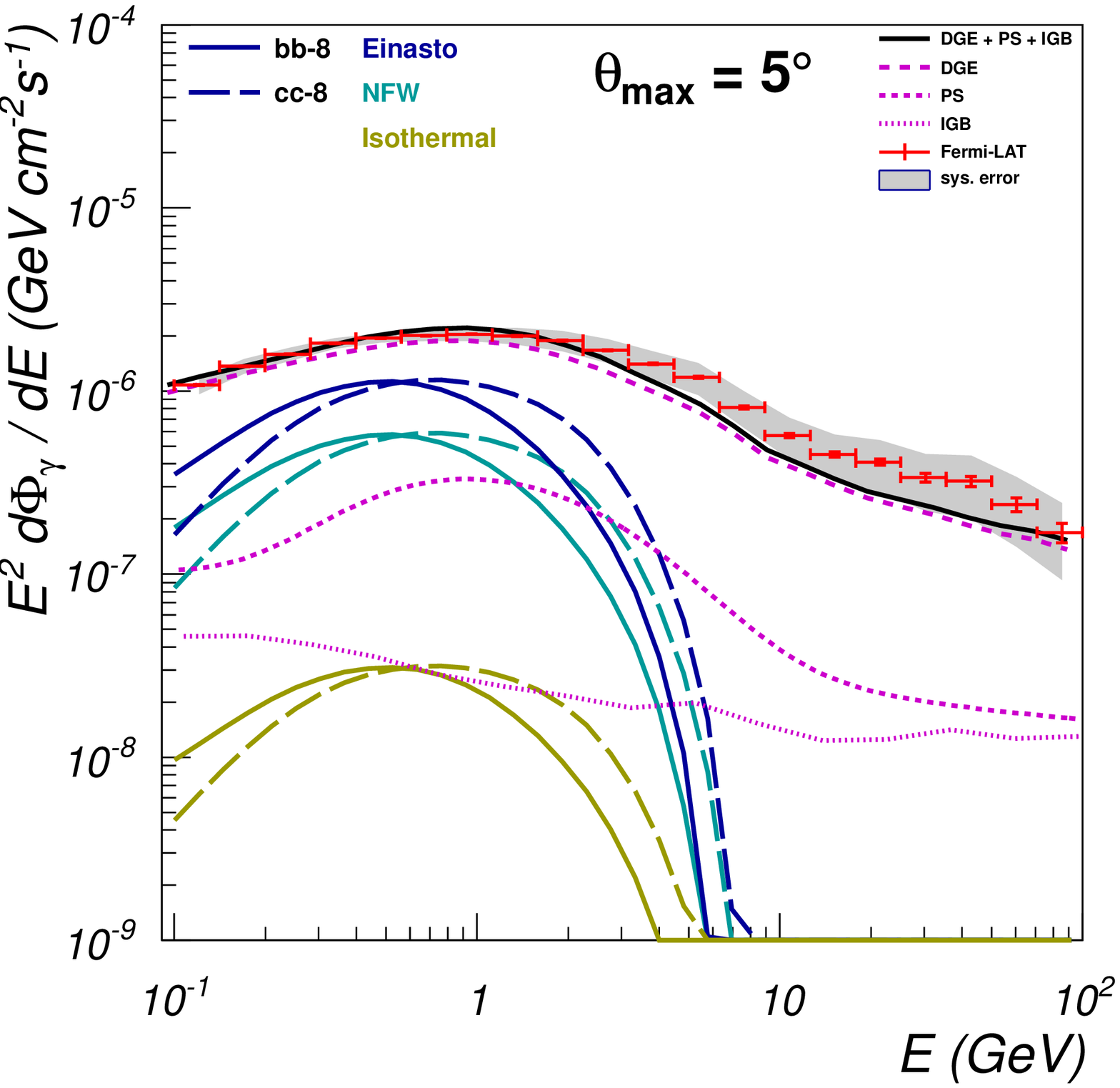,width=8.cm}
  \epsfig{file=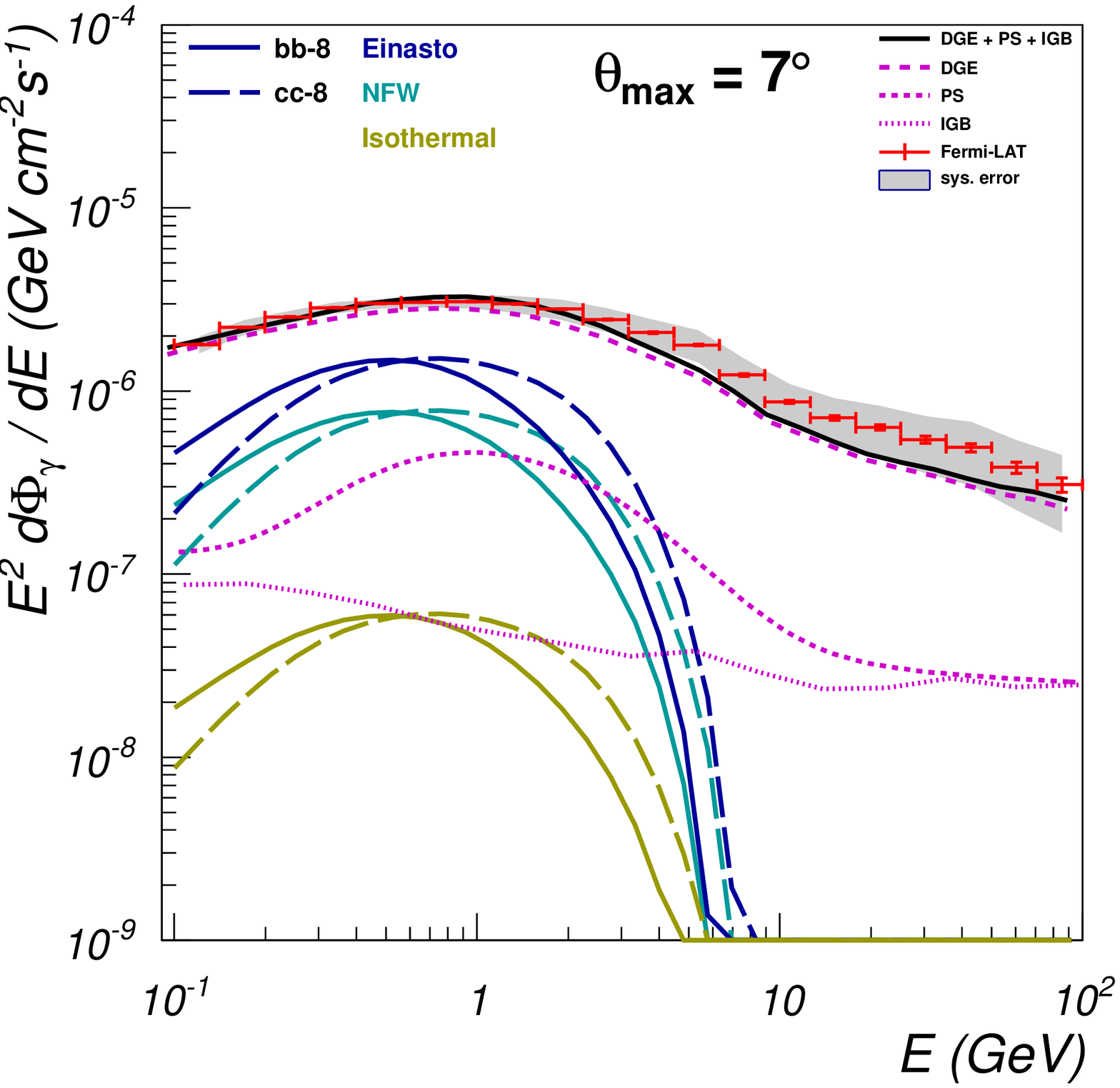,width=8.cm}
  \epsfig{file=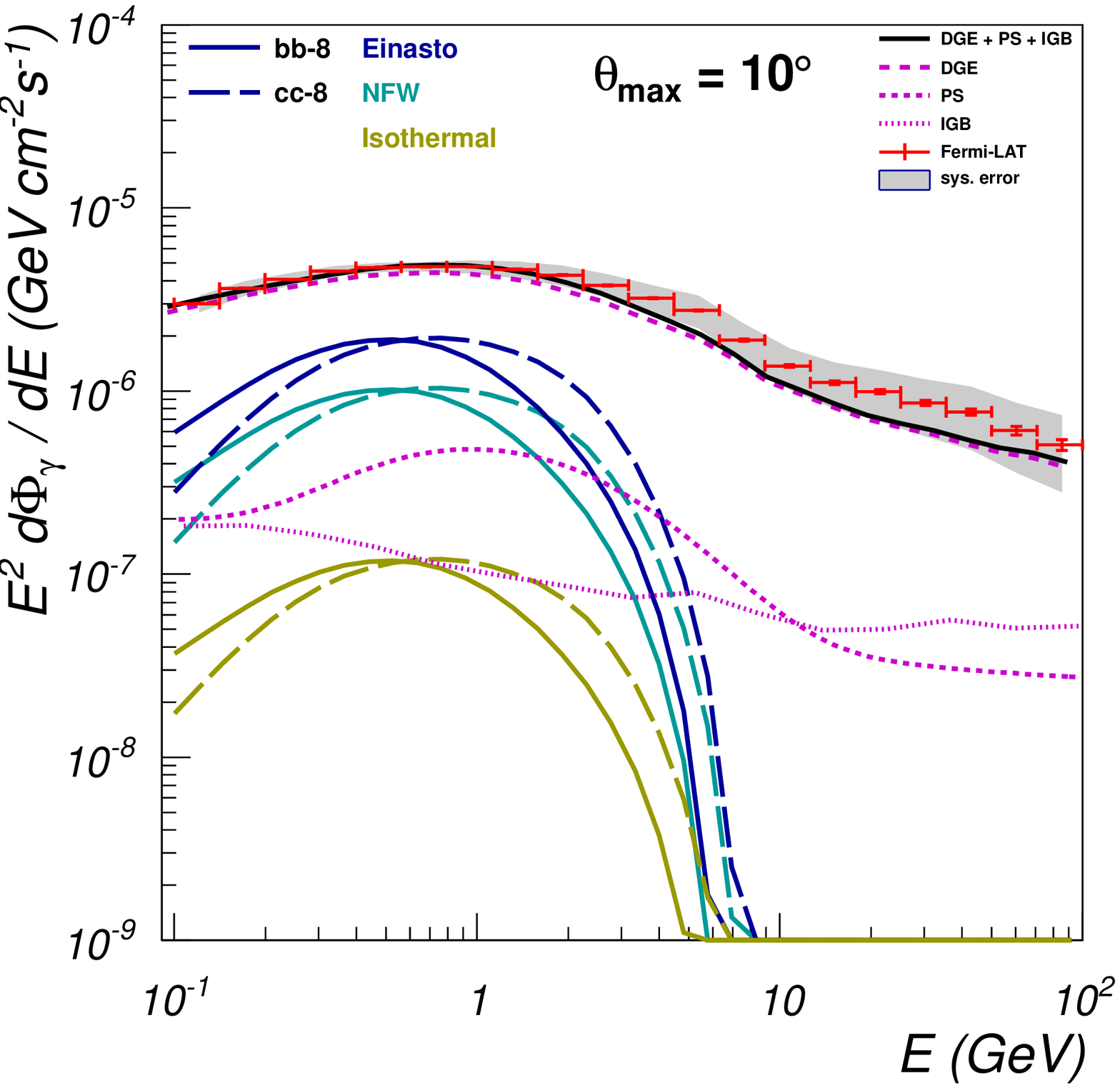,width=8.cm}
  \captions{Expected gamma ray flux in bb-8) and cc-8). ROI with radii 1$^\circ$
  (top left), 5$^\circ$ (top right), 7$^\circ$ (bottom left) and 10$^\circ$
  (bottom right) are used.}
  \label{fig:fluxbbcc}
\end{figure}

Let us first consider the case in which RH sneutrinos
annihilate into a pair of fermions. As we explained
in Section\, \ref{sec:verylight}, the dominant channel is either $b\bar b$
or $c\bar c$,
depending on the choice of parameters, while the $\tau\bar\tau$ channel is always
negligible. We have chosen $\snmassr=8$
GeV which is compatible with the CoGeNT result. The benchmarks where either annihilation into $b\bar b$ or $c\bar c$ dominates are thus labelled bb-8) and cc-8), respectively.

The predicted gamma ray flux is represented as a function of the energy in Fig. \ref{fig:fluxbbcc} for the various choices of ROI.
As we can observe, the 
flux with Einasto or NFW profile is larger than the flux from
resolved point sources.
However, the expected flux calculated for
$5^\circ$, $7^\circ$ and $10^\circ$ 
is still smaller than DGE, and thus still consistent
with the observed data after taking the systematic uncertainties
into account.

Notice that the predicted flux for $1^\circ$ is comparable to the
DGE component. However, in this region, a significant discrepancy between fluxes from
background model and observed flux is present. Indeed, our
knowledge of the background in this region is still poor.
In particular, the suppressed flux observed at low energy cannot be explained
by point sources near the GC which are not resolved yet.

\subsection{$\snr\snr \to \phiggsl\phiggsl$}

\begin{figure}[t!]
  \epsfig{file=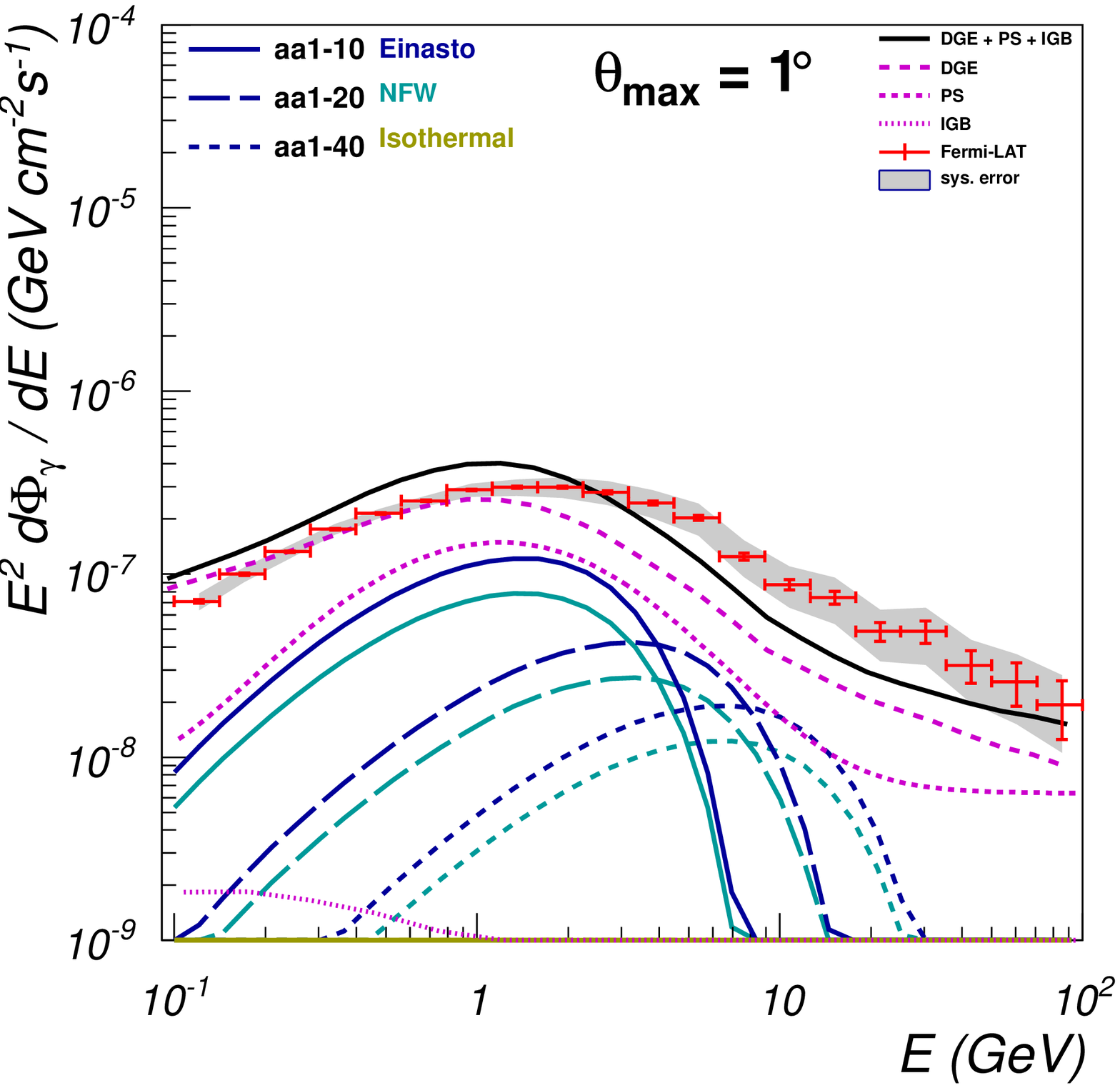,width=8.cm}
  \epsfig{file=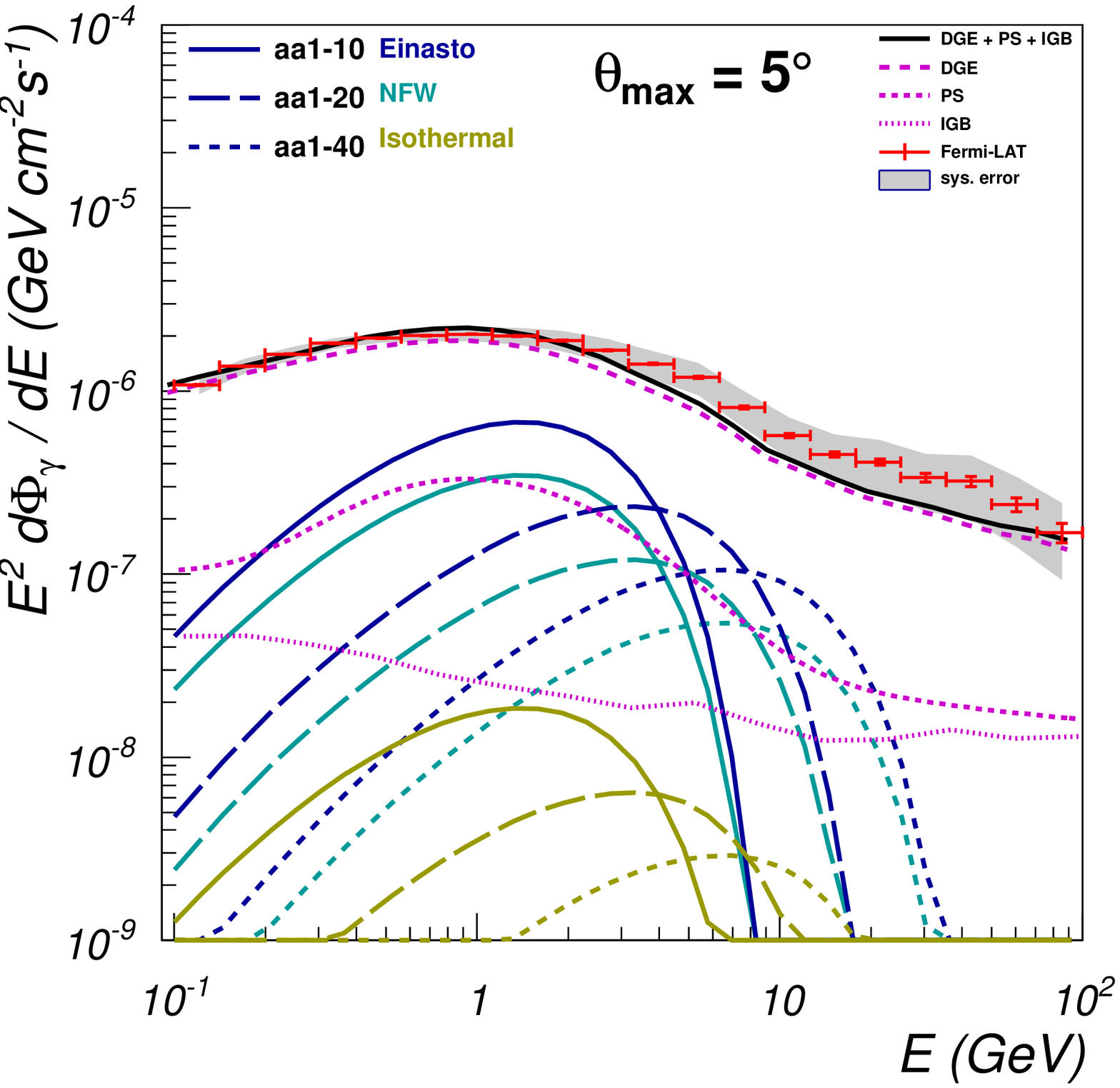,width=8.cm}
  \epsfig{file=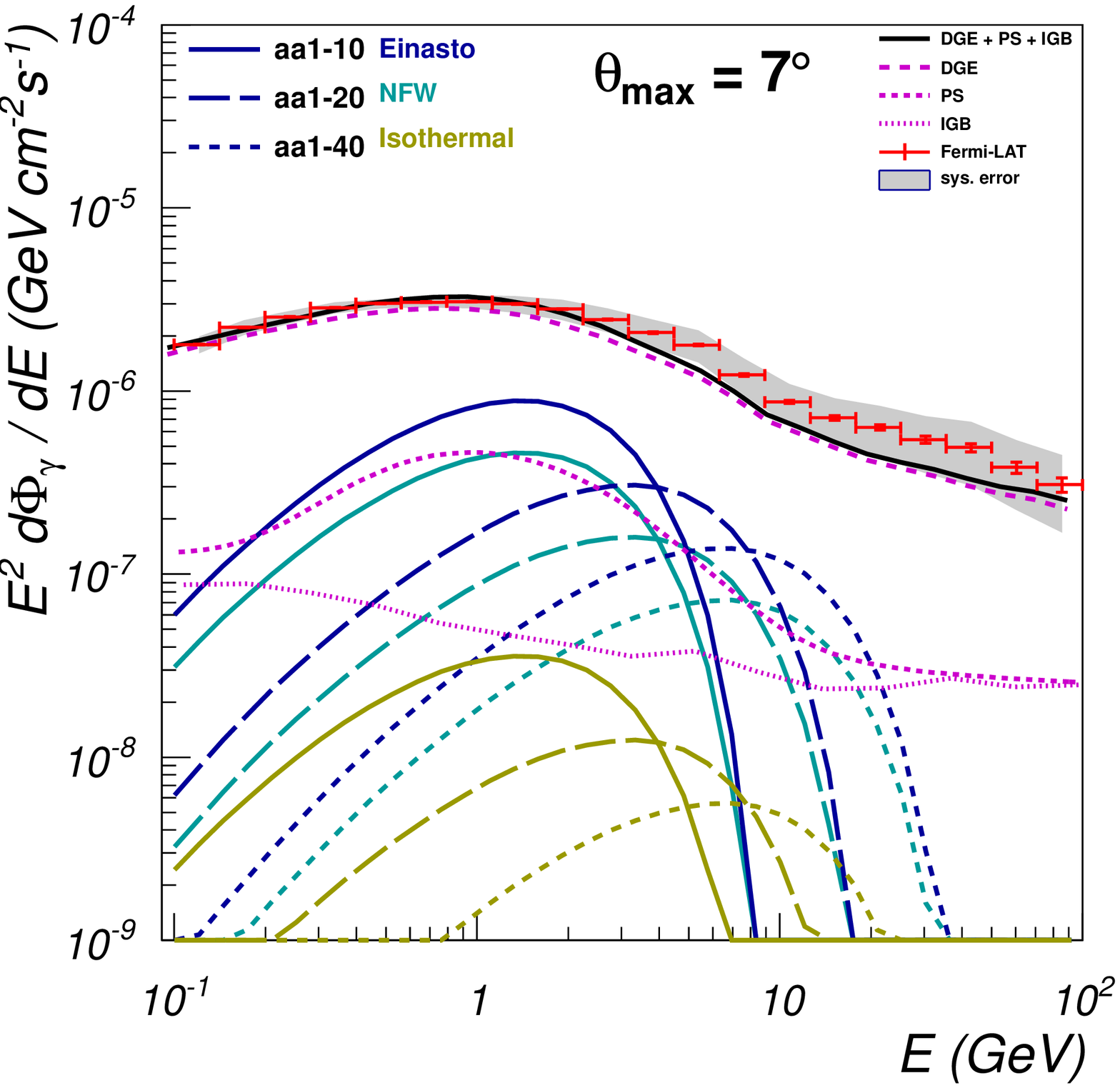,width=8.cm}
  \epsfig{file=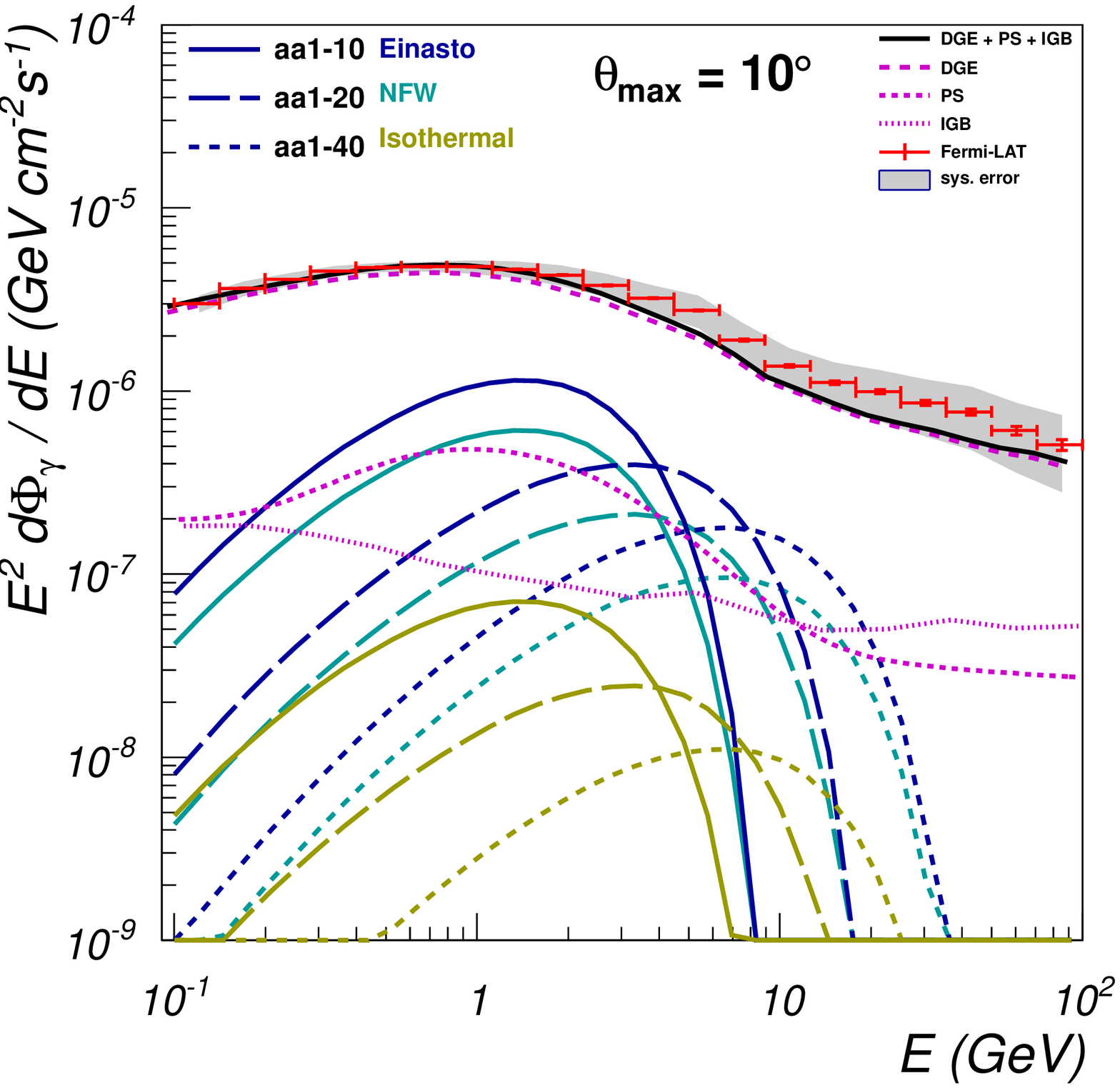,width=8.cm}
  \captions{Expected gamma ray flux in aa1-10), aa1-20) and aa1-40). ROI with radii 1$^\circ$
  (top left), 5$^\circ$ (top right), 7$^\circ$ (bottom left) and 10$^\circ$
  (bottom right) are used.}
  \label{fig:fluxaa1}
\end{figure}

\begin{figure}[t!]
  \epsfig{file=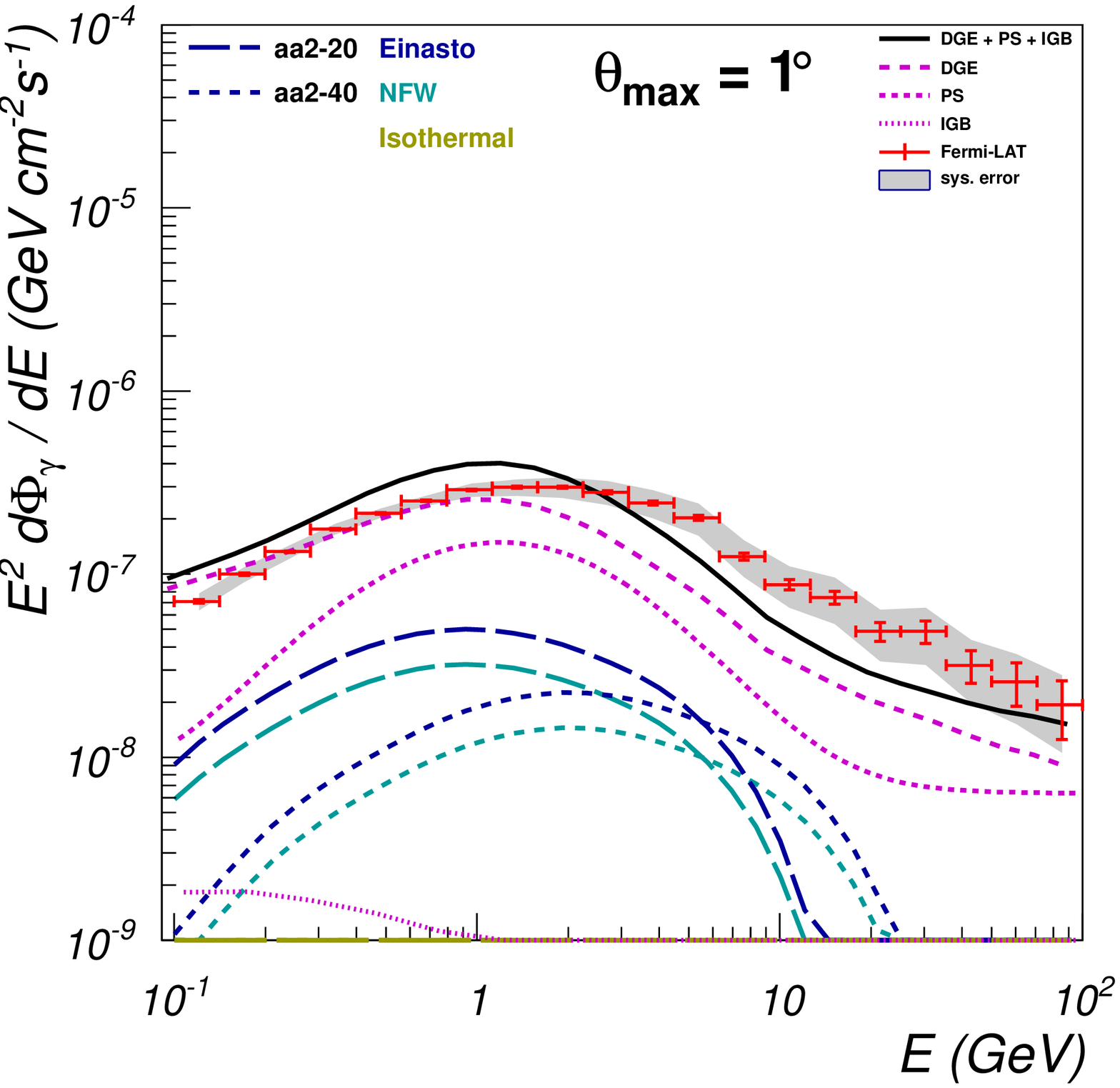,width=8.cm}
  \epsfig{file=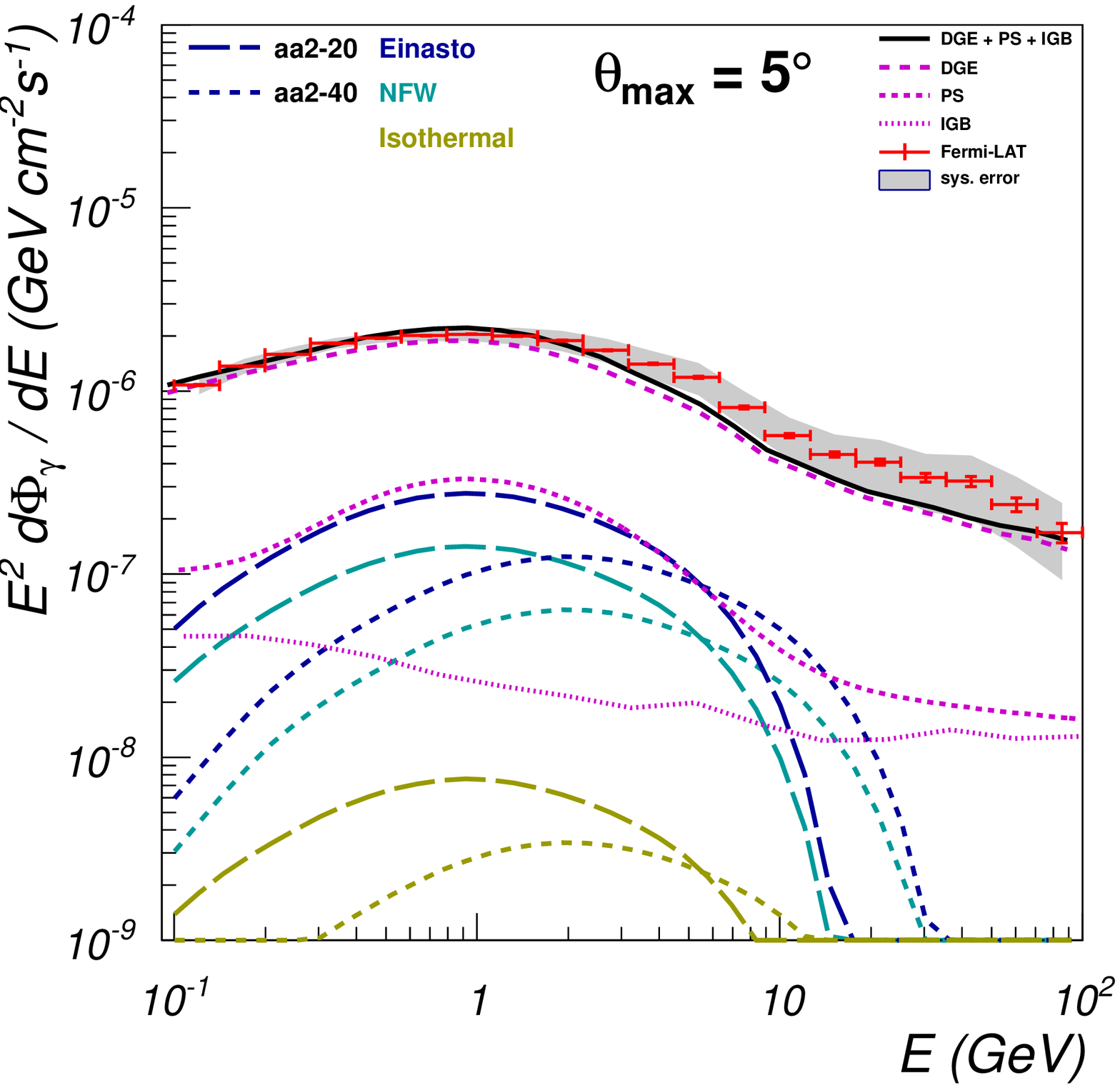,width=8.cm}
  \epsfig{file=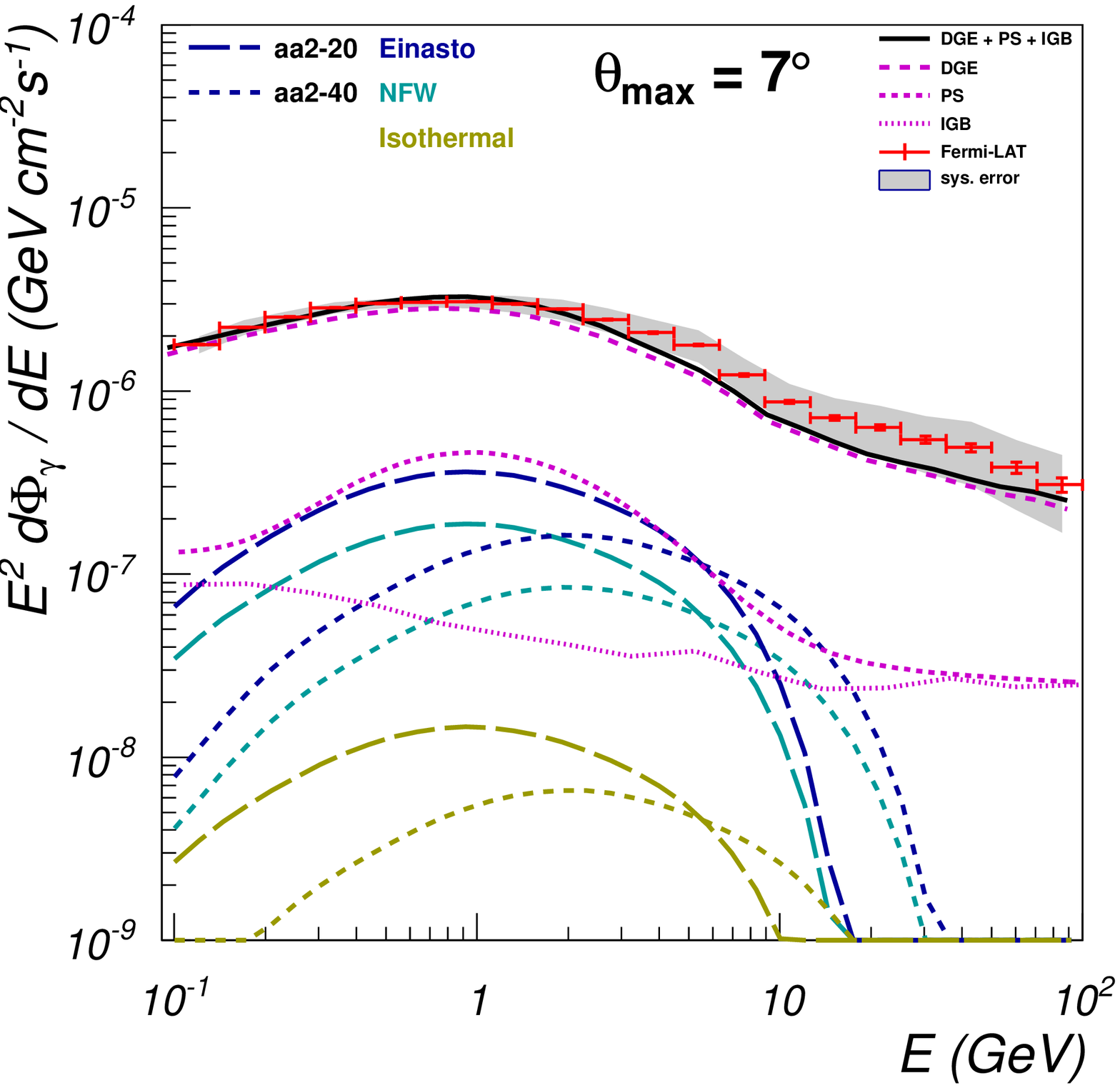,width=8.cm}
  \epsfig{file=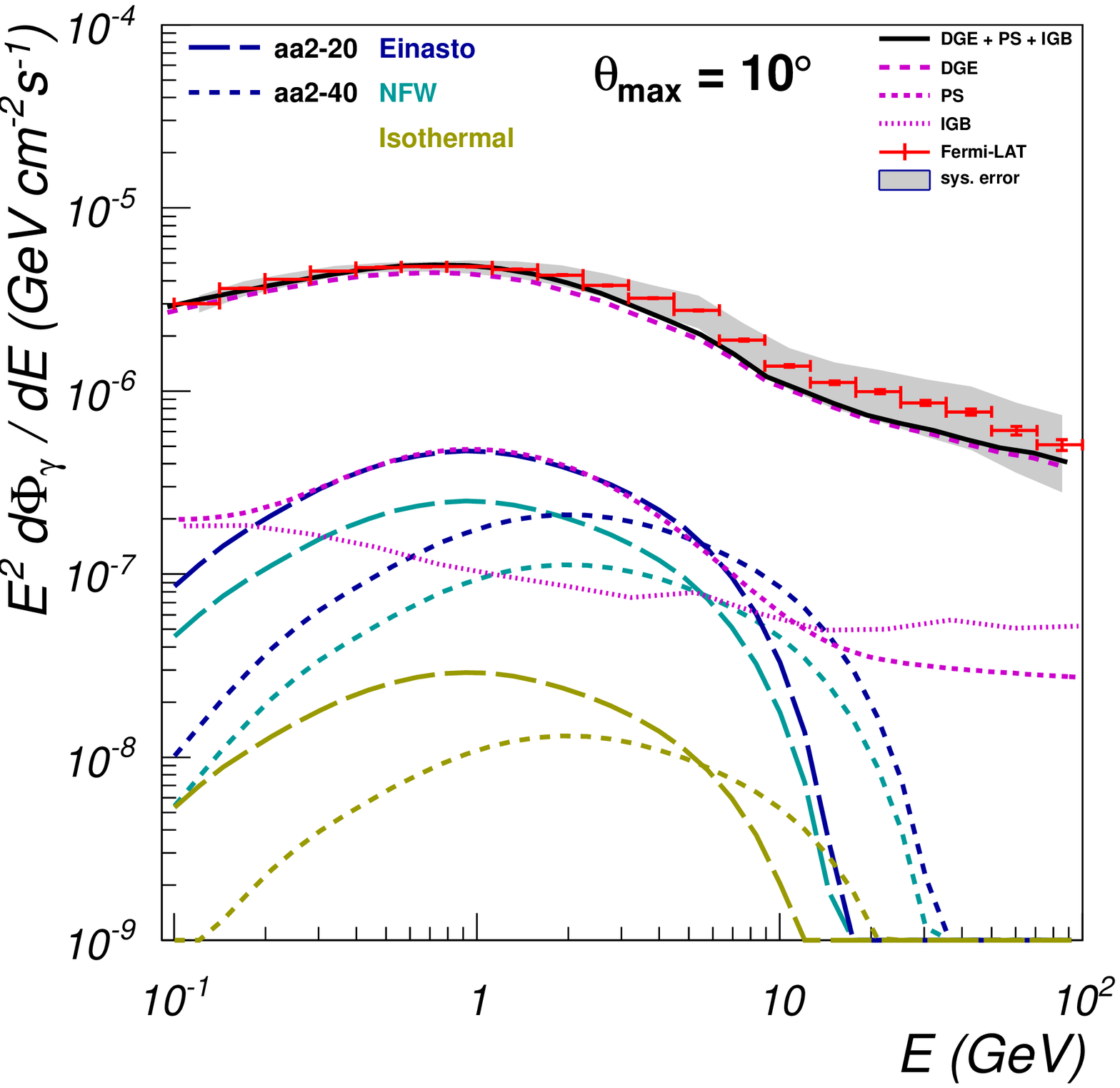,width=8.cm}
  \captions{Expected gamma ray flux in aa2-10), aa2-20) and aa2-40). ROI with radii 1$^\circ$
  (top left), 5$^\circ$ (top right), 7$^\circ$ (bottom left) and 10$^\circ$
  (bottom right) are used.}
  \label{fig:fluxaa2}
\end{figure}

In the second set of benchmark points, we focus on examples in which the RH sneutrino annihilates into
$A^0_1A^0_1$. We consider two cases aa1) and aa2) defined in Section\,
\ref{sec:verylight} and fix the RH sneutrino mass to $\snmassr=10$, $20$
and $40$ GeV for aa1), and $\snmassr=20$ and $40$ GeV for aa2). The various benchmark points are labelled accordingly in Table.\,\ref{tab:BenchID}.

The main difference in the predicted gamma ray spectra for scenarios aa1)
and aa2) is the mass of the lightest pseudoscalar Higgs.
In cases aa2) the pseudoscalar is heavy enough to decay into
$b\bar b$, whereas in cases aa1) this is not possible
As shown in Fig. \ref{fig:NAgamma}, the presence of the $b \bar b$ mode
leads to a softer spectrum. This is further evidenced in Figs.\,\ref{fig:fluxaa1} and \ref{fig:fluxaa2}, where the predicted gamma ray fluxes for the different benchmark points are represented. 
Contrary to the $f\bar f$ scenarios, we have varied the mass of RH sneutrinos in the
relatively large range from $10$ GeV to $50$ GeV in these cases. The number
density of dark matter with a fixed halo model is inversely proportional to the
RH sneutrino mass-squared, thus so is the gamma ray flux. 
  
These results are similar to those obtained in the $f\bar f$ scenarios analysed previously.
The expected gamma ray fluxes are too small to be observed except for
the case where a ROI of $1^\circ$ is considered.
Therefore, without
a significant improvement of our understanding of the background,
we cannot constrain the relevant parameter space of these examples with
gamma ray flux from the GC.

\subsection{$\snr\snr\to\rhn\rhn$}

\begin{figure}[t!]
  \epsfig{file=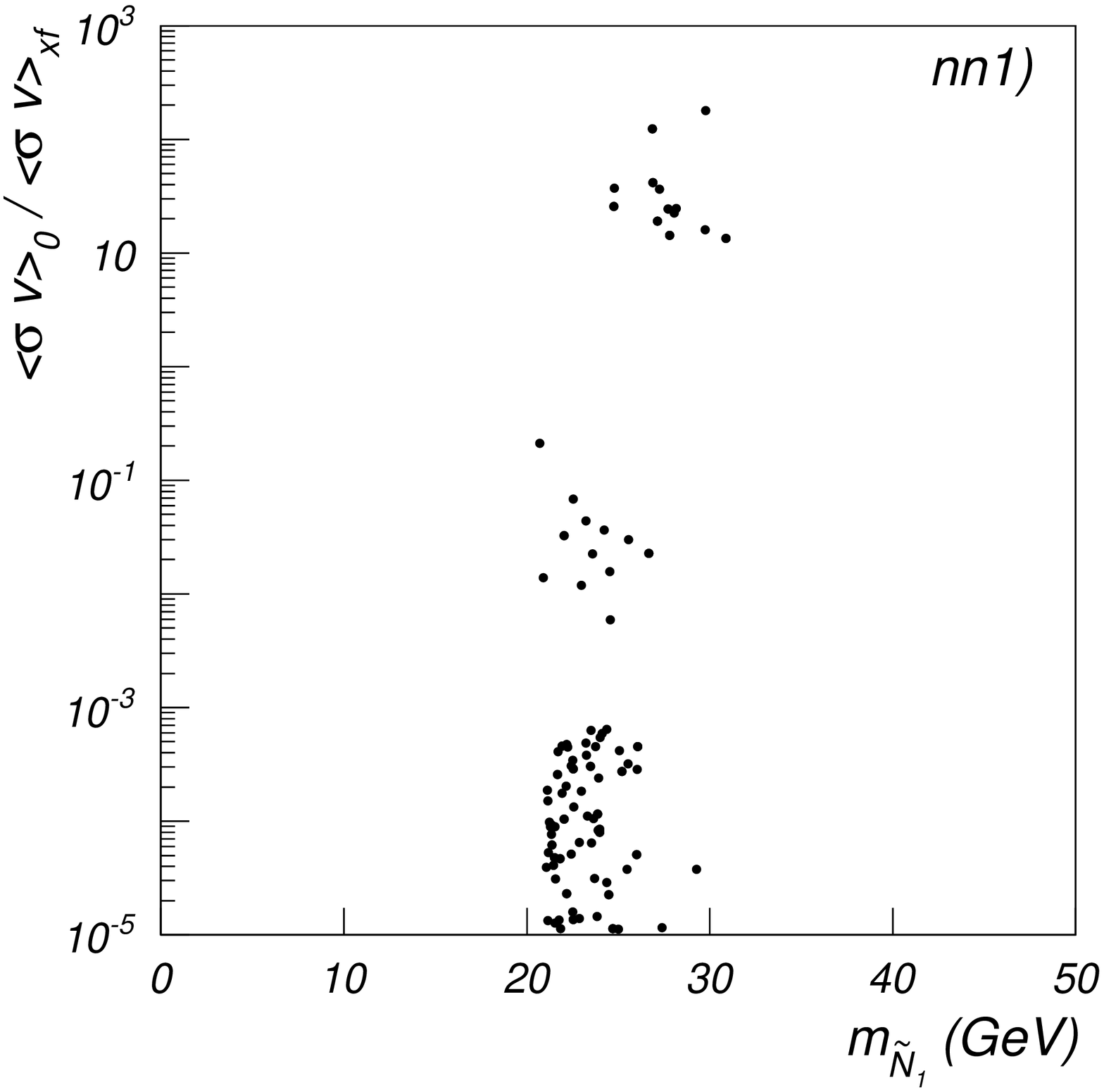,width=8.cm}
  \epsfig{file=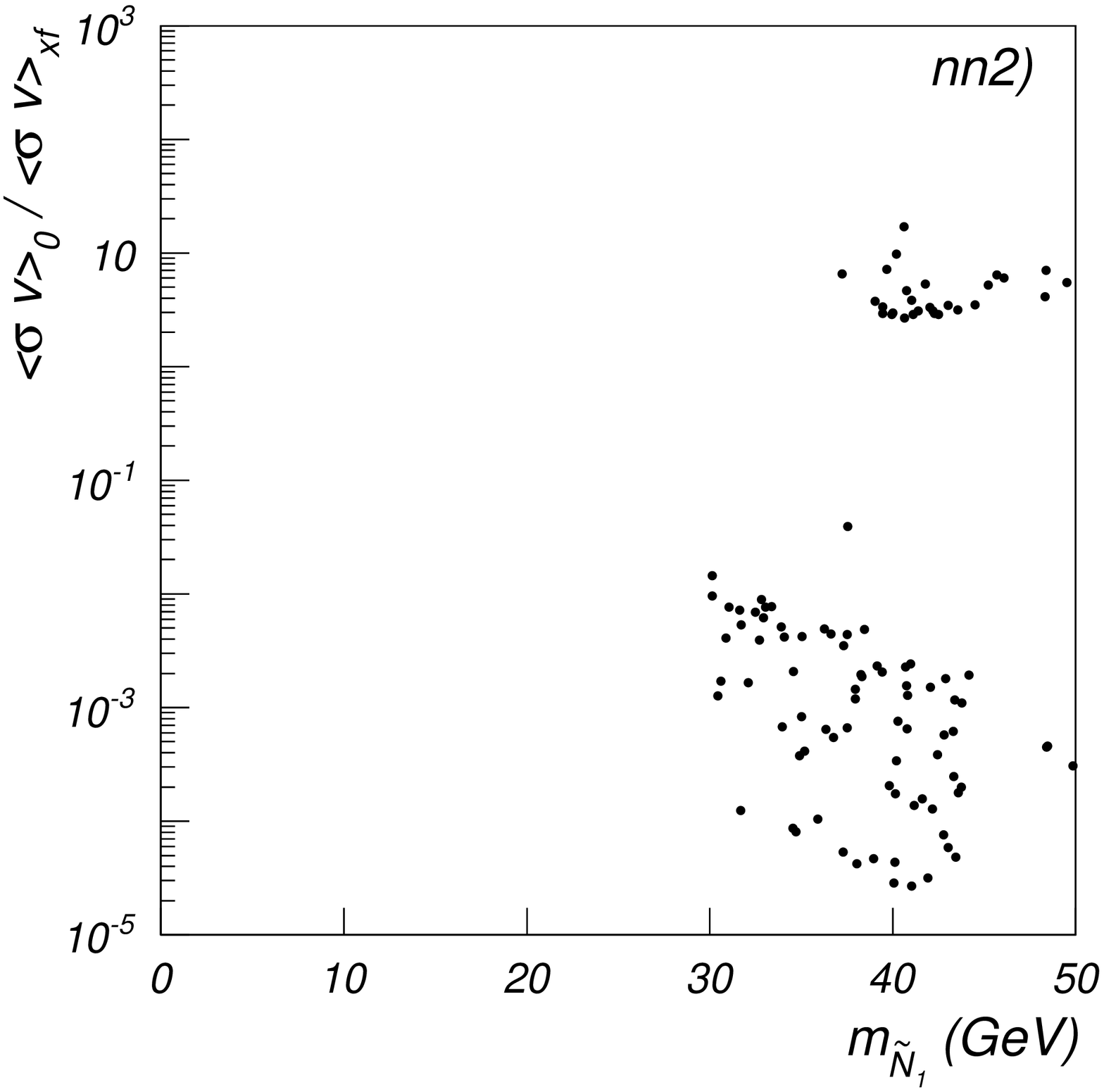,width=8.cm}
  \captions{The distributions of (de-)enhancement factor $\langle\sigma
  v\rangle_{x=0}/\langle\sigma v\rangle_{x=x_f}$ in RH sneutrino
  mass for nn1) case (left panel) and nn2) case (right panel). }
  \label{fig:EF}
\end{figure}

The last class of benchmark points corresponds to scenarios in which RH sneutrino into $NN$ is dominant. 
These cases are potentially very interesting since the RH neutrino subsequently decays 
into three fermions ($\rhn\to ll\nu_L$ or $\rhn\to lq q$). This leads to a distinctive gamma ray spectrum. 

Moreover, as explained in Section\,\ref{sec:relicnn}, the correct RH sneutrino relic abundance in these cases is only obtained through the CP-even Higgs resonance. In such a case, there can be a significant enhancement (or
suppression) of its annihilation in the dark matter halo and consequently in the gamma ray flux. This so-called Breit-Wigner
enhancement has been studied in various models and contexts
\cite{Feldman:2008xs,Ibe:2008ye,Vasquez:2011js}.

To illustrate this effect we have represented in Fig.\,\ref{fig:EF} 
the enhancement factors of the annihilation cross
section, defined as $\langle\sigma
  v\rangle_{x=0}/\langle\sigma v\rangle_{x=x_f}$ 
for the points with the correct RH sneutrino relic density in examples nn1) and nn2) of Table\,\ref{tab:cases}. 
As we observe, the enhancement factor ranges from
$10^{-5}$ to
$10^{2}$, depending on the difference
between $2m_{\tilde N}$ and $m_H$, and the decay width of corresponding
CP-even Higgs.
Consequently, these benchmark points cannot be represented
with a single fixed cross section. We have therefore defined cases
nn-23), nn-30) and nn-40) with a canonical annihilation cross
section $3\times10^{-26}$ cm$^3$/s, while $3\times10^{-28}$ cm$^3$/s
and $3\times10^{-25}$ cm$^3$/s are used for the nnB-22) and nnB-25) cases, respectively, to
study the impact of the enhancement/suppression on the gamma ray flux.

\begin{figure}[t!]
  \epsfig{file=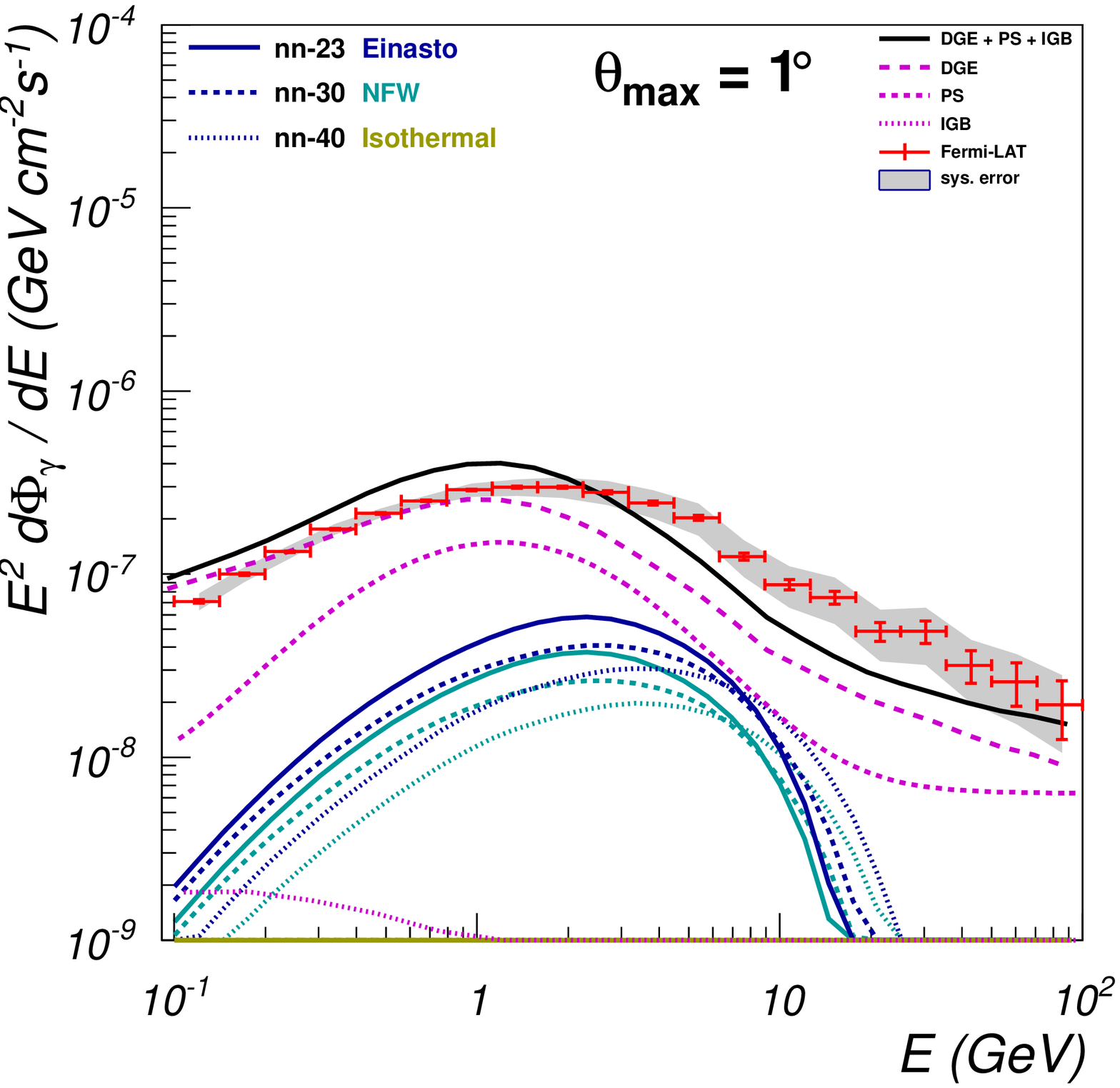,width=8.cm}
  \epsfig{file=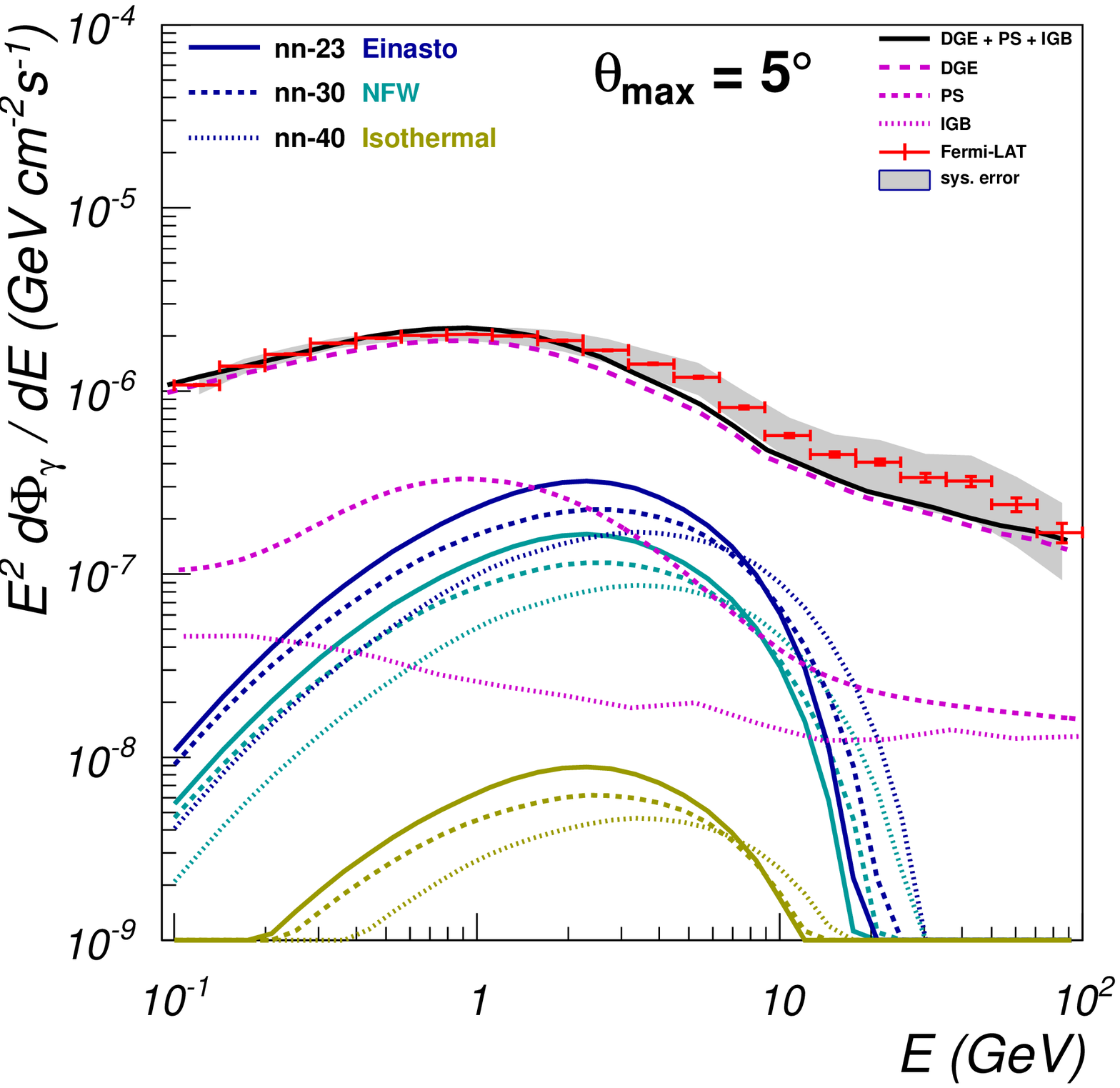,width=8.cm}
  \epsfig{file=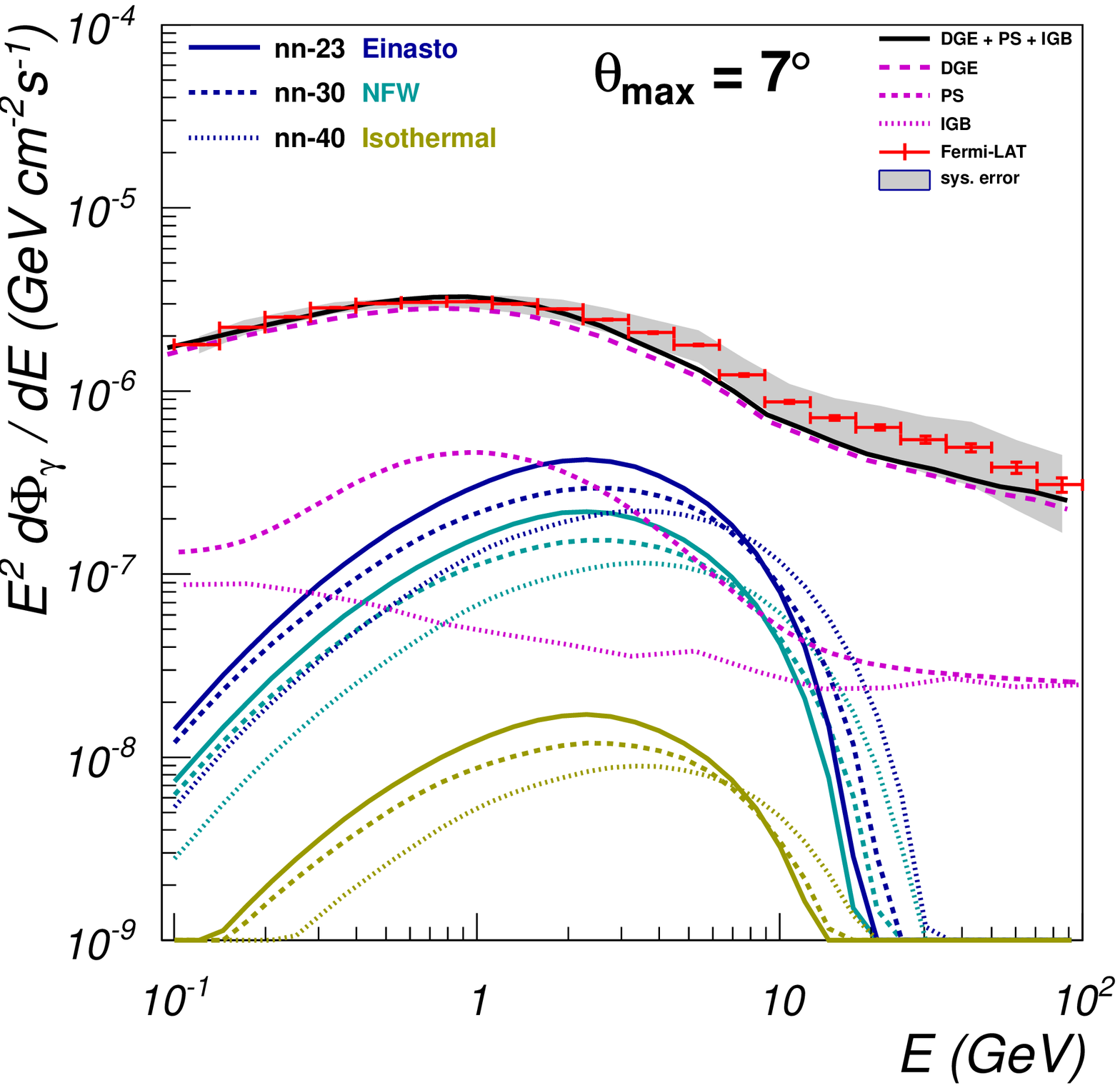,width=8.cm}
  \epsfig{file=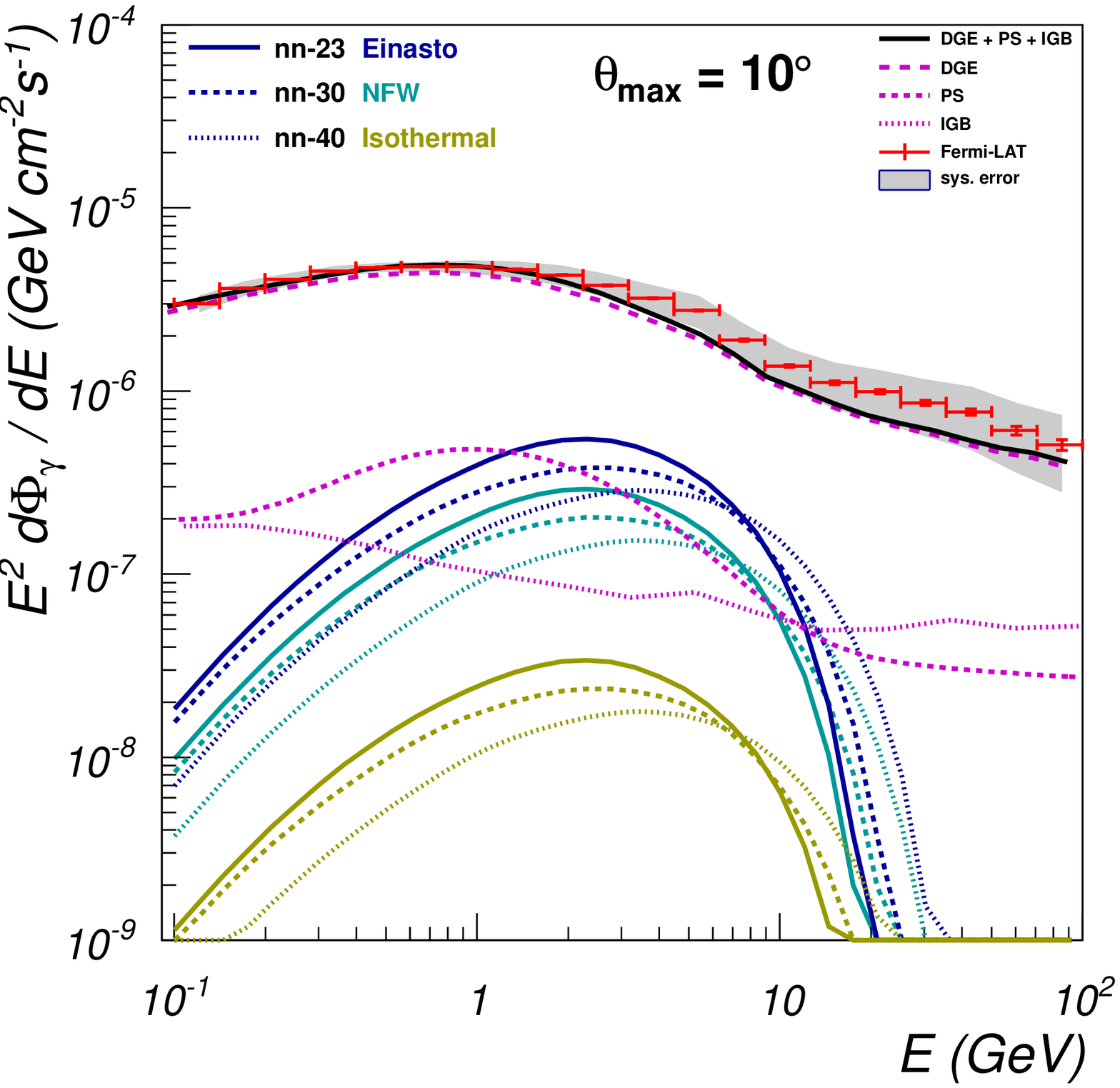,width=8.cm}
  \captions{Expected gamma ray flux in nn-23), nn-30) and nn-40). ROI with radii 1$^\circ$
  (top left), 5$^\circ$ (top right), 7$^\circ$ (bottom left) and 10$^\circ$
  (bottom right) are used.}
  \label{fig:fluxnn}
\end{figure}

\begin{figure}[t!]
  \epsfig{file=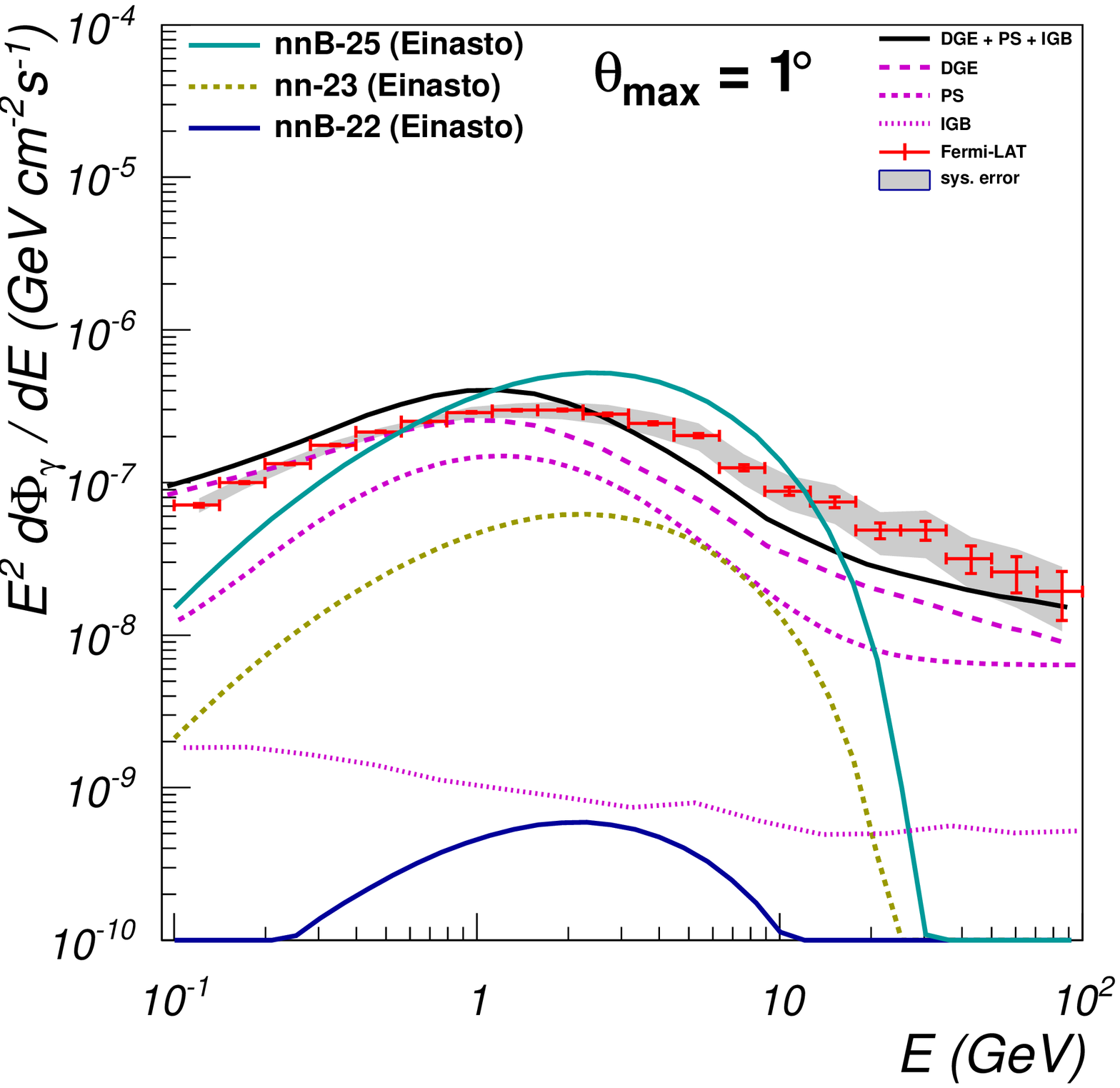,width=8.cm}
  \epsfig{file=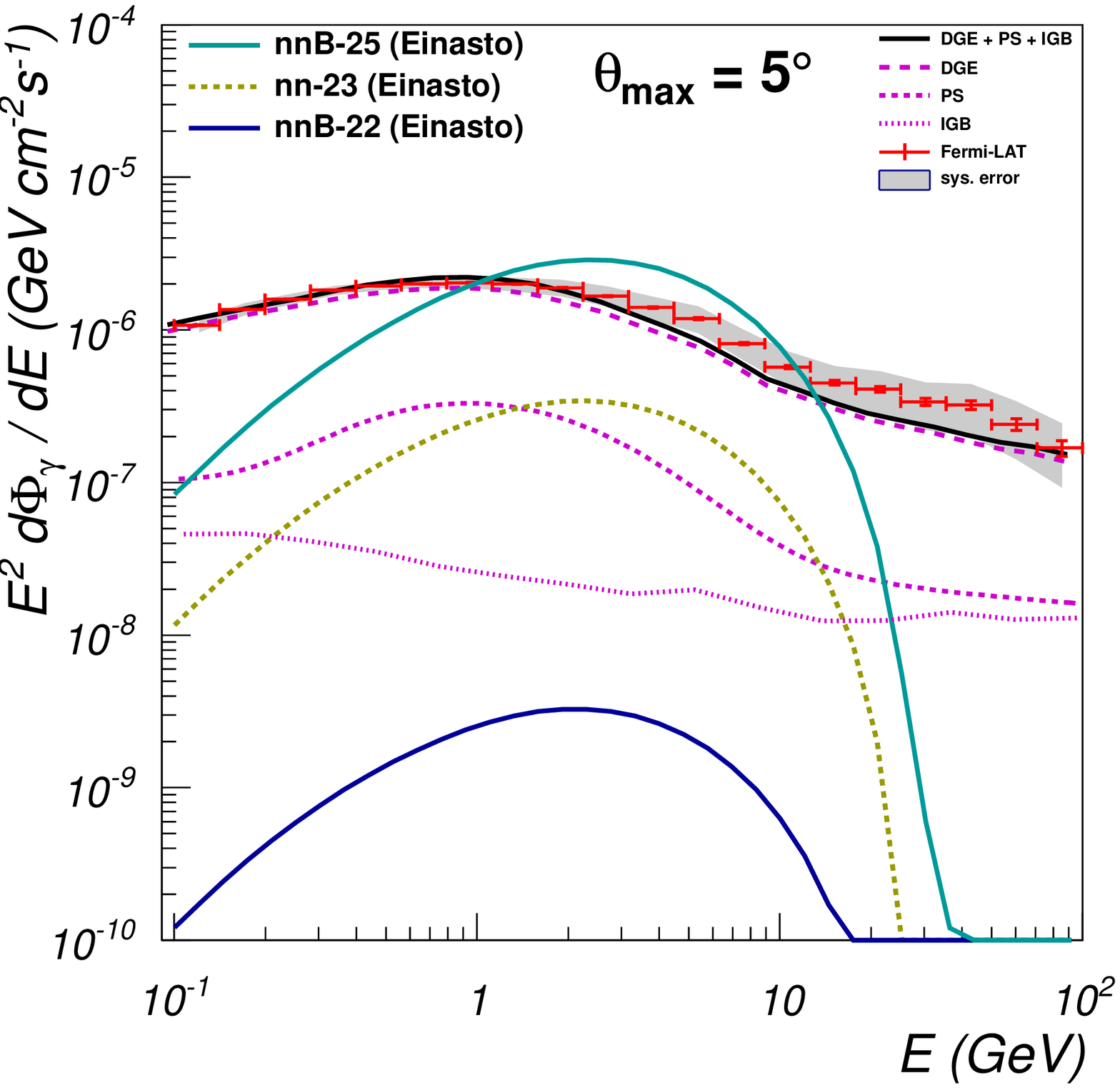,width=8.cm}
  \epsfig{file=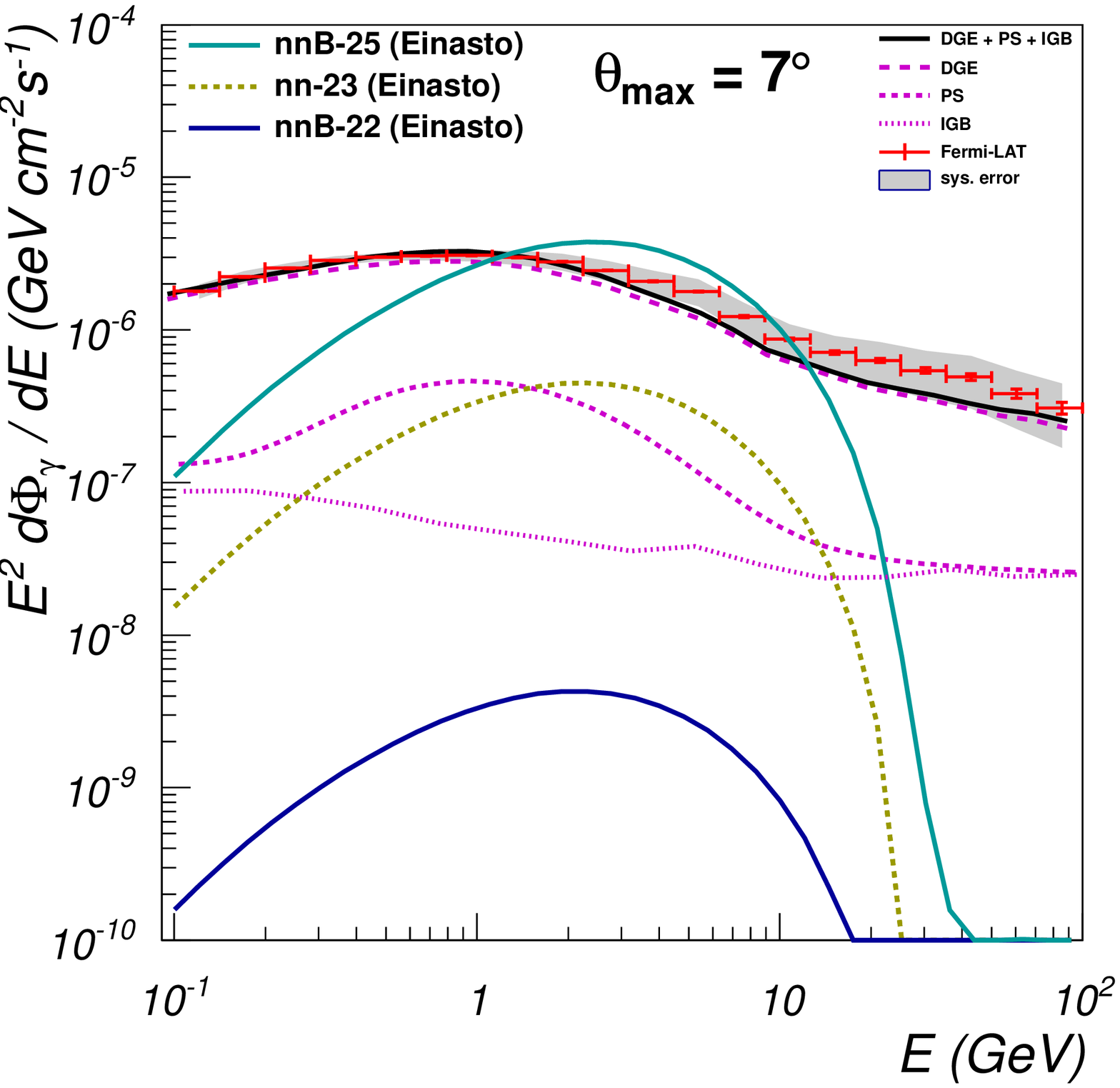,width=8.cm}
  \epsfig{file=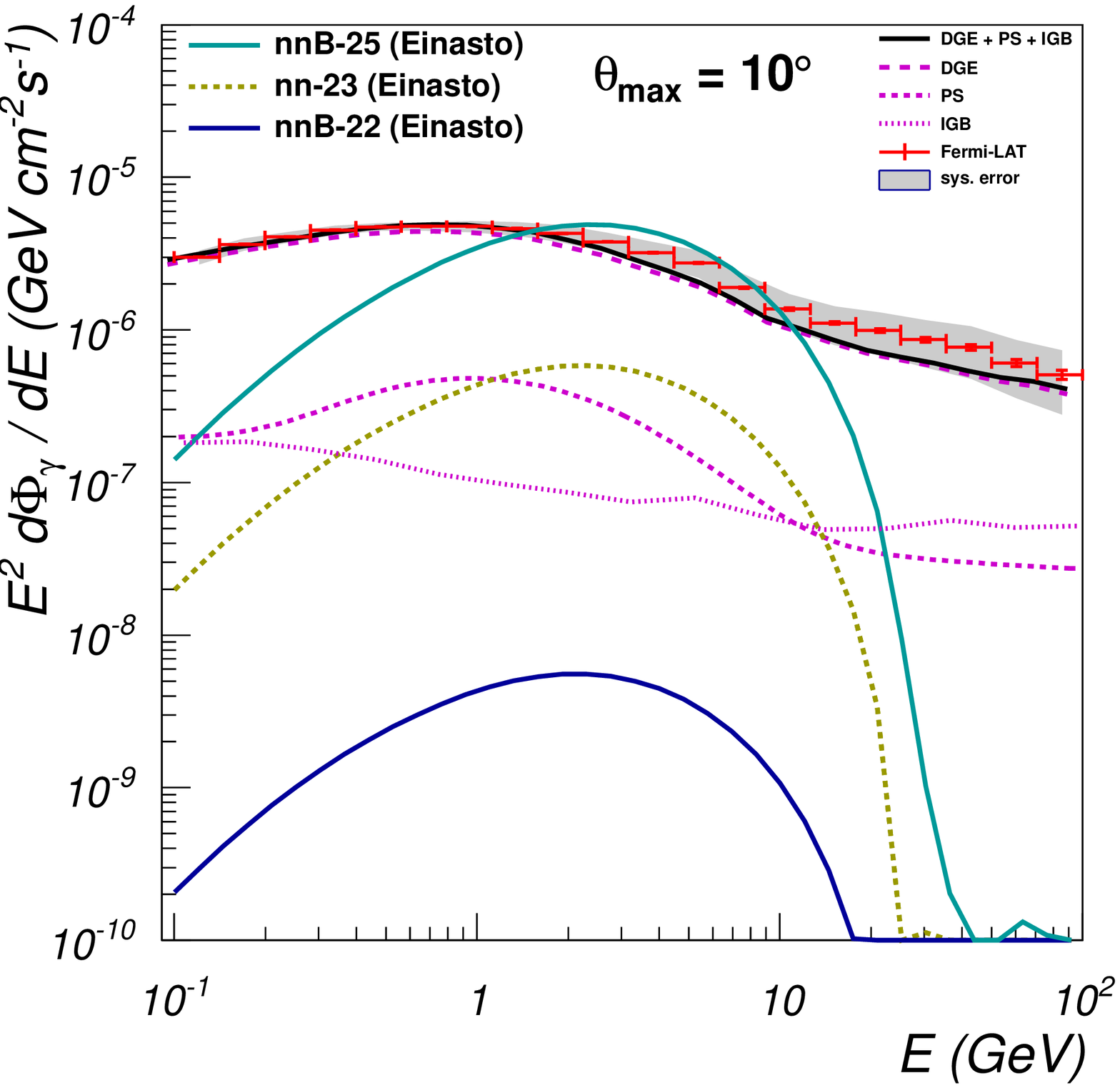,width=8.cm}
  \captions{Expected gamma ray flux in nn-23), nnB-22) and nnB-25). ROI with radii 1$^\circ$
  (top left), 5$^\circ$ (top right), 7$^\circ$ (bottom left) and 10$^\circ$
  (bottom right) are used.}
  \label{fig:fluxnnB}
\end{figure}

The corresponding predictions for the gamma ray flux are represented 
in Fig. \ref{fig:fluxnn} for benchmark points nn-23), nn-30) and nn-40).
We can see again that the expected gamma ray flux
is too small to constrain RH sneutrino annihilation in the halo in these cases
except for a ROI of  $1^\circ$. However, these gamma ray fluxes are
calculated under
the assumption that RH sneutrinos annihilate with fixed annihilation cross
section with value of $3\times10^{-26}$ cm$^3$/s, while Fig. \ref{fig:EF}
shows that the annihilation cross section in the halo can be at least one order of magnitude larger or 
smaller. Therefore, the fluxes shown in
Fig. \ref{fig:fluxnn} should be adjusted by the appropriate factors.

The impact of this Breit-Wigner enhancement is illustrated in Fig.\,\ref{fig:fluxnnB}, where two cases with the suppression factor of $10^{-2}$ for nnB-22)
and the enhancement factor of $10$ for nnB-25) are represented.
With the Einasto model, the expected fluxes of nnB-25) are
larger than the observed fluxes. In other words, the gamma ray flux
from the GC region observed by Fermi-LAT is already constraining
a portion of the parameter space of our model. Nevertheless, a thorough analysis
requires scanning over full parameter space of the model, and we leave it for a
future study.

\section{Conclusions}
\label{sec:conclusions}

In this paper we have investigated the viability of very light RH sneutrinos in the NMSSM and analysed the implications for direct dark matter detection, the potential effects on Higgs phenomenology and the prospects for indirect detection through gamma rays. 
The model contains a new singlet superfield that provides RH neutrinos and sneutrinos 
and three input parameters, a soft scalar mass $\mn$, a coupling constant $\ln$ and the associated trilinear parameter $\aln$.

First we have studied in detail the conditions under which RH sneutrinos in the NMSSM can be very light and reproduce the correct value for the relic abundance. We have found that this is possible in three different scenarios, namely when RH sneutrino annihilates mainly in fermions ($\tilde N \tilde N \to f\bar f$), in a pair of very light pseudoscalar Higgses ($\tilde N \tilde N \to A^0_1 A^0_1$), or in RH neutrinos ($\tilde N \tilde N \to NN$). 

The first case  ($\tilde N \tilde N \to f\bar f$) is possible
in the presence of a light (singlet-like) scalar Higgs or with
an increase of the coupling constant $\ln$. In both cases, the
diagrams contributing to the sneutrino annihilation in the
early Universe and those describing their scattering cross
section off quarks are correlated. We observe that this
correlation implies that if the correct relic density for RH
sneutrinos is imposed, their predictions for direct detection
are compatible with the WIMP interpretation of the CoGeNT results. We derive an analytical approximation to make this correlation explicit. 
This scenario has a deep impact in the predicted invisible decay width of the Higgs bosons. We show that the lightest Higgs is mostly invisible since it decays in a pair of RH sneutrinos. Interestingly, this applies to both a very light singlet-like Higgs and a SM-like Higgs with mass of order $114$~GeV. 
In fact, we also observe that the second lightest Higgs in this scenario can also have a significant invisible decay width, although whether or not the invisible modes dominate is much more dependent on the specific input parameters. 

The second case  ($\tilde N \tilde N \to A^0_1 A^0_1$) is possible when the pseudoscalar Higgs is light and singlet-like.
There is now no correlation between the annihilation cross section and scattering cross section off nuclei and as a consequence the theoretical predictions for RH sneutrino direct detection cannot account for the CoGeNT signal and are much smaller, of order $10^{-7}-10^{-10}$~pb. However, they are consistent with the exclusion regions set by the CDMS and XENON experiments. In some cases they could be within he reach of future experiments such as SuperCDMS. 
The resulting Higgs phenomenology is slightly different than in the previous scenario. 
The invisible decay width of the lightest Higgs is not necessarily large for the lightest Higgs (which is SM-like), but it is typically sizable for the second lightest Higgs (although much more dependent on the specific input parameters of the model).

The third possibility ($\tilde N \tilde N \to NN$) is more constrained but potentially very interesting. 
We described the stringent conditions under which this diagram can be dominant, which is only possible in the resonance with a light singlet-like Higgs with a very small $H_d$ component. 
Once more there is no correlation between the annihilation cross section and scattering cross section off nuclei and the resulting $\crosssec$ is very small, of order $10^{-8}-10^{-10}$ for RH sneutrinos with masses $\snmassr\gsim20$~GeV and featuring RH neutrinos with masses above $\rhnmass\gsim8$~GeV. Although it cannot explain the CoGeNT result, it would be compatible with the bounds set by CDMS and XENON.
This scenario leaves a potentially characteristic signal in colliders, namely the lightest Higgs decay into a RH neutrino pair whose subsequent (late) decay leaves two displaced vertices. 

Last we have investigated the theoretical predictions for the gamma ray flux from the GC and compared it with the results from the Fermi satellite. 
In this respect, the cases $\tilde N \tilde N \to f\bar f$ and
$\tilde N \tilde N \to A^0_1 A^0_1$ lead to the conventional
results obtained for other DM models.
Much more interesting is the new annihilation mode ($\tilde N \tilde N \to NN$), due to the subsequent decay of the RH neutrino.
We have characterised this new possibility finding that the
resulting spectrum is an intermediate case between those of
annihilation into $b\bar b$ and $\tau\bar\tau$.
Furthermore, since the annihilation occurs at the Higgs resonance, a Breit-Wigner enhancement of the resulting annihilation cross section in the dark matter halo can result in a boost factor as large as a factor 100 (or a suppression factor of several orders of magnitude).

\noindent {\bf Acknowledgements}

\noindent 
We have greatly benefited from conversations with B.~Ca\~nadas, J.~Fidalgo,
G.~G\'omez-Vargas, V.~Mart\'in Lozano and C.~B.~Park.
D.G.C. is supported by the Ram\'on y Cajal program of the Spanish MICINN. 
O.S. is partially supported by the scientific research grants from
Hokkai-Gakuen
and would like to thank the IFT for their hospitality in the early stages of this work.
This work was supported by the Spanish MICINNs  Consolider-Ingenio 2010 Programme under grant MultiDark CSD2009-00064. We also thank the support of the MICINN under grant FPA2009-08958, the Community of Madrid under grant HEPHACOS S2009/ESP-1473, and the European Union under the Marie Curie-ITN program PITN-GA-2009-237920.

\end{document}